\RequirePackage{fix-cm} 
\documentclass[a4paper, twoside, reqno, 12pt]{amsart}
\usepackage{fixltx2e}   

\usepackage{etex}

\usepackage[latin1]{inputenc}
\usepackage[T1]{fontenc}

\usepackage{esint}
\usepackage{dsfont}
\usepackage{xspace}
\usepackage{amsgen}
\usepackage{amsthm}
\usepackage{amssymb}
\usepackage{amsmath}
\usepackage{wasysym}
\usepackage{upgreek}
\usepackage{amsfonts}
\usepackage{stmaryrd}
\usepackage{mathtools}
\usepackage{textgreek}

\usepackage{cases}
\usepackage{graphics}
\usepackage{epsfig}

\usepackage{relsize}
\usepackage[mathcal, mathscr]{euscript}

\usepackage{mathrsfs}
\DeclareMathAlphabet{\mathscrbf}{OMS}{mdugm}{b}{n}

\usepackage{a4wide}

\usepackage[dvipsnames, table]{xcolor}
\definecolor{bckg}{RGB}{20.8, 20.8, 20.8}
\definecolor{oneblue}{rgb}{0.0, 0.0, 0.85}
\definecolor{Lightblue}{RGB}{214, 214, 214}
\definecolor{bluepigment}{rgb}{0.2, 0.2, 0.6}
\definecolor{charcoal}{rgb}{0.21, 0.27, 0.31}
\definecolor{denimblue}{rgb}{0.08, 0.38, 0.74}
\definecolor{Lightgray}{rgb}{0.89, 0.89, 0.89}
\definecolor{darkgrey}{rgb}{0.273, 0.281, 0.30}
\definecolor{darkelectricblue}{rgb}{0.33, 0.41, 0.47}

\usepackage[sort&compress, comma, square, numbers]{natbib}

\usepackage{psfrag}
\usepackage{graphicx}
\usepackage{subfigure}
\usepackage{morefloats}
\usepackage{indentfirst}

\usepackage{acronym}
\usepackage{microtype}
\usepackage[labelsep=period,%
            labelfont={bf,sf,color=bluepigment},%
            justification=raggedright]{caption}

\usepackage[perpage, symbol]{footmisc}

\usepackage[usenames, dvipsnames, pdf]{pstricks}
\usepackage{epsfig}
\usepackage{pst-grad} 
\usepackage{pst-plot} 

\usepackage[colorlinks,
           urlcolor=oneblue,
           linkcolor=denimblue,
           citecolor=NavyBlue,
           bookmarksopen=false,
           pdfpagemode=UseNone,
           pagebackref]{hyperref}

\usepackage{algorithm}
\usepackage[]{algpseudocode}

\usepackage[explicit]{titlesec}

\titleformat{\section}[block]
  {\color{NavyBlue}\Large\sffamily\bfseries}
  {}
  {0.0em}
  {\colorbox{bckg!5}{\strut\parbox{\dimexpr\linewidth-2\fboxsep\relax}{\thesection. #1}}}
  [\vspace*{0.33em}]

\titleformat{name=\section,numberless}[block]
  {\color{NavyBlue}\Large\sffamily\bfseries}
  {}
  {0.0em}
  {\colorbox{bckg!5}{\strut\parbox{\dimexpr\linewidth-2\fboxsep\relax}{#1}}}
  [\vspace*{0.33em}]

\titleformat{\subsection}
  {\color{NavyBlue}\large\sffamily\bfseries}
  {}
  {0.0em}
  {\colorbox{bckg!5}{\parbox{\dimexpr\linewidth-2\fboxsep\relax}{\thesubsection. #1}}}
  [\vspace*{0.33em}]

\titleformat{name=\subsection,numberless}
  {\color{NavyBlue}\Large\sffamily\bfseries}
  {}
  {0em}
  {\colorbox{bckg!5}{\parbox{\dimexpr\linewidth-2\fboxsep\relax}{#1}}}
  [\vspace*{0.33em}]

\titleformat{\subsubsection}
  {\color{bluepigment}\sffamily\normalsize\bfseries}
  {\thesubsubsection}
  {0.5em}
  {#1}
  [\vspace*{0.33em}]

\titleformat{\paragraph}[runin]
  {\color{bluepigment}\sffamily\small\bfseries}
  {}
  {0em}
  {#1}

\titlespacing{\section}{0.0em}{1.5em plus 2pt minus 2pt}%
{1.0em plus 2pt minus 2pt}[0em]
\titlespacing{\subsection}{0.5em}{1.5em plus 2pt minus 2pt}%
{1.0em}[0em]
\titlespacing{\subsubsection}{0.5em}{1.5em plus 2pt minus 2pt}%
{1.0em plus 2pt minus 2pt}[0em]

\usepackage{titletoc}

\contentsmargin{0.5em}
\newlength{\tocsep} 
\titlecontents{section}[\tocsep]
  {\addvspace{10pt}\bfseries\sffamily}
  {\contentslabel[\thecontentslabel]{\tocsep}}
  {}
  {\ \titlerule*[0.75pc]{.}\ \thecontentspage}
  []
\titlecontents{subsection}[\tocsep]
  {\addvspace{8pt}\sffamily}
  {\contentslabel[\thecontentslabel]{\tocsep}}
  {}
  {\ \titlerule*[0.5pc]{.}\ \thecontentspage}
  []
\titlecontents*{subsubsection}[\tocsep]
  {\addvspace{2pt}\footnotesize\sffamily}
  {}
  {}
  {\ \titlerule*[0.35pc]{.}\ \thecontentspage}
  [\\*]

\makeatletter
\def\@setauthors{%
  \begingroup
  \def\thanks{\protect\thanks@warning}%
  \trivlist
  \centering\footnotesize \@topsep30\p@\relax
  \advance\@topsep by -\baselineskip
  \item\relax
  \author@andify\authors
  \def\\{\protect\linebreak}%
  \textsc{\normalsize\textcolor{darkelectricblue}{\authors}}%
  \ifx\@empty\contribs
  \else
    ,\penalty-3 \space \@setcontribs
    \@closetoccontribs
  \fi
  \endtrivlist
  \endgroup
}
\def\@settitle{\begin{center}%
  \baselineskip14\p@\relax
    \bfseries
    \textsc{\Large\textcolor{charcoal}{\@title}}
  \end{center}%
}
\makeatother

\usepackage{enumitem}
\setlist[description]{%
  topsep=30pt,               
  itemsep=5pt,               
  font={\bfseries\sffamily\color{NavyBlue}}, 
}

\usepackage{fancyhdr}
\usepackage{lastpage}

\newcommand*\Title{\textcolor{bluepigment}{Relaxed formulations for dispersive wave equations}}
\newcommand*\Authors{\textcolor{bluepigment}{J.-P.~Chehab \& D.~Dutykh}}
\newcommand*{\plogo}{\textcolor{gray}{{\texttt{arXiv.org} / \textsc{hal}}}} 

\numberwithin{equation}{section}

\newtheorem{remark}{Remark}

\newcommand{\up}[1]{$^{\mathrm{\small\textsf{#1}}}$} 

\newcommand{\R}{\mathds{R}}
\newcommand{\D}{\mathcal{D}}
\newcommand{\Id}{\mathds{I}}
\newcommand{\M}{\mathcal{M}}

\newcommand{\ue}{\mathrm{e}}
\newcommand{\ui}{\mathrm{i}}

\renewcommand{\O}{\mathcal{O}}

\renewcommand{\eta}{\text{\texteta}}
\renewcommand{\beta}{\text{\textbeta}}
\renewcommand{\alpha}{\text{\textalpha}}

\renewcommand{\Re}{\operatorname{Re}}

\newcommand{\ie}{\emph{i.e.}\xspace}
\newcommand{\eg}{\emph{e.g.}\xspace}

\newcommand{\sech}{\mathrm{sech}}

\newcommand{\od}[2]{\frac{\mathrm{d}\/ #1}{\mathrm{d}\/#2}}

\newcommand{\eqdef}{\mathop{\stackrel{\,\mathrm{def}}{:=}\,}}

\newcommand{\half}{{\textstyle{1\over2}}}
\newcommand{\third}{{\textstyle{1\over3}}}

\newcommand{\fiveighth}{{\textstyle{5\over8}}}

\usepackage{acronym}
\acrodef{cs}[CS]{Compact Schemes}
\acrodef{kdv}[KdV]{\textsc{Korteweg}--\textsc{de Vries}}
\acrodef{sgn}[SGN]{\textsc{Serre}--\textsc{Green}--\textsc{Naghdi}}
\acrodef{bbm}[BBM]{\textsc{Benjamin}--\textsc{Bona}--\textsc{Mahony}}
\acrodef{cfl}[CFL]{\textsc{Courant}--\textsc{Friedrichs}--\textsc{Lewy}}

\begin{document}

\title[\Title]{On time relaxed schemes and formulations for dispersive wave equations}

\author[J.-P. Chehab]{Jean-Paul Chehab}
\address{\textbf{J.-P. Chehab:} Universit\'e de Picardie Jules Verne, LAMFA CNRS UMR 7352, 33, rue Saint-Leu, 80039 Amiens, France}
\email{Jean-Paul.Chehab@u-picardie.fr}
\urladdr{http://www.lamfa.u-picardie.fr/chehab/\vspace*{0.5em}}

\author[D.~Dutykh]{Denys Dutykh$^{\,\star}$}
\address{\textbf{D.~Dutykh:} Univ. Grenoble Alpes, Univ. Savoie Mont Blanc, CNRS, LAMA, 73000 Chamb\'ery, France and LAMA, UMR 5127 CNRS, Universit\'e Savoie Mont Blanc, Campus Scientifique, 73376 Le Bourget-du-Lac Cedex, France}
\email{Denys.Dutykh@univ-savoie.fr}
\urladdr{http://www.denys-dutykh.com/}
\thanks{$^*$ Corresponding author}



\begin{titlepage}
\thispagestyle{empty} 
\noindent
{\Large Jean-Paul \textsc{Chehab}}\\
{\it\textcolor{gray}{LAMFA, Universit\'e de Picardie Jules Verne, France}}
\\[0.02\textheight]
{\Large Denys \textsc{Dutykh}}\\
{\it\textcolor{gray}{CNRS--LAMA, Universit\'e Savoie Mont Blanc, France}}
\\[0.08\textheight]

\vspace*{1.5cm}

\colorbox{Lightblue}{
  \parbox[t]{1.0\textwidth}{
    \centering\huge\sc
    \vspace*{0.7cm}
    
    \textcolor{bluepigment}{On time relaxed schemes and formulations for dispersive wave equations}
    
    \vspace*{0.7cm}
  }
}

\vfill 

\raggedleft     
{\large \plogo} 
\end{titlepage}


\newpage
\thispagestyle{empty} 
\par\vspace*{\fill}   
\begin{flushright} 
{\textcolor{denimblue}{\textsc{Last modified:}} \today}
\end{flushright}


\newpage
\maketitle
\thispagestyle{empty}


\begin{abstract}

The numerical simulation of nonlinear dispersive waves is a central research topic of many investigations in the nonlinear wave community. Simple and robust solvers are needed for numerical studies of water waves as well. The main difficulties arise in the numerical approximation of high order derivatives and in severe stability restrictions on the time step, when explicit schemes are used. In this study we propose new relaxed system formulations which approximate the initial dispersive wave equation. However, the resulting relaxed system involves first order derivatives only and it is written in the form of an evolution problem. Thus, many standard methods can be applied to solve the relaxed problem numerically. In this article we illustrate the application of the new relaxed scheme on the classical \textsc{Korteweg}--\textsc{de Vries} equation as a prototype of stiff dispersive PDEs.


\bigskip
\noindent \textbf{\keywordsname:} dispersive wave equations; shallow water flows; relaxation; quasi-compressibility. \\


\end{abstract}


\newpage
\tableofcontents
\thispagestyle{empty}


\newpage
\section{Introduction}

Nonlinear dispersive waves arise in various fields of science including the solid \cite{Porubov2006} and fluid mechanics \cite{Johnson1997}. Below we focus on models stemming from the free surface hydrodynamics mainly due to scientific interests of the authors. In this field, analytical solutions (such as solitary or cnoidal waves \cite{Clamond2003, Chen2007}) are available only for some simplified models. That is why the numerical simulation remains one of the main tools in nonlinear dispersive wave studies. In this work the authors have an ambition to propose a novel strategy to tackle numerically these problems.

Let us review first the modern numerical approaches to this problem. Historically, the finite differences were applied to dispersive wave problems \cite{Ismail2000}. Today, the trend is to apply \textsc{Galerkin}--type methods (\eg FEM) in smooth situations \cite{Mitsotakis2014, Antonopoulos2010} and finite volumes \cite{Dutykh2010e, Dutykh2011e} or coupled finite volumes (for the hyperbolic part)/finite differences (for the dispersive part) via the operator splitting \cite{Bonneton2011}. Notice that there is an effort to develop discontinuous \textsc{Galerkin}--type methods for dispersive wave equations as well \cite{Eskilsson2005, YS}, which combine the robustness of finite volumes with the accuracy of finite elements. However, the `price to pay' (\ie the CPU time) remains excessively high turning these methods into a luxury limousine rather than a working horse of numerical analysis. Finally, for periodic situations one can apply highly efficient \textsc{Fourier}-type pseudo-spectral methods \cite{Dutykh2011a}. On the borderline between finite differences and pseudo-spectral methods there exist compact finite difference schemes proposed by \textsc{Lele} (1992) with spectral-like resolution. To give an example, these schemes were applied to the \textsc{Serre} equations in \cite{CBB1}.

In the present study we propose to change the numerical strategy in contrast to studies described above. The main idea consists in modifying the governing equations by slightly perturbing them with some \emph{ad-hoc} terms. In this way, we gain the structure suitable for numerical simulations and we hope that solutions of the perturbed and un-perturbed problems will remain close provided that perturbation parameter is small. This idea is inspired by pseudo-compressibility methods proposed for the numerical simulation of incompressible \textsc{Navier}--\textsc{Stokes} equations (see \eg \cite{Kameyama2005}). The main philosophy of this approach is as follows. When we solve numerically a differential equation, some errors are introduced due to the underlying discretization process. So, perhaps, for the sake of convenience, one can perturb also the governing equations within the same discretization error which is already introduced. And if everything is done judiciously, the end user will not even see the difference between the relaxed and the original problems solutions. It makes a lot of `ifs', but nevertheless we undertake this programme below for some well-known dispersive wave equations. We would like to mention here also another relaxation technique based on the application of local time-averaging operators \cite{Antuono2009}. A relaxation scheme for the KdV equation was proposed in \cite{Benkhaldoun2008}. However, their formulation contains 2\up{nd} order derivatives. Below, we will propose an alternative formulation which has an advantage to involve only the 1\up{st} order derivatives. The idea we employ can be traced back at least to various \emph{local}\footnote{In the case of continuous \textsc{Galerkin} methods this formulation is also sometimes called the \emph{modified} one.} formulations used in continuous \cite{Walkley2002, Walkley1999} and discontinuous \cite{Levy2004, YS} \textsc{Galerkin} methods. However, we push it one step further towards the so-called relaxed local formulations.

The present study is organized as follows. In Section~\ref{sec:models} we propose the relaxed formulations for several well-known dispersive wave models. Some mathematical properties of a relaxed formulation for the \acs{kdv} equation are discussed in Section~\ref{sec:prop}. Our numerical approach based on this relaxed formulation of the \acs{kdv} equation is outlined in Section~\ref{sec:num}. The first numerical results are presented in the same Section. Finally, in Section~\ref{sec:disc} we outline the main conclusions and perspectives of our study. The relaxation of some other dispersive wave equations is discussed in Appendix~\ref{app:a}.


\section{Mathematical models}
\label{sec:models}

Instead of working out the most general situation, in the present study we follow the so-called \textsc{Gelfand}\footnote{Here we mean I.~M.~\textsc{Gelfand} to avoid any confusion.} principle which states that a theory should be illustrated on the simplest non-trivial example. Below we consider one such example and two others are treated in Appendix~\ref{app:a}. All cases which steem from the water wave theory \cite{Stoker1957}. For each model we develop the corresponding relaxed formulation.


\subsection{Korteweg--de Vries equation}

The celebrated \acf{kdv} equation was derived independently by J.~\textsc{Boussinesq} \cite{Boussinesq1877} and D.~\textsc{Korteweg} \& G.~\textsc{de Vries} \cite{KdV} at the end of the XIX\up{th} century and in the scaled form it reads:
\begin{equation}\label{eq:kdv}
  u_{\,t}\ +\ u\,u_{\,x}\ +\ u_{\,xxx}\ =\ 0\,.
\end{equation}
The variable $u\,(x,\,t)$ may be interpreted physically as the free surface elevation or the horizontal depth-integrated fluid velocity. There exist transformations in order to obtain the canonical form \eqref{eq:kdv} \cite{Johnson1997}. We shall rewrite the last equation in the conservative form in order to obtain the flux of the evolution variable $u\,$:
\begin{equation}\label{eq:kdvcons}
  u_{\,t}\ +\ \bigl(\half\, u^{\,2}\ +\ u_{\,xx}\bigr)_x\ =\ 0\,.
\end{equation}
Now, the order of derivatives has to be lowered. It can be done similarly to the ODE theory by introducing additional variables:
\begin{align*}
  u_{\,t}\ +\ \bigl(\half\, u^{\,2}\ +\ w\bigr)_x\ &=\ 0\,, \\
  u_{\,x}\ &=\ v\,, \\
  v_{\,x}\ &=\ w\,. \\
\end{align*}
At this step we are compatible with local formulations used in the framework of discontinuous \textsc{Galerkin} methods \cite{Levy2004, YS}. It is obvious that the last system is completely equivalent to equation \eqref{eq:kdvcons} (at least for the smooth solutions). However, the last system is not of the evolution type. It is here that the relaxation comes into the play in order to correct this shortcoming:
\begin{align}\label{eq:kdv1}
  u_{\,t}\ +\ \bigl(\half\, u^{\,2}\ +\ w\bigr)_x\ &=\ 0\,, \\
  \delta\, v_{\,t}\ +\ v\ -\ u_{\,x}\ &=\ 0\,, \label{eq:kdv2} \\
  \delta\, w_{\,t}\ +\ w\ -\ v_{\,x}\ &=\ 0\,, \label{eq:kdv3}
\end{align}
where $\delta\ \ll\ 1$ is a small parameter. It can be seen that in the limit $\delta\ \to\ 0$ we recover the original formulation \eqref{eq:kdv} (at least formally).

Equations \eqref{eq:kdv2}, \eqref{eq:kdv3} allow also to assess the physical meaning of the parameter $\delta\,$. Indeed, the terms $v$ and $\delta\,v_{\,t}$ (and $w$ with $\delta\,w_{\,t}$) need to have the same units. Thus, the multiplication by $\delta$ has to compensate the derivative with respect to time. Henceforth, from dimensional arguments, $\delta$ has the unit of time.


\section{Properties}
\label{sec:prop}

Above we presented relaxed formulations for three well-known equations. However, the properties of new formulations have to be studied. From now on we focus on the \acs{kdv} equation only \eqref{eq:kdv1} -- \eqref{eq:kdv3} in order to understand the relaxation effects exhaustively. Other equations will be tackled in following studies.


\subsection{Dispersion relation analysis}

In order to study the dispersive properties, first we have to linearize equations \eqref{eq:kdv1} -- \eqref{eq:kdv3}:
\begin{align}\label{eq:kdv4}
  u_{\,t}\ +\ w_{\,x}\ &=\ 0\,, \\
  \delta\, v_{\,t}\ +\ v\ -\ u_{\,x}\ &=\ 0\,, \\
  \delta\, w_{\,t}\ +\ w\ -\ v_{\,x}\ &=\ 0\,. \label{eq:kdv6}
\end{align}
Now, we look for plane-wave solutions of the form
\begin{equation*}
  u\,(x,\,t)\ =\ u_{\,0}\cdot\ue^{\,\ui\,(k\,x\ -\ \omega\,t)}\,, \quad
  v\,(x,\,t)\ =\ v_{\,0}\cdot\ue^{\,\ui\,(k\,x\ -\ \omega\,t)}\,, \quad
  w\,(x,\,t)\ =\ w_{\,0}\cdot\ue^{\,\ui\,(k\,x\ -\ \omega\,t)}\,,
\end{equation*}
where $k$ is the wavenumber, $\omega$ is the frequency and $\bigl\{u_{\,0},\,v_{\,0},\,w_{\,0}\bigr\}\ \in\ \R$ are some real amplitudes. By substituting\footnote{This operation can be seen as the \textsc{Fourier} transform in both space and time.} the plane-wave ansatz\"a into a linear system \eqref{eq:kdv4} -- \eqref{eq:kdv6}, we obtain a system of linear algebraic equations, where unknowns are $\bigl\{u_{\,0},\,v_{\,0},\,w_{\,0}\bigr\}\,$:
\begin{align*}
  -\ui\,\omega\,u_{\,0}\ +\ \ui\,k\,w_{\,0}\ &=\ 0\,, \\
  -\ui\,\delta\,\omega\,v_{\,0}\ +\ v_{\,0}\ -\ \ui\,k\,u_{\,0}\ &=\ 0\,, \\
  -\ui\,\delta\,\omega\,w_{\,0}\ +\ w_{\,0}\ -\ \ui\,k\,v_{\,0}\ &=\ 0\,. \\
\end{align*}
The necessary condition to have non-trivial solutions to the above linear system of equations is that its determinant is equal to zero:
\begin{equation*}
  \det\begin{vmatrix}
    \,-\,\ui\,\omega & 0 & \ui\,k\, \\
    \,-\,\ui\,k & 1\ -\ \ui\,\delta\,\omega & 0\, \\
    \,0 & -\,\ui\,k & 1\ -\ \ui\,\delta\,\omega\,
  \end{vmatrix}\ =\ 0\,.
\end{equation*}
After some simplifications we obtain the desired \emph{dispersion relation} between $\omega$ and $k\,$:
\begin{equation}\label{eq:drel}
  \omega\ +\ k^{\,3}\ -\ 2\,\ui\,\delta\,\omega^{\,2}\ -\ \delta^{\,2}\,\omega^{\,3}\ =\ 0\,.
\end{equation}
So, one can see that by taking the limit $\delta\ \to\ 0$ in the last equation, we recover the classical \acs{kdv} dispersion relation $\omega\ =\ -\,k^{\,3}\,$. This polynomial equation \eqref{eq:drel} in wave frequency $\omega$ admits in general three complex roots. Their general expression is rather cumbersome. However, we can construct their asymptotic expansions:
\begin{align}
  \omega_{\,1}\,(k)\ &=\ -\,k^{\,3}\ +\ 2\,\ui\,\delta\,k^{\,6}\ +\ 7\,k^{\,9}\,\delta^{\,2}\ +\ \O\,(\delta^{\,3})\,,\label{eq:b1} \\
  \omega_{\,2,\,3}\,(k)\ &=\ -\;\frac{\ui}{\delta}\ \pm\ \ui\,\sqrt{-\,\ui}\;\frac{k^{\,3/2}}{\sqrt{\delta}}\ +\ \half\,k^{\,3}\ \pm\ \fiveighth\,\sqrt{-\,\ui}\,k^{\,9/2}\,\sqrt{\delta}\ -\ \ui\,k^{\,6}\,\delta\ +\ \O\,(\delta^{\,3/2})\,.\label{eq:b23}
\end{align}
These expansions \eqref{eq:b1}, \eqref{eq:b23} might be used to interpret the behaviour of numerical solutions. Since we deal with singular perturbations in equation~\eqref{eq:drel}, the passage to the limit $\delta\ \to\ 0$ in solutions $\omega_{\,2,\,3}\,(k)$ is far from being trivial. However, from the expansion \eqref{eq:b1} we learn that parameter $\delta$ plays the both r\^oles of dissipation and dispersion since it enters into real and imaginary parts of the regular branch $\omega_{\,1}\,(k)\,$. We would like to highlight the fact that the relaxation parameter $\delta$ affects the dispersive properties of the regular branch (\ie $\Re\omega_{\,1}\,(k)$) only to the second order in $\delta\,$. Thus, we can conclude that the proposed relaxation approach is very careful with respect to the dispersion relation of the original \acs{kdv} equation.

\begin{remark}
Taking the limit in the polynomial equation \eqref{eq:drel} turns out to be an easy task, as it often happens, than taking the limit $\delta\ \to\ 0$ in the solutions $\omega_{\,2,\,3}\,(k)\,$. On the other hand, the branch $\omega_{\,1}\,(k)$ seems to be completely regular from this point of view. Two other branches $\omega_{\,2,\,3}\,(k)$ are artificial modes introduced by relaxation.
\end{remark}


\section{Numerical methods and results}
\label{sec:num}

In this Section we are going to study numerically the relaxed formulations. For this purpose we propose three different discretisations of the \acs{kdv} equation. One of them is based on the relaxed formulation proposed above and two others are popular finite difference schemes (\textsc{Crank}--\textsc{Nicolson} and \textsc{Sanz}--\textsc{Serna}) based on the original \acs{kdv} equation. The \emph{numerical} advantages of the relaxed formulation are illustrated below.

Before considering the time schemes, we focus on the discretization in space which is realized with finite differences. The use of compact schemes as presented in \cite{Lele1992} allows to reach a high order of accuracy, with a spectral-like resolution, while preserving the capabilities of the finite differences approach, including the implementation on a general \textsc{Cartesian} grids and for a variety of boundary conditions. When dealing with surface wave equations, the high level of accuracy is an important property to capture correctly the active frequencies of the solution without working on very fine grids. Hence, the computation can be done with a reasonable computational time.

At first we recall briefly the principle of the \acf{cs} and we give the discretization matrices that will be used for the simulation of the models presented in Section~\ref{sec:models}.

In two words, the compact schemes consist in approaching a linear operator (differentiation as well as interpolation) by a rational (instead of polynomial-like) finite differences scheme. These schemes are then implicit, this allows to increase the accuracy and mimic the spectral global dependence, see \cite{Lele1992} for more details. The approximations at grid points of the operators applied to a regular function $u$ is realized as follows.

Let $U\ \eqdef\ (U_{\,1},\,\ldots,\,U_{\,N})^{\,\top}$ be a vector whose the components are the approximations $u$ at (regularly spaced) grid points $x_{\,i}\ =\ i\,h\,$, $i\ =\ 1,\,\ldots,\,N\,$, here $h\ =\ \dfrac{1}{N}$ is the spatial step-size. We compute approximations of $V_{\,i}\ =\ \mathcal{L}\,(u)\,(x_{\,i})$ as solution of a system
\begin{equation*}
  P \cdot V\ =\ Q\, U\,,
\end{equation*}
the approximation matrix is then formally $B\ =\ P^{\,-1}\cdot Q\,$. For simplicity we present here fourth order accurate \acs{cs}. When considering periodic boundary conditions, the matrices $P$ and $Q$ are for the first order derivative in space:
\begin{equation*}
  P\ \eqdef\ \begin{pmatrix}
    1 & \frac{1}{4} &   &   \frac{1}{4}\\ 
    \frac{1}{4} & 1 & \ddots &  \\ 
    & \ddots & \ddots & \frac{1}{4} \\ 
    \frac{1}{4}  &   & \frac{1}{4} & 1
  \end{pmatrix}\,, \qquad
  Q\ \eqdef\ \dfrac{1}{2\,h}\;\begin{pmatrix}
    0 &  \frac{3}{2}  & &  &  -\frac{3}{2}   \\ 
    -\frac{3}{2} & 0 & \frac{3}{2} &   &   \\ 
    & \ddots & \ddots & \ddots &   \\ 
    &   & -\frac{3}{2} & 0 & \frac{3}{2} \\ 
    \frac{3}{2}   &  & & -\frac{3}{2} & 0
  \end{pmatrix}\,,
\end{equation*}
For the second order derivative in space we have the following pair of matrices:
\begin{equation*}
  P\ =\ \begin{pmatrix}
    1 & \frac{1}{10} &   &   &  \frac{1}{10}  \\ 
    \frac{1}{10} & 1 & \frac{1}{10} &   &   \\ 
    & \ddots & \ddots & \ddots &   \\ 
    &  & \frac{1}{10} & 1 & \frac{1}{10} \\
    \frac{1}{10}  &  &  & \frac{1}{10} & 1
  \end{pmatrix}\,, \qquad
  Q\ =\ \frac{1}{h^{\,2}}\;\begin{pmatrix}
    \frac{12}{5} & -\frac{6}{5} & & & &   & -\frac{6}{5}  \\ 
    -\frac{6}{5} & \frac{12}{5} & -\frac{6}{5} &   &   &   &   \\ 
    & -\frac{6}{5} & \frac{12}{5} & -\frac{6}{5} &   &   &   \\ 
    &   & \ddots & \ddots & \ddots &   &   \\ 
    &   &   & -\frac{6}{5} & \frac{12}{5} & -\frac{6}{5} &   \\ 
    &   &   &   & -\frac{6}{5} & \frac{12}{5} & -\frac{6}{5} \\ 
    -\frac{6}{5}&   & & & & -\frac{6}{5} & \frac{12}{5}
  \end{pmatrix}\,.
\end{equation*}
Finally, for the third order derivative in space we have
\begin{equation*}
  P\ =\ \begin{pmatrix}
    1 & \frac{1}{2} &   &   &  \frac{1}{2}  \\ 
    \frac{1}{2} & 1 & \frac{1}{2} &   &   \\ 
    & \ddots & \ddots & \ddots &   \\ 
    &  & \frac{1}{2} & 1 & \frac{1}{2} \\ 
    \frac{1}{2}  &  &  & \frac{1}{2} & 1
  \end{pmatrix}\,, \qquad
  Q\ =\ \frac{2}{h^{\,3}}\;\begin{pmatrix}
    0 &  -1 & \frac{1}{2} & & & &  -\frac{1}{2} & 1\\ 
    1 & 0 &  -1 & \frac{1}{2} & & & &  -\frac{1}{2} \\
    &  & \ddots & \ddots & \ddots & &  \\ 
    \frac{1}{2}   &&&  -\frac{1}{2}  & 1 & 0 & -1\\ 
    -1&\frac{1}{2}    & -\frac{1}{2} & 1 & 0
  \end{pmatrix}\,.
\end{equation*}

We now present the semi-implicit relaxed time scheme we will use for the simulation, stressing out its enhanced stability as compared to classical semi-implicit schemes (\textsc{Crank}--\textsc{Nicolson}'s for the linear terms and forward \textsc{Euler}'s for the nonlinear ones) while giving comparable results to the ones obtained by fully nonlinear \textsc{Sanz}--\textsc{Serna}'s which is second order accurate in time and unconditionally stable.

Let $\D_{\,x}\,$, $\D_{\,xx}$ and $\D_{\,xxx}$ be the discretization matrices of the first, second and third order derivative on a $N$ regularly spaced grid points, with periodic boundary conditions; these matrices will be obtained in practice with compact schemes as presented above; $\Id_{\,N}$ will denote the $N\times N$ identity matrix. The relaxed scheme produces the iterations
\begin{align*}
  \frac{U^{\,(k+1)}\ -\ U^{\,(k)}}{\Delta t}\ +\ \frac{1}{2}\;\D_{\,x}\,\bigl[\,(W^{\,(k+1)}\ +\ W^{\,(k)})\ +\ (U^{\,(k)})^2\,\bigr]\ &=\ 0\,,\\
  \delta\; \frac{V^{\,(k+1)}\ -\ V^{\,(k)}}{\Delta t}\ +\ \frac{1}{2}\;\bigl[\,(V^{\,(k+1)}\ +\ V^{\,(k)})\ -\ \D_{\,x}\, (U^{\,(k+1)}\ +\ U^{\,(k)})\,\bigr]\ &=\ 0\,,\\
  \delta\; \frac{W^{\,(k+1)}\ -\ W^{\,(k)}}{\Delta t}\ +\ \frac{1}{2}\;\bigl[\,(W^{\,(k+1)}\ +\ W^{\,(k)})\ -\ \D_{\,x}\, (V^{\,(k+1)}\ +\ V^{\,(k)})\,\bigr]\ &=\ 0\,.
\end{align*}
We rewrite this coupled system by blocs as
\begin{multline*}
\begin{pmatrix}
  \Id_{\,N} & 0 &\frac{\Delta t}{2} \\
  -\,\frac{\Delta t}{2}\; \D_{\,x} & (\delta\ +\ \frac{\Delta t}{2})\, \Id_{\,N} & 0 \\
  0 &  -\,\frac{\Delta t}{2}\, \D_x & (\delta\ +\ \frac{\Delta t}{2})\,\Id_{\,N}
\end{pmatrix}
\cdot
\begin{pmatrix}
U^{\,(k+1})\\
V^{\,(k+1})\\
W^{\,(k+1})
\end{pmatrix}
\ = \\
  \begin{pmatrix}
    \Id_{\,N} & 0 &-\,\frac{\Delta t}{2} \\
    \frac{\Delta t}{2}\;\D_{\,x} & (\delta\ -\ \frac{\Delta t}{2})\,\Id_{\,N} & 0 \\
    0 & \frac{\Delta t}{2}\;\D_{\,x} & (\delta\ -\ \frac{\Delta t}{2})\,\Id_{\,N} 
  \end{pmatrix}
  \cdot
  \begin{pmatrix}
    U^{\,(k)}\\
    V^{\,(k)}\\
    W^{\,(k)}
  \end{pmatrix}\ +\ 
  \begin{pmatrix}
    -\,\frac{\Delta t}{2}\; \D_{\,x}\,(U^{\,(k)})^{\,2} \\
    0 \\
    0
\end{pmatrix}\,.
\end{multline*}
We can now introduce the iteration matrix
\begin{multline*}
  \M_{\,r}\ \eqdef\ (M^{\,i}_{\,r})^{\,-1}\cdot M^{\,e}_{\,r}\ =\\ 
  \begin{pmatrix}
    \Id_{\,N} & 0 & \frac{\Delta t}{2} \\
    -\,\frac{\Delta t}{2}\;\D_{\,x} & (\delta\ +\ \frac{\Delta t}{2})\,\Id_{\,N} & 0 \\
    0 & -\,\frac{\Delta t}{2}\,\D_{\,x} & (\delta\ +\ \frac{\Delta t}{2})\,\Id_{\,N}
  \end{pmatrix}^{\,-1}
  \cdot
  \begin{pmatrix}
   \Id_{\,N} & 0 & -\,\frac{\Delta t}{2} \\
   \frac{\Delta t}{2}\,\D_{\,x} & (\delta\ -\ \frac{\Delta t}{2})\,\Id_{\,N} & 0\\
   0 & \frac{\Delta t}{2}\;\D_{\,x} & (\delta\ -\ \frac{\Delta t}{2})\,\Id_{\,N} \\
  \end{pmatrix}\,.
\end{multline*}

We can resume the relaxed scheme as follows
\begin{center}
  \begin{algorithm}
    \caption{\small\em Relaxed Scheme for the \acs{kdv} equation.}
    \label{RelaxedKdV}
    \begin{algorithmic}[1]
        \State $U^{\,(0)}$ is given
        \State {\bf Set} $V^{\,(0)}\ =\ \D_{\,x}\,U^{\,(0)}\,$, $W^{\,(0)}\ =\ \D_{\,x}\,V^{\,(0)}$
        \State {\bf Set} $Z^{\,(0)}\ =\ \bigl(U^{\,(0)},\,V^{\,(0)}\,,\ W^{\,(0)}\bigr)^{\,\top}$
        \For{$k\,=\,0,\, 1,\, \ldots$}
          \State {\bf Set} $F^{\,(k)}\ =\ \bigl(-\,\frac{\Delta t}{2}\, \D_{\,x}\,(U^{\,(k)})^{\,2}\,,\, 0,\, 0\bigr)^{\,\top}$
          \State {\bf Solve} $M^{\,i}_{\,r}\cdot Z^{\,(k\,+\,1)}\ =\ M^{\,e}_{\,r}\cdot Z^{\,(k)}\ +\ F^{\,(k)}$
          \State {\bf Set} $U^{\,(k+1)}\, =\, Z^{\,(k\,+\,1)}\,(1:N)\,,\ V^{\,(k\,+\,1)}\ =\ Z^{\,(k\,+\,1)}\,(N\,+\,1\,:\,2\,N)\,$,\par $W^{\,(k\,+\,1)}\ =\ Z^{\,(k\,+\,1)}\,(2\,N\,+\,1\,:\,3\,N)$
        \EndFor
    \end{algorithmic}
    \end{algorithm}
\end{center}

We now present the two references schemes to which we will compare the relaxed one. We set 
\begin{equation*}
  \M^{\,i}_{\,\mathrm{CN}}\ \eqdef\ \Id_{\,N}\ +\ \frac{\Delta t}{2}\;\D_{\,x\,x\,x}\,, \qquad
  \M^{\,e}_{\,\mathrm{CN}}\ \eqdef\ \Id_{\,N}\ -\ \frac{\Delta t}{2}\;\D_{\,x\,x\,x}
\end{equation*}
and we define the corresponding iteration matrix $\M_{\,\mathrm{CN}}$ as
\begin{equation*}
  \M_{\,\mathrm{CN}}\ \eqdef\ (\M^{\,i}_{\,\mathrm{CN}})^{\,-1}\cdot \M^{\,e}_{\,\mathrm{CN}}\,.
\end{equation*}

\begin{center}
  \begin{algorithm}
    \caption{\small\em Fully nonlinear \textsc{Sanz}--\textsc{Serna}'s scheme.}
    \label{SanzSernaKdV}
    \begin{algorithmic}[1]
        \State $U^{\,(0)}$ is given\\
            \For{$k\,=\,0,\,1,\,\ldots$}
             \State {\bf Solve} $\M^{\,i}_{\,\mathrm{CN}}\,U^{\,(k\,+\,1)}\ =\ \M^{\,e}_{\,\mathrm{CN}}\,U^{\,(k)}\ -\ \frac{\Delta t}{2}\; \D_{\,x}\,\Bigl(\frac{U^{\,(k\,+\,1)}\ +\ U^{\,(k)}}{2}\Bigr)^{\,2}$
            \EndFor
    \end{algorithmic}
    \end{algorithm}
\end{center}
We will use also
\begin{center}
  \begin{algorithm}
    \caption{\small\em Semi-implicit \textsc{Crank}--\textsc{Nicolson} scheme.}
    \label{SemiCNKdV}
    \begin{algorithmic}[1]
        \State $U^{\,(0)}$ is given\\
            \For{$k\,=\,0,\,1,\,\ldots$}
             \State {\bf Solve} $\M^{\,i}_{\,\mathrm{CN}}\,U^{\,(k\,+\,1)}\ =\ \M^{\,e}_{\,\mathrm{CN}}\,U^{\,(k)}\ -\ \frac{\Delta t}{2}\; \D_{\,x}\,\Bigl(U^{\,(k)}\Bigr)^{\,2}$
            \EndFor
    \end{algorithmic}
    \end{algorithm}
\end{center}


\subsection{Linear stability analysis}

\begin{figure}
  \subfigure[$\Delta t\ =\ 10^{\,-3}$ CN Matrices spectra]{\includegraphics[width=0.48\textwidth]{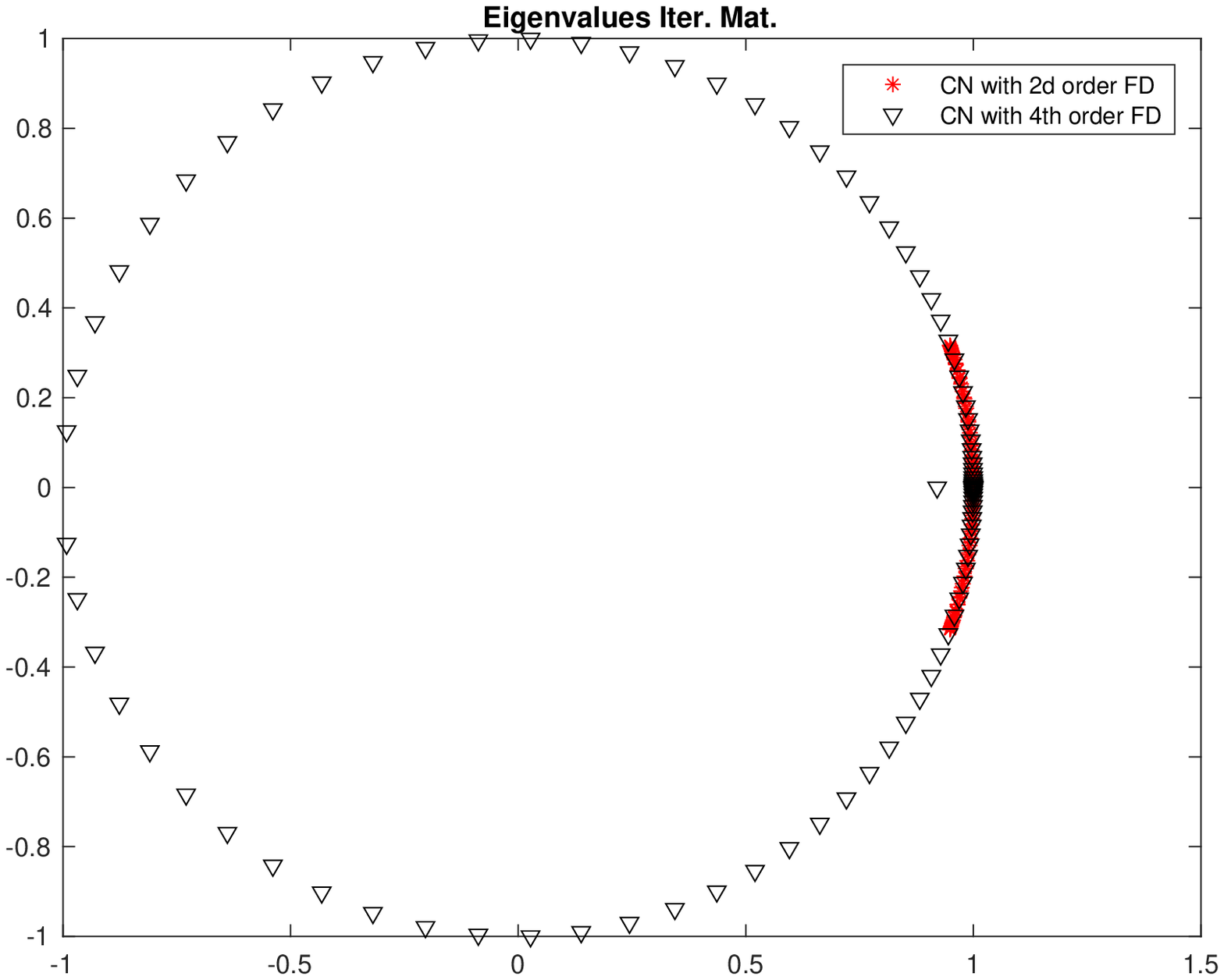}}
  \subfigure[$\Delta t\ =\ 10^{\,-3}$ \textsc{Fourier} symbol]{\includegraphics[width=0.48\textwidth]{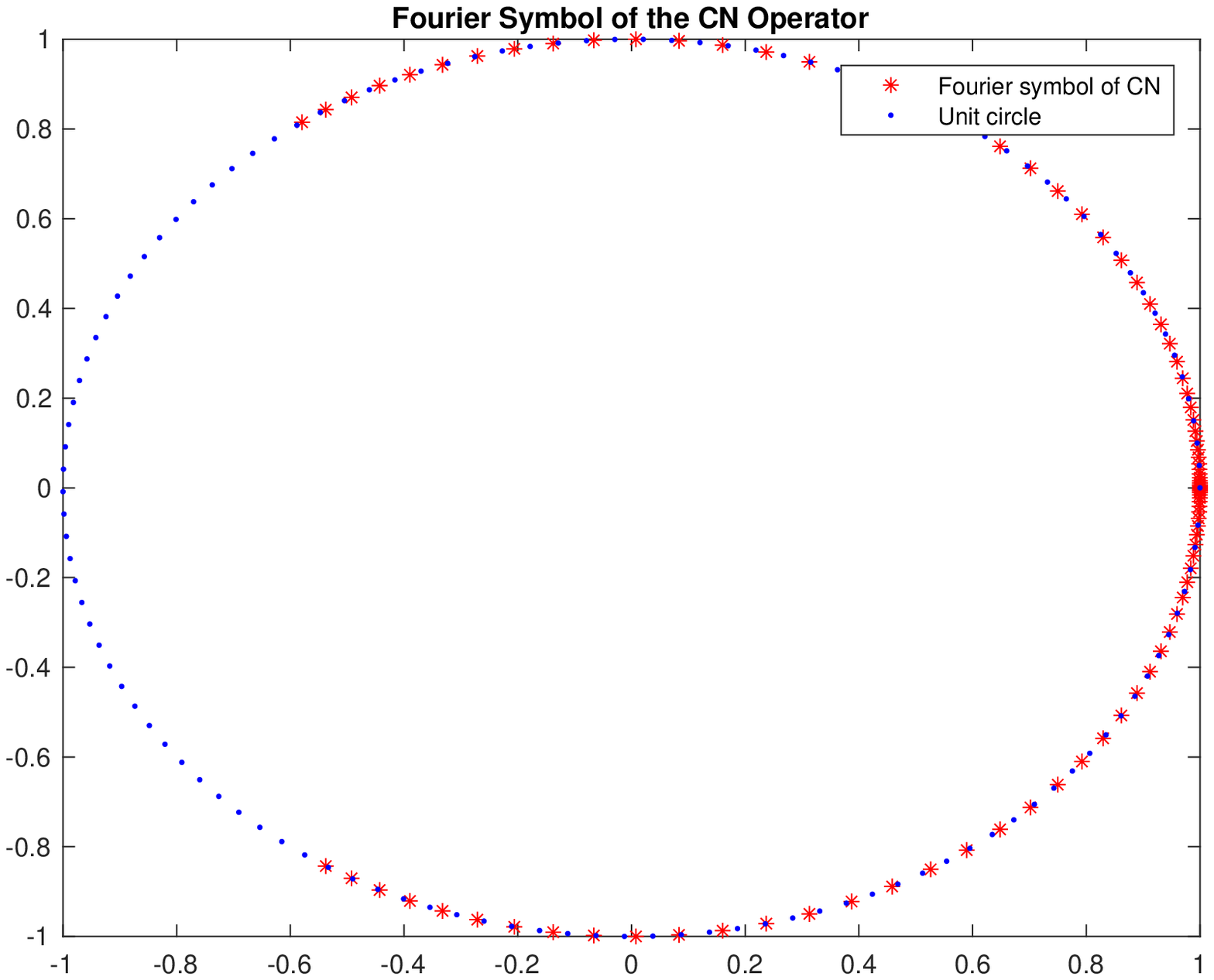}}
  \subfigure[$\Delta t\ =\ 10^{\,-2}$ CN Matrices spectra]{\includegraphics[width=0.48\textwidth]{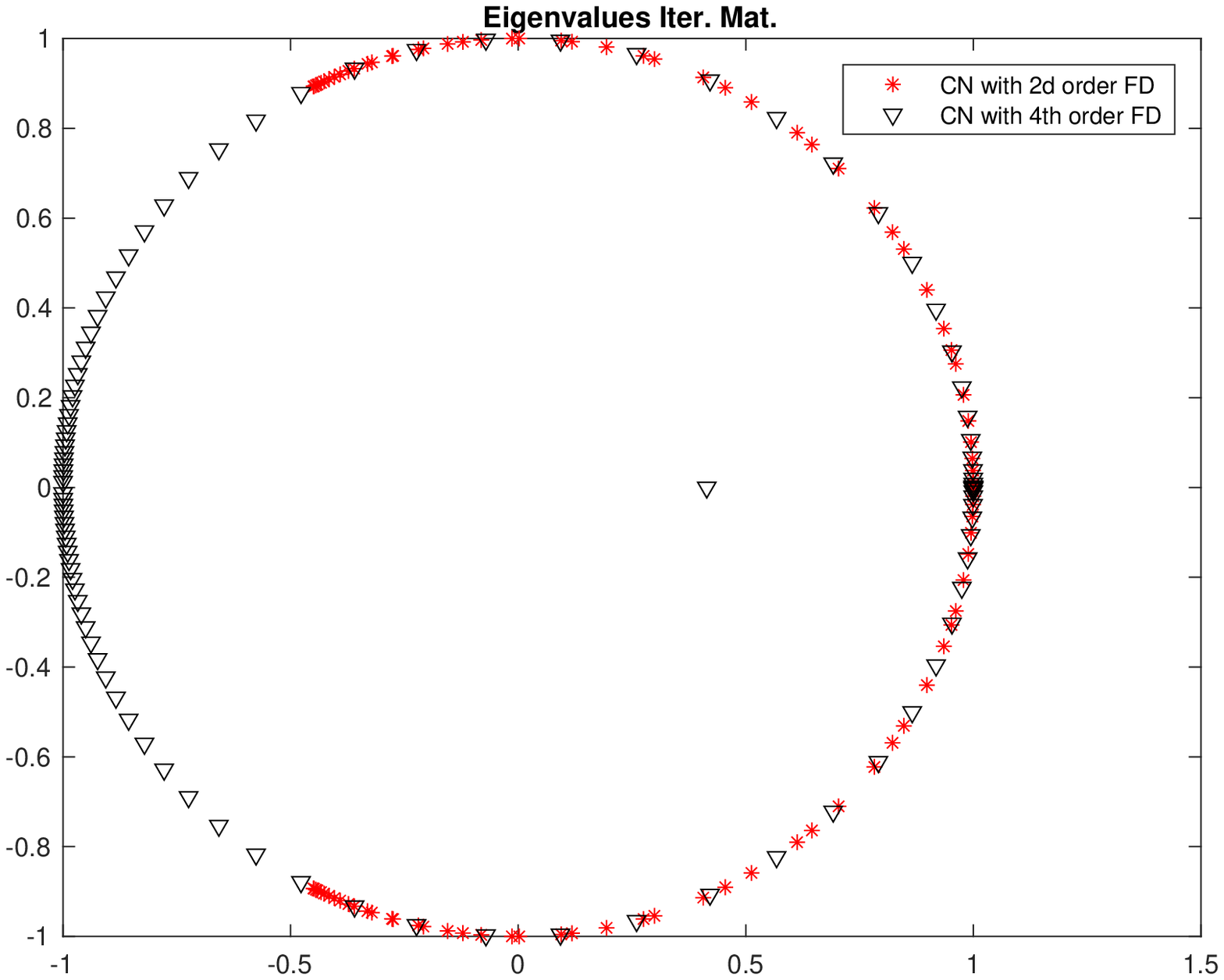}}
  \subfigure[$\Delta t\ =\ 10^{\,-2}$ \textsc{Fourier} symbol]{\includegraphics[width=0.48\textwidth]{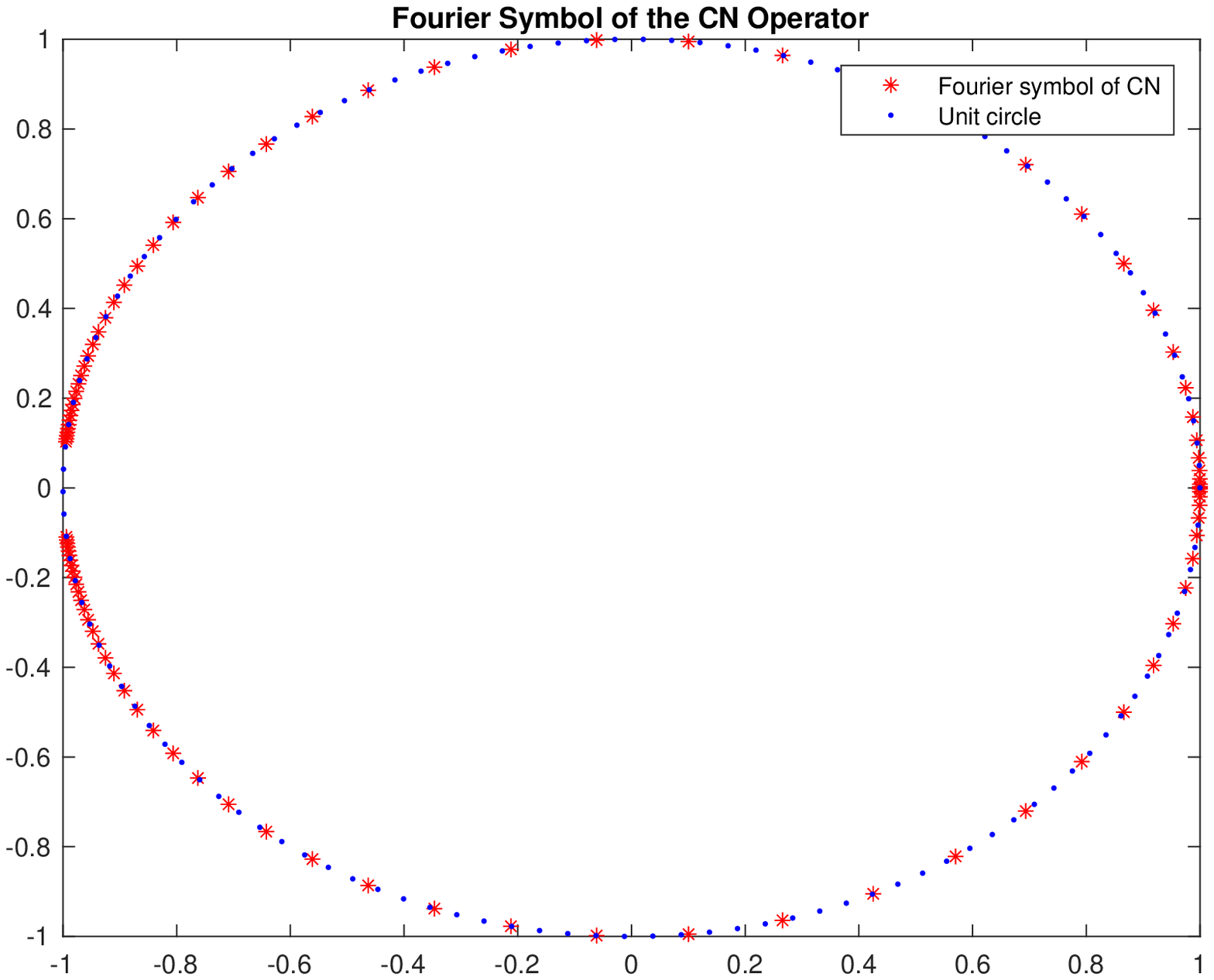}}
  \caption{\small\em Spectra of the $\M_{\,\mathrm{CN}}$ matrices for the 2\up{nd} (red markers) and 4\up{th} order (black markers) finite differences schemes: $\ell=\ 20\,$, $N\ =\ 100\,$, $\Delta t\ =\ 10^{\,-3}$ and $\Delta t\ =\ \ 10^{\,-2}\,$.}
  \label{Fig1:comp.spect}
\end{figure}

\begin{figure}
  \subfigure[$\Delta t\ =\ 10^{\,-3}$ CN Matrices spectra]{\includegraphics[width=0.48\textwidth]{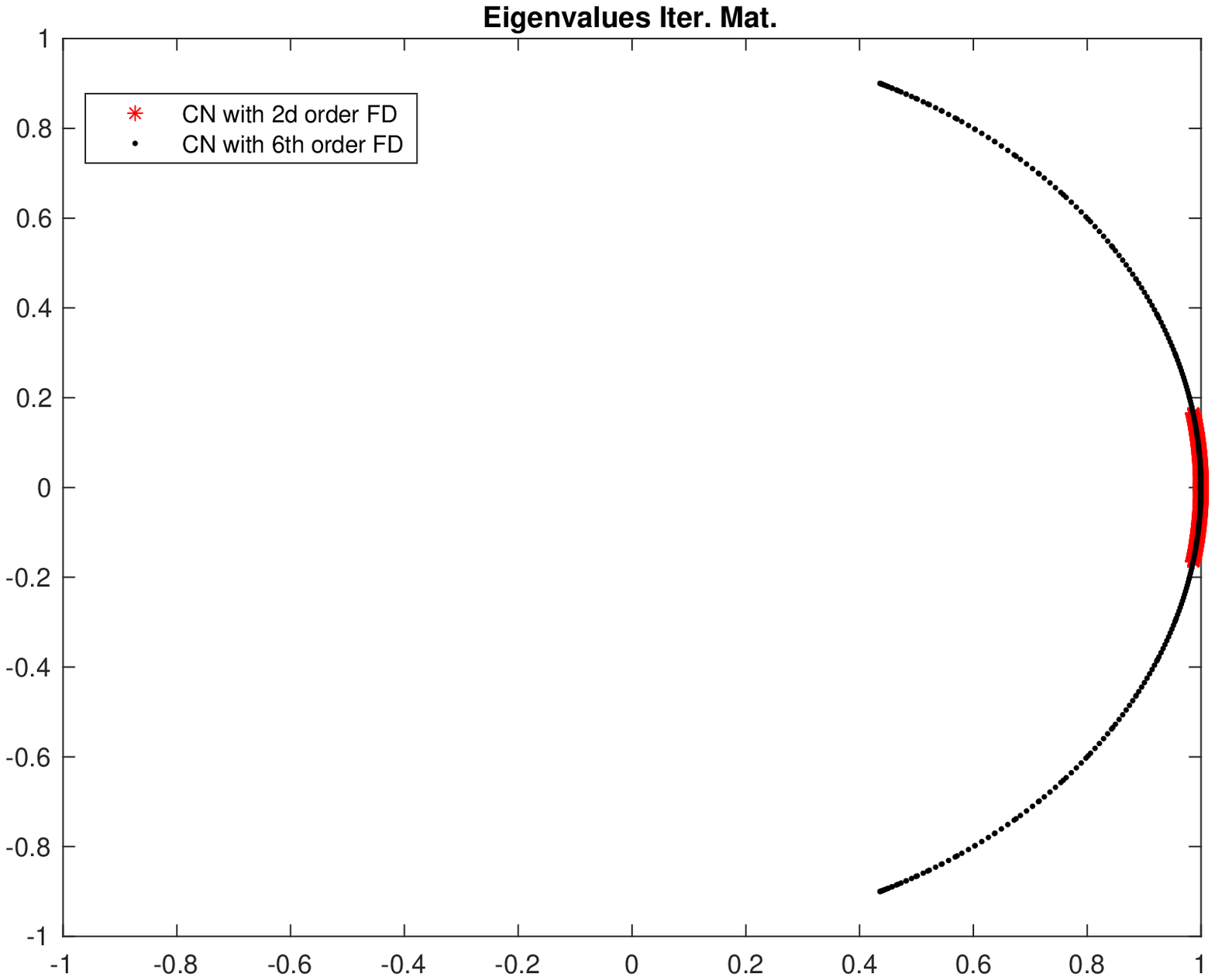}}
  \subfigure[$\Delta t\ =\ 10^{\,-3}$ \textsc{Fourier} symbol]{\includegraphics[width=0.48\textwidth]{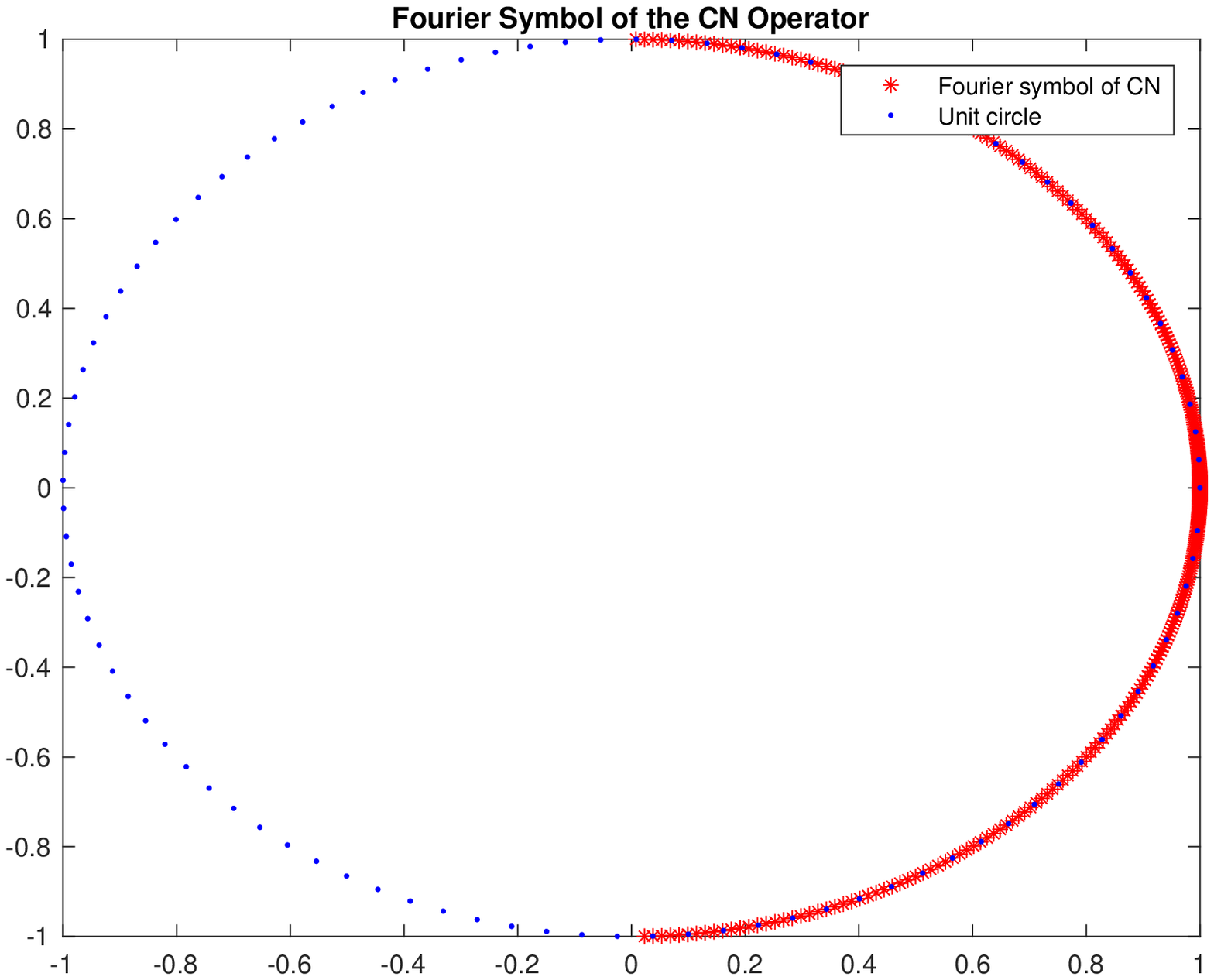}}
  \subfigure[$\Delta t\ =\ 10^{\,-2}$ CN Matrices spectra]{\includegraphics[width=0.48\textwidth]{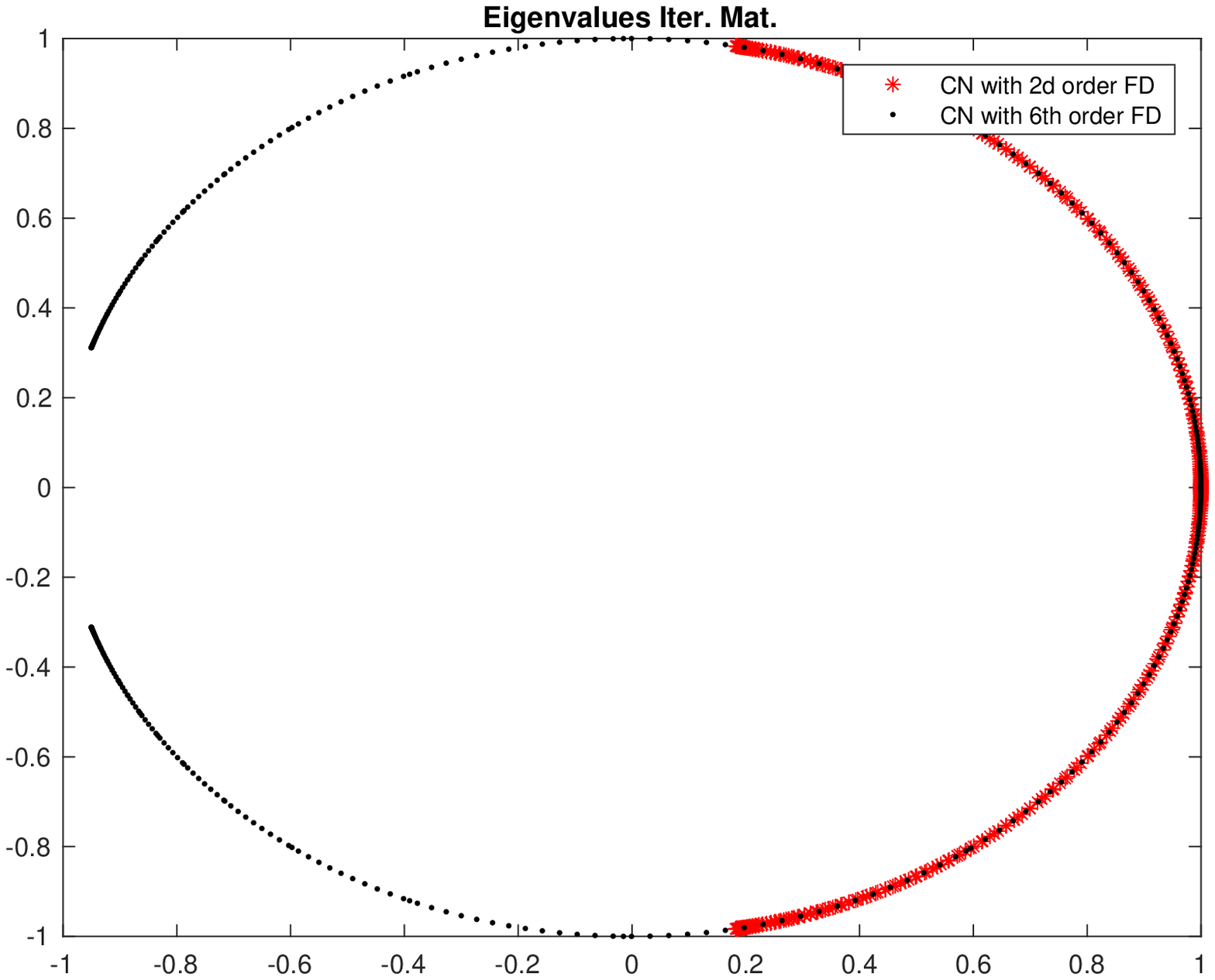}}
  \subfigure[$\Delta t\ =\ 10^{\,-2}$ \textsc{Fourier} symbol]{\includegraphics[width=0.48\textwidth]{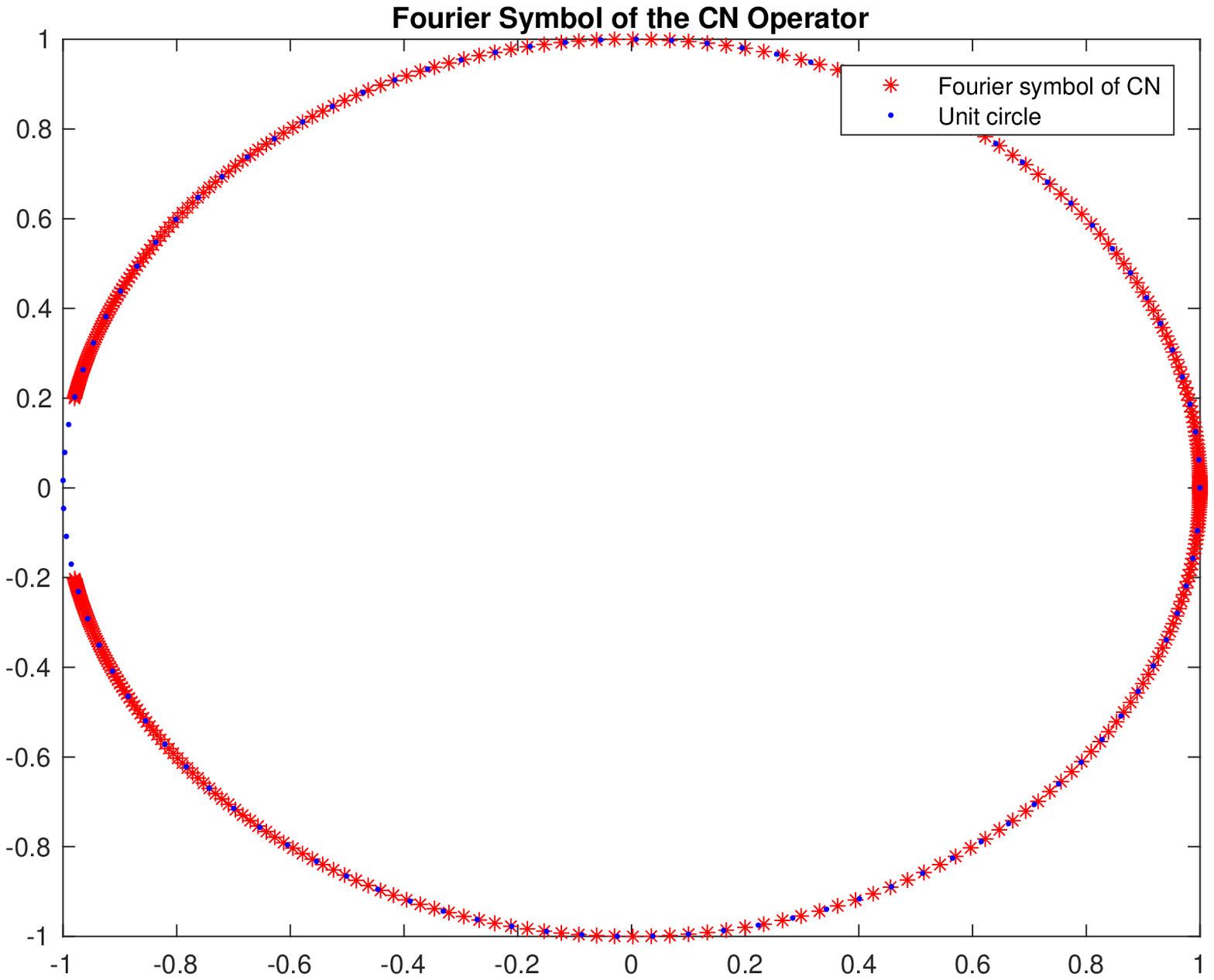}}
  \caption{\small\em Spectra of the $\M_{\,\mathrm{CN}}$ matrices for the 2\up{nd} (red markers) and 6\up{th} order (black markers) finite differences schemes: $\ell=\ 100\,$, $N\ =\ 400\,$, $\Delta t\ =\ 10^{\,-3}$ and $\Delta t\ =\ \ 10^{\,-2}\,$.}
  \label{Fig2:comp.spect}
\end{figure}

Before comparing the stability properties of the classical CN and of the relaxed schemes, we give hereafter a simple but instructive result illustrating the advantage of the compact schemes in a context in which the spectral properties of the operators must be restituted at the discrete level. Writing the CN scheme for the \textsc{Airy} equation, we have the following induction relation among \textsc{Fourier} coefficients attached to the $m$\up{th} frequency
\begin{align*}
  \frac{{\hat u}^{\,(k\,+\,1)}_{\,m}\ -\ {\hat u}^{\,(k)}_{\,m}}{\Delta t}\ -\ \frac{\ui}{2}\;m^{\,3}\;\bigl({\hat u}^{\,(k\,+\,1)}_{\,m}\ +\ {\hat u}^{\,(k)}_{\,m}\bigr)\ =\ 0\,,
\end{align*}
hence
\begin{align*}
  {\hat u}^{\,(k\,+\,1)}_{\,m}\ =\ \frac{1\ +\ \ui\;\dfrac{\Delta t}{2}\,m^{\,3}}{1\ -\ \ui\,\dfrac{\Delta t}{2}\;m^{\,3}}\;{\hat u}^{\,(k)}_{\,m}\,.
\end{align*}
The \textsc{Fourier} symbol of the CN operator is $\dfrac{1\ +\ \ui\;\dfrac{\Delta t}{2}\;m^{\,3}}{1\ -\ \ui\;\dfrac{\Delta t}{2}\;m^{\,3}}$ and its values are all displayed on the unit circle, for any $\Delta t\ >\ 0\,$. It is then an important property to be captured by the spectrum of the \textsc{Crank}--\textsc{Nicolson} matrix, when dealing with finite differences. We give hereafter in Figure~\ref{Fig1:comp.spect} the spectra of the CN matrix when the 2\up{nd} order FD and the 4\up{th} order compact FD are used. We observe that the compact scheme allows to capture numerically the spectral properties of the linear operator as in the analytical \textsc{Fourier} analysis, while the 2\up{nd} order finite differences succeed to capture correctly only a small part of the spectrum. Comparable results are obtained in Figure~\ref{Fig2:comp.spect} when considering 6\up{th} order compact FD: they illustrate the capabilities of the compact schemes to mimmic the spectral properties of the linear propagation operator.

\subsection{Enhanced stability analysis}

\begin{figure}
  \centering
  \subfigure[$\Delta t\ =\ 10^{\,-3}$]{\includegraphics[width=0.48\textwidth]{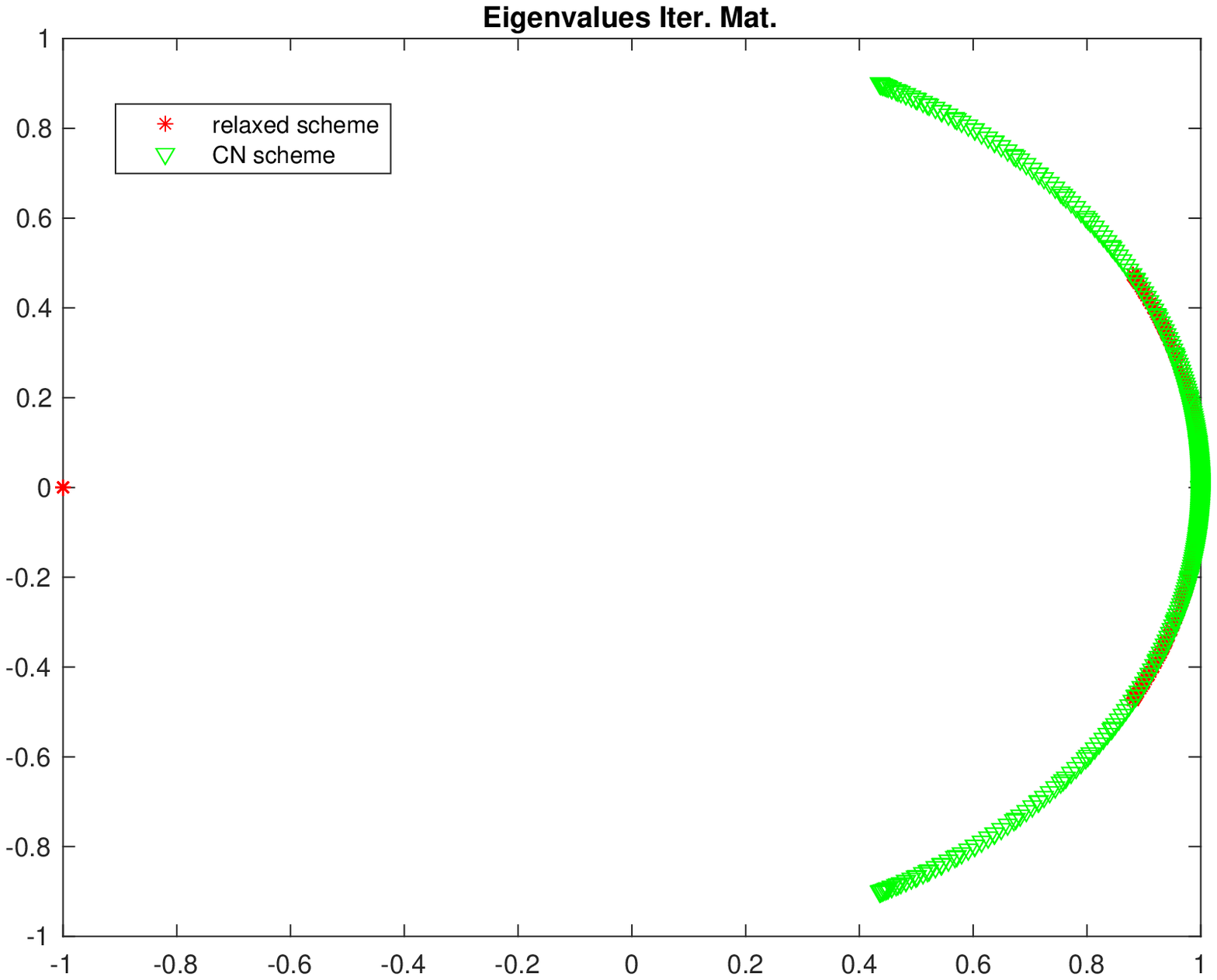}}
  \subfigure[$\Delta t\ =\ 10^{\,-2}$]{\includegraphics[width=0.48\textwidth]{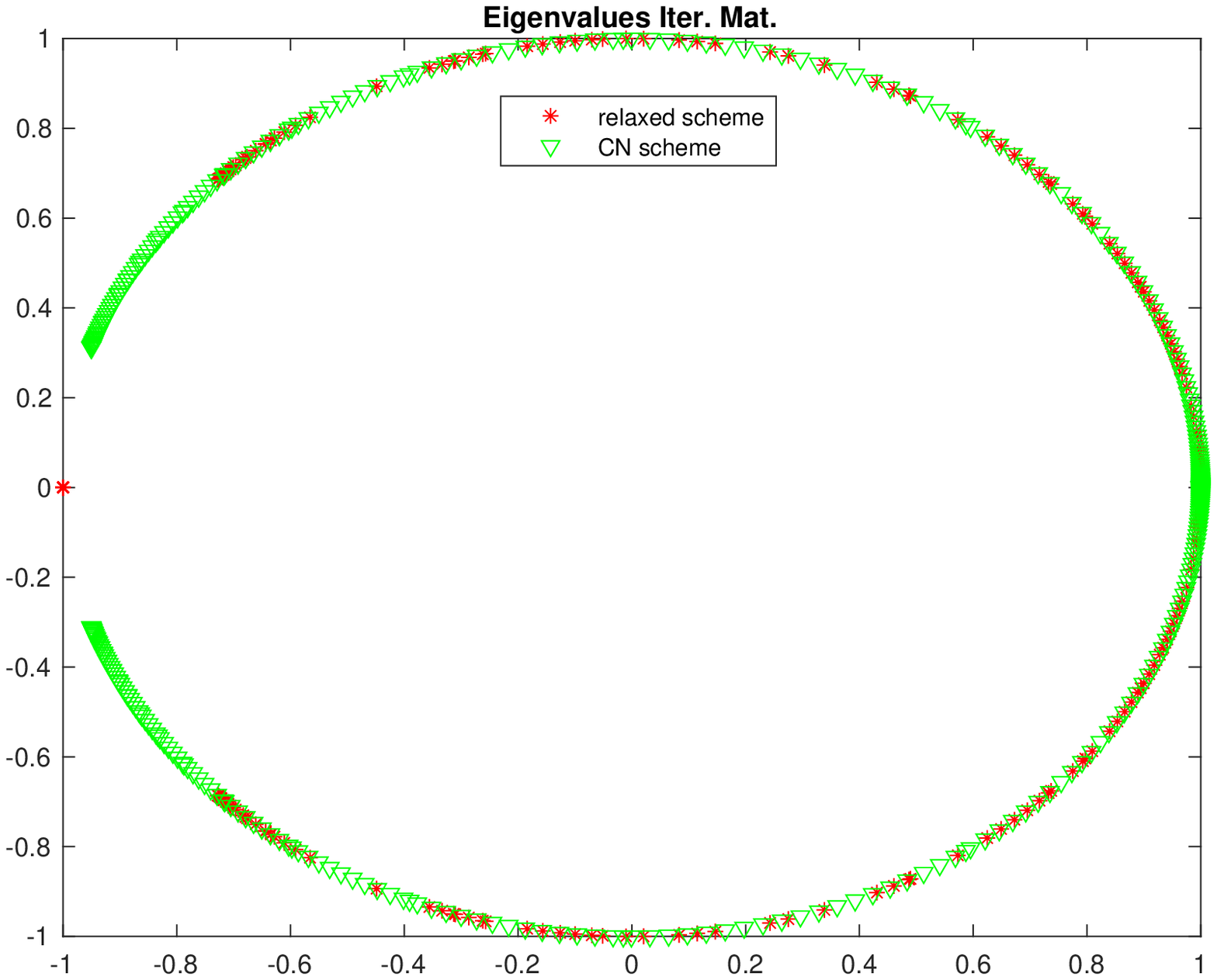}}
  \subfigure[$\Delta t\ =\ 10^{\,-1}$]{\includegraphics[width=0.48\textwidth]{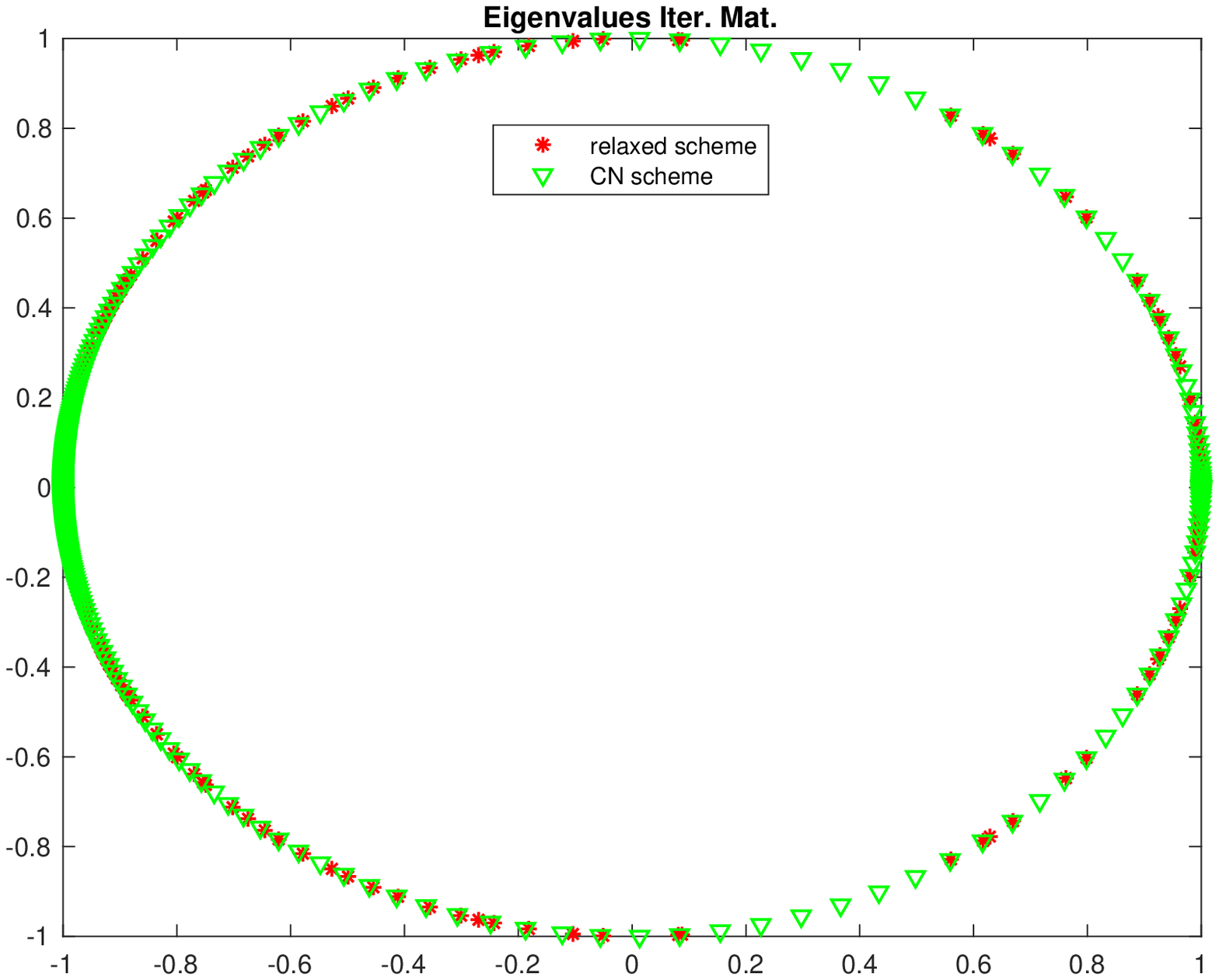}}
  \subfigure[$\Delta t\ =\ 5\times 10^{\,-1}$]{\includegraphics[width=0.48\textwidth]{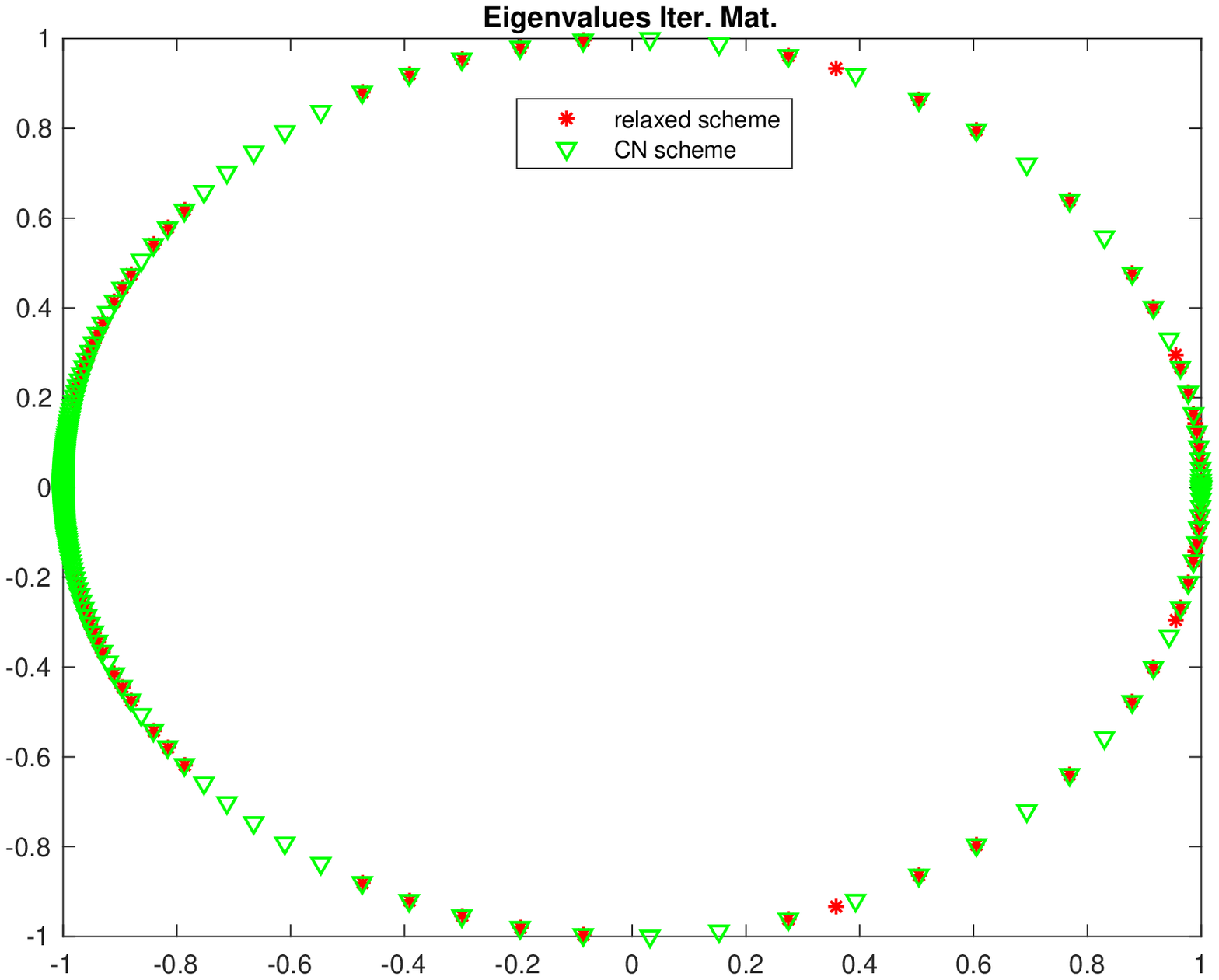}}
  \caption{\small\em Spectra of $\M_{\,r}$ and $\M_{\,\mathrm{CN}}$ matrices. For the spatial discretisation we use $\ell= \ 100$, $N\ =\ 400$ points and the relaxation parameter $\delta\ =\ 10^{-17}\,$. Various values of the time step are taken.}
  \label{Fig5:comp.spect}
\end{figure}

\begin{figure}
  \centering
  \subfigure[$\Delta t\ =\ 10^{\,-3}$]{\includegraphics[width=0.48\textwidth]{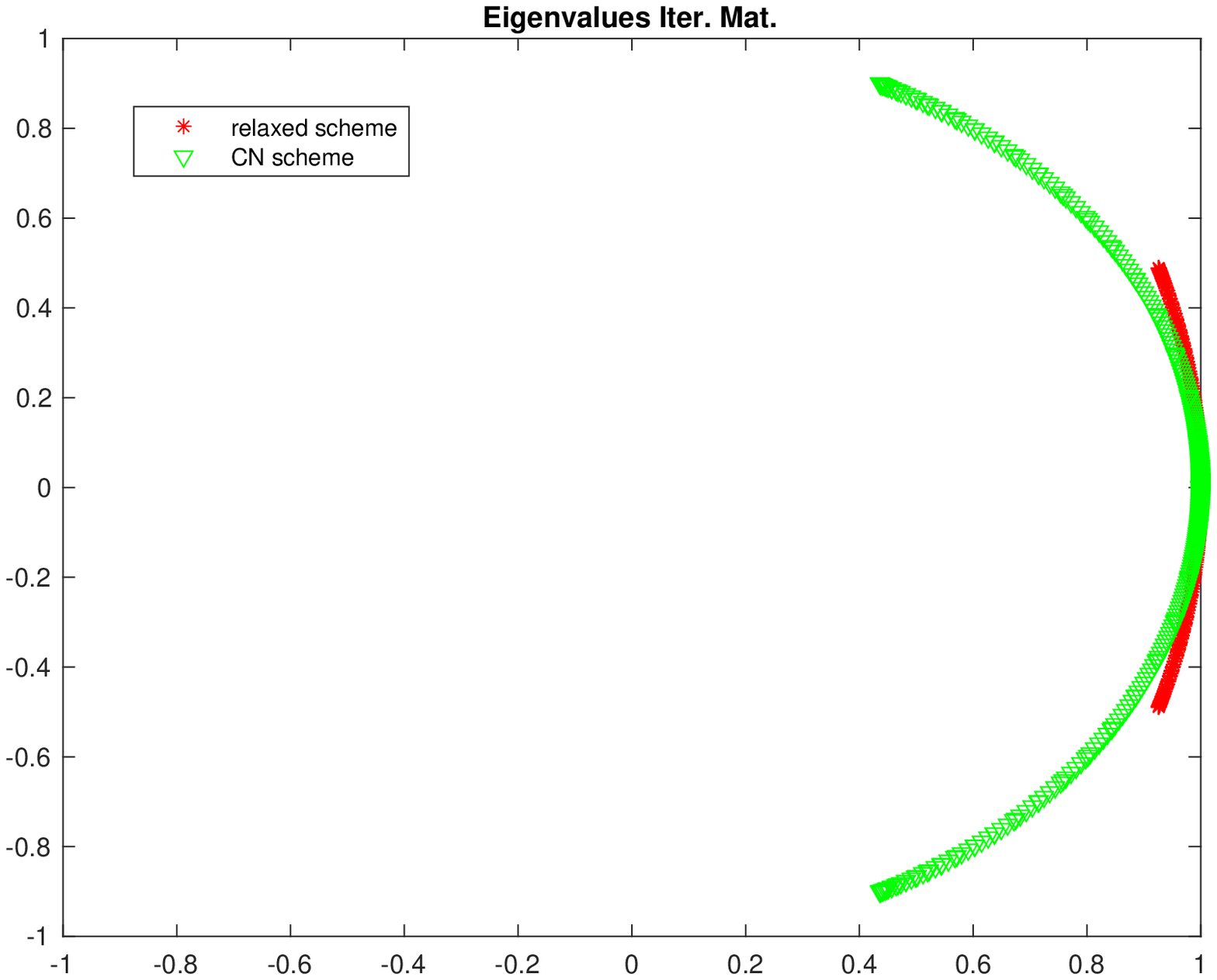}}
  \subfigure[$\Delta t\ =\ 10^{\,-2}$]{\includegraphics[width=0.48\textwidth]{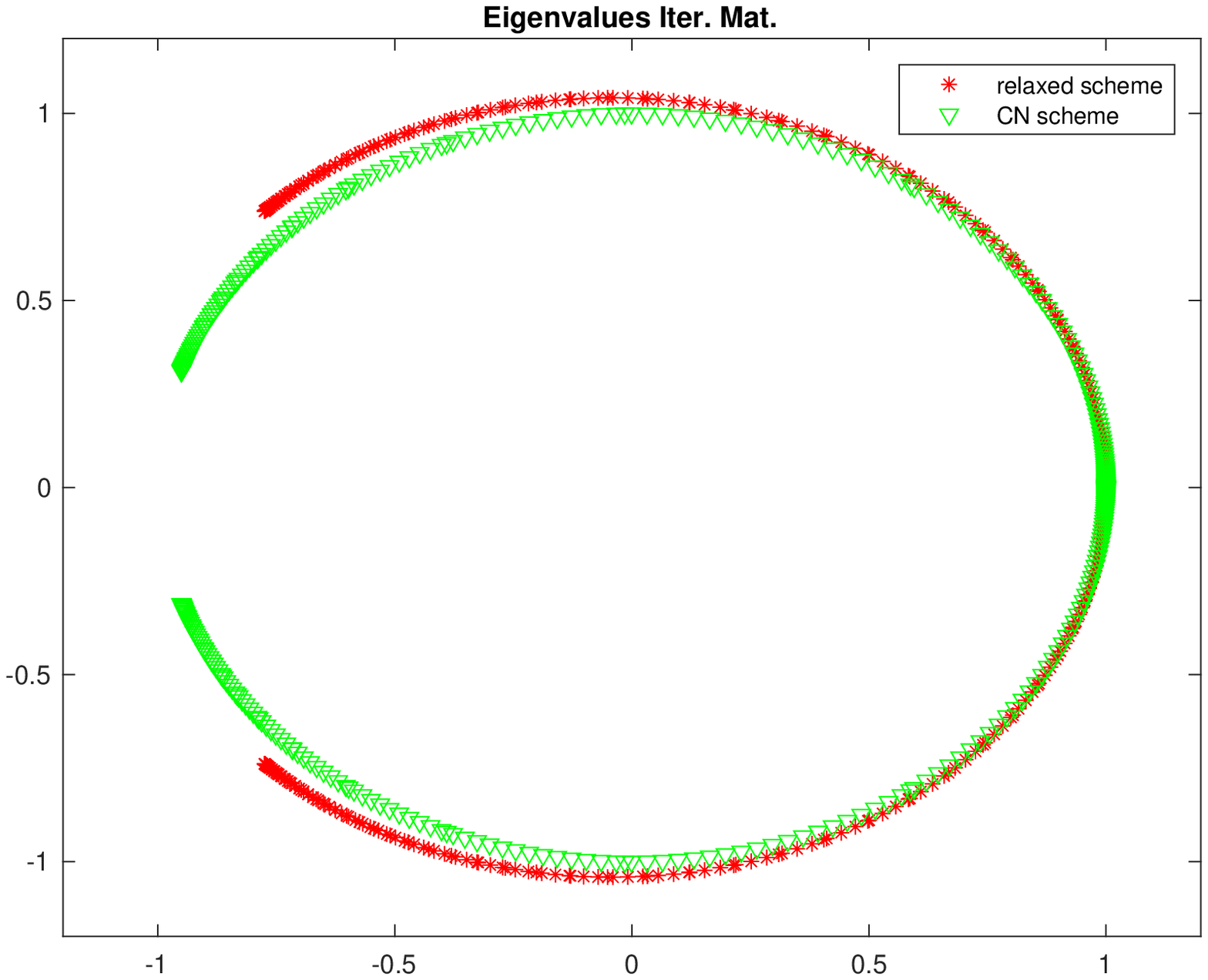}}
  \subfigure[$\Delta t\ =\ 10^{\,-1}$]{\includegraphics[width=0.48\textwidth]{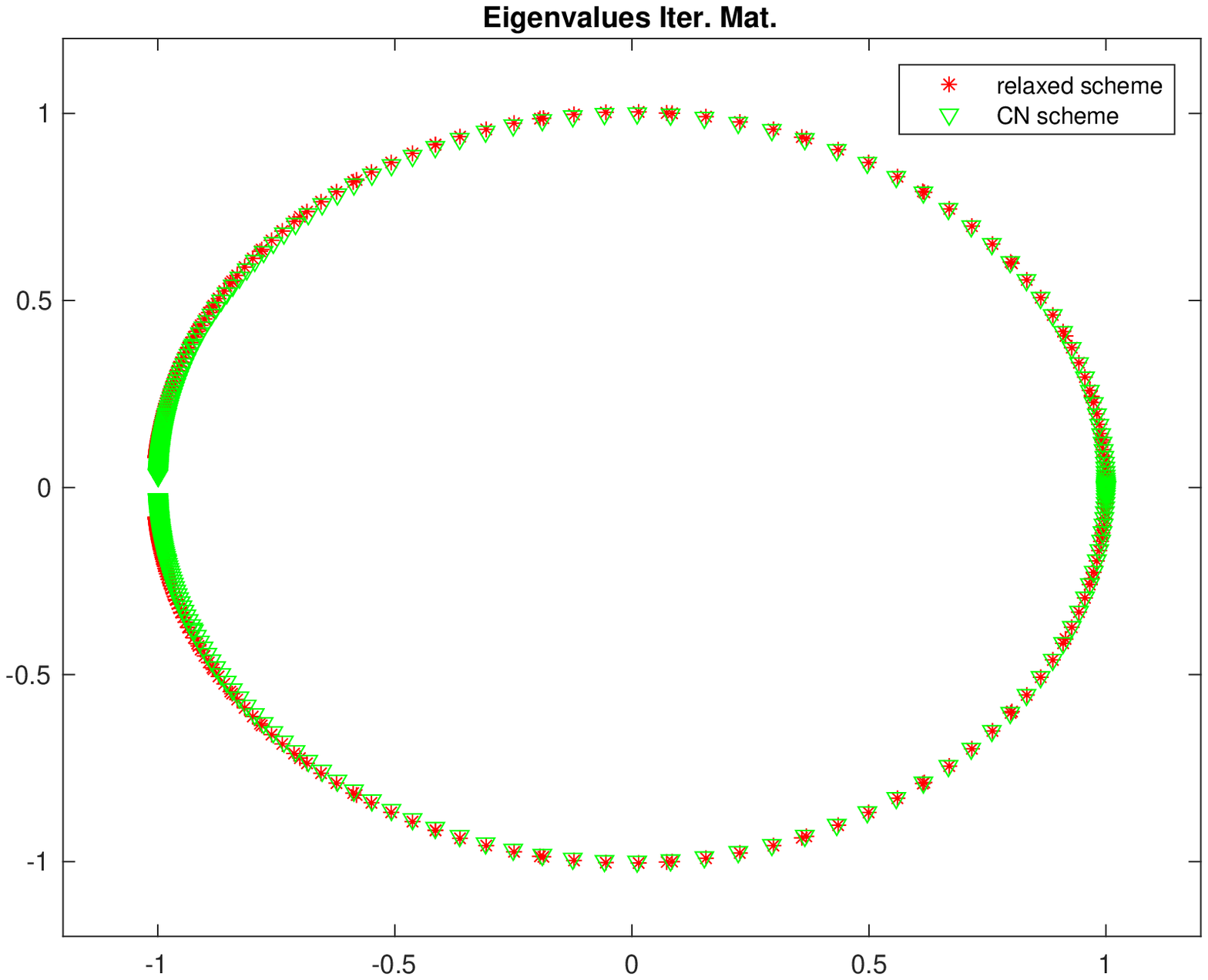}}
  \subfigure[$\Delta t\ =\ 5\times 10^{\,-1}$]{\includegraphics[width=0.48\textwidth]{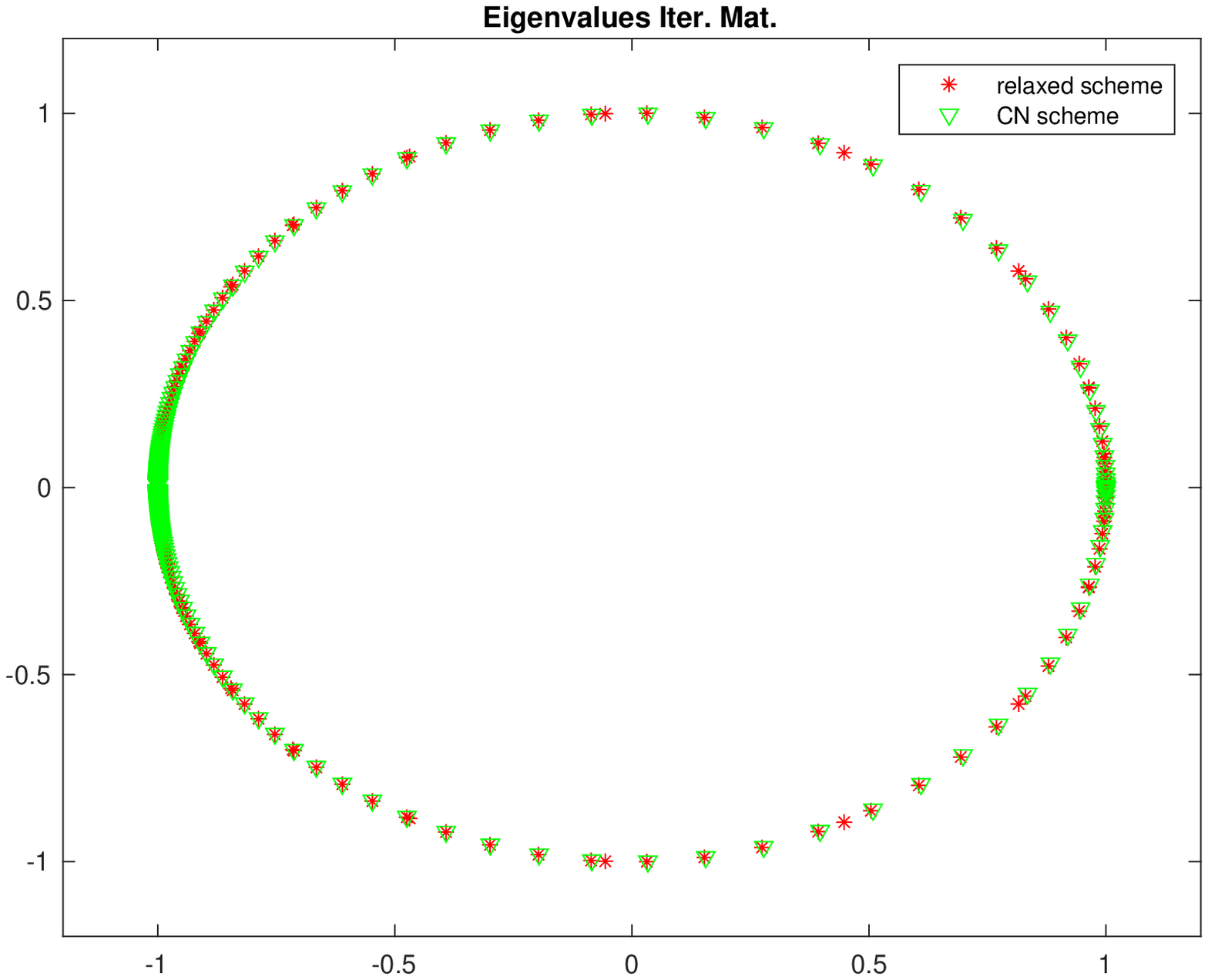}}
  \caption{\small\em Spectra of $\M_{\,r}$ and $\M_{\,\mathrm{CN}}$ matrices. For the spatial discretisation we use $\ell= \ 100$, $N\ =\ 400$ points and the relaxation parameter $\delta\ =\ 10^{-4}\,$. Various values of the time step are taken.}
  \label{Fig6:comp.spect}
\end{figure}

\begin{figure}
  \centering
  \subfigure[$\Delta t\ =\ 10^{\,-3}$]{\includegraphics[width=0.48\textwidth]{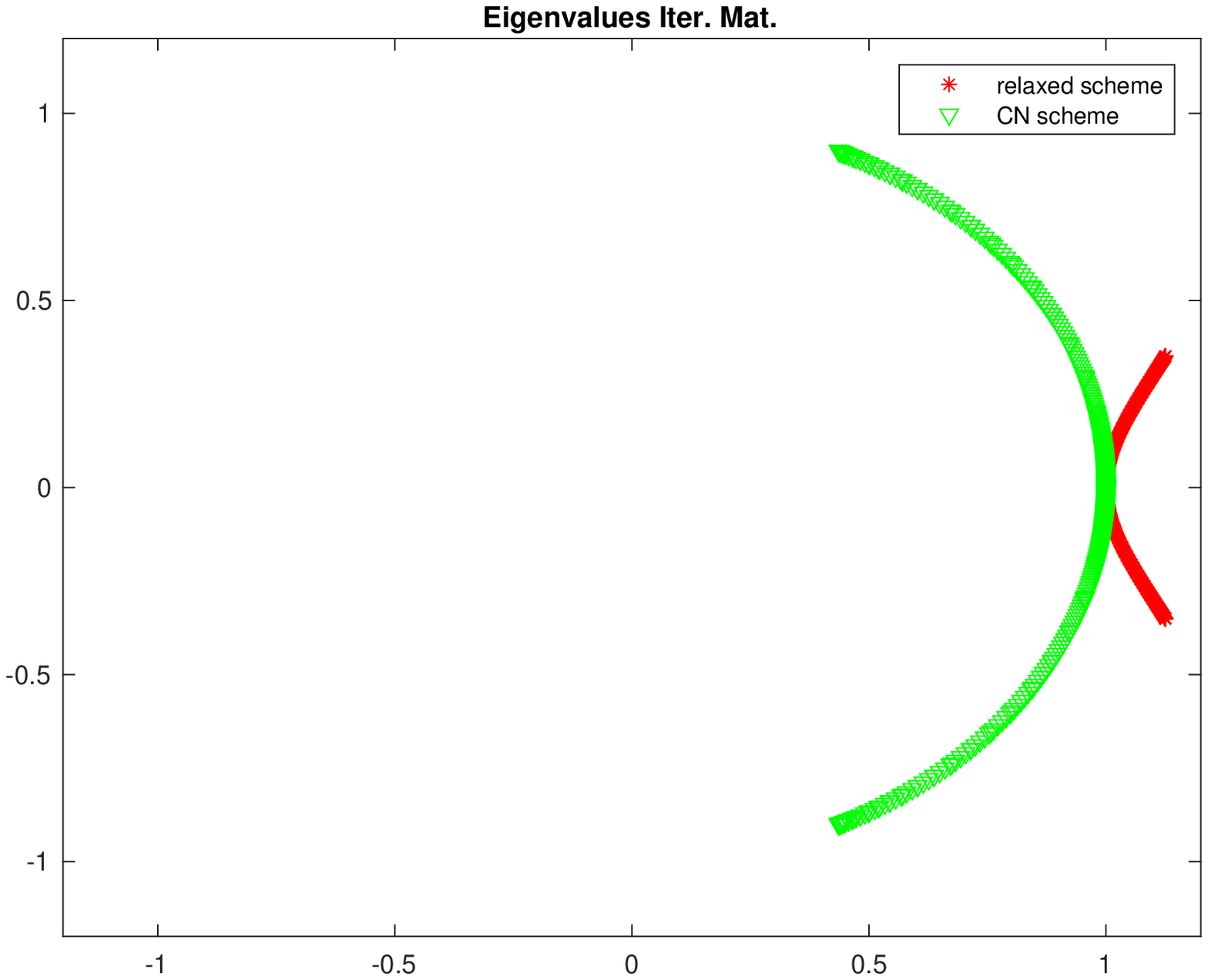}}
  \subfigure[$\Delta t\ =\ 10^{\,-2}$]{\includegraphics[width=0.48\textwidth]{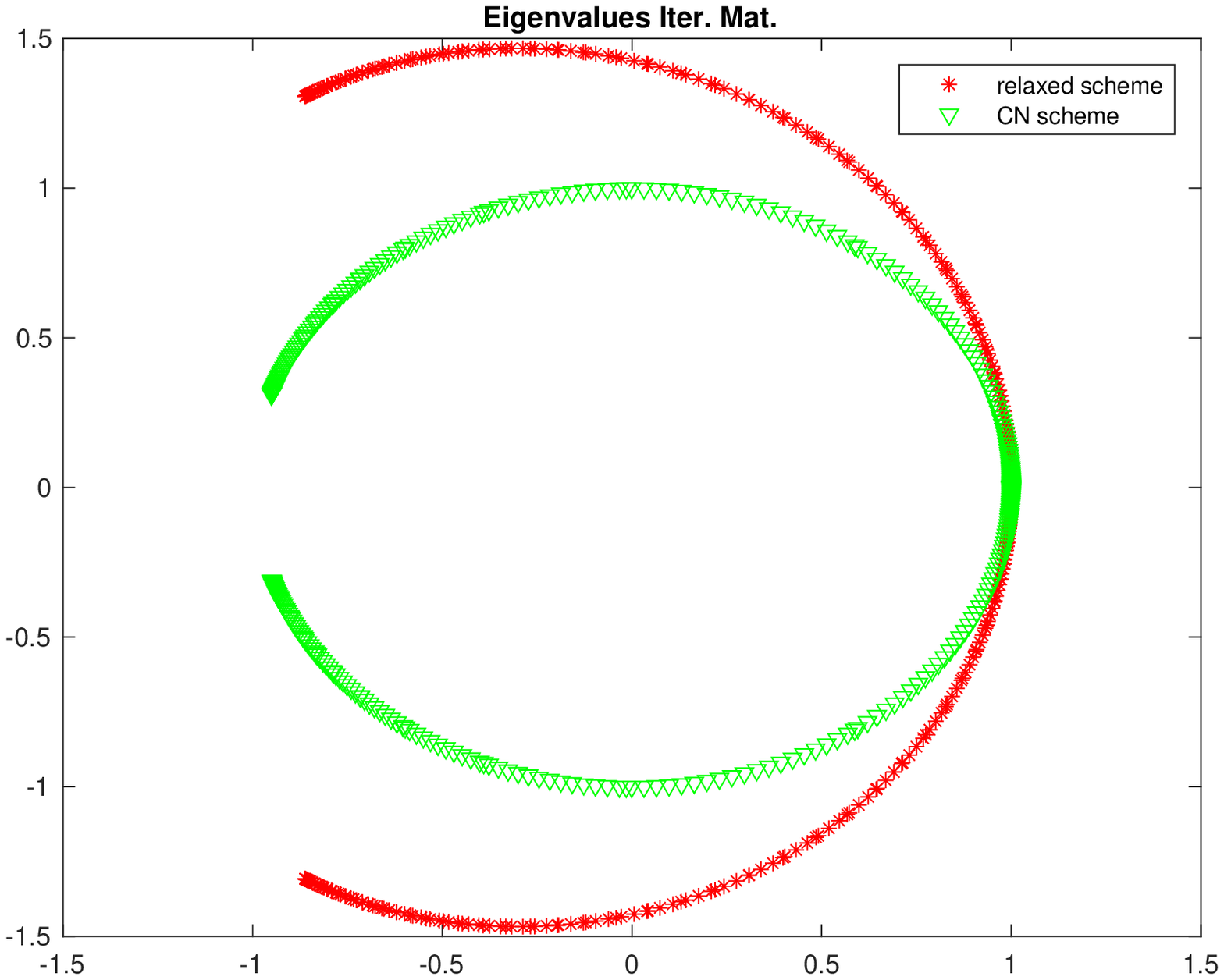}}
  \subfigure[$\Delta t\ =\ 10^{\,-1}$]{\includegraphics[width=0.48\textwidth]{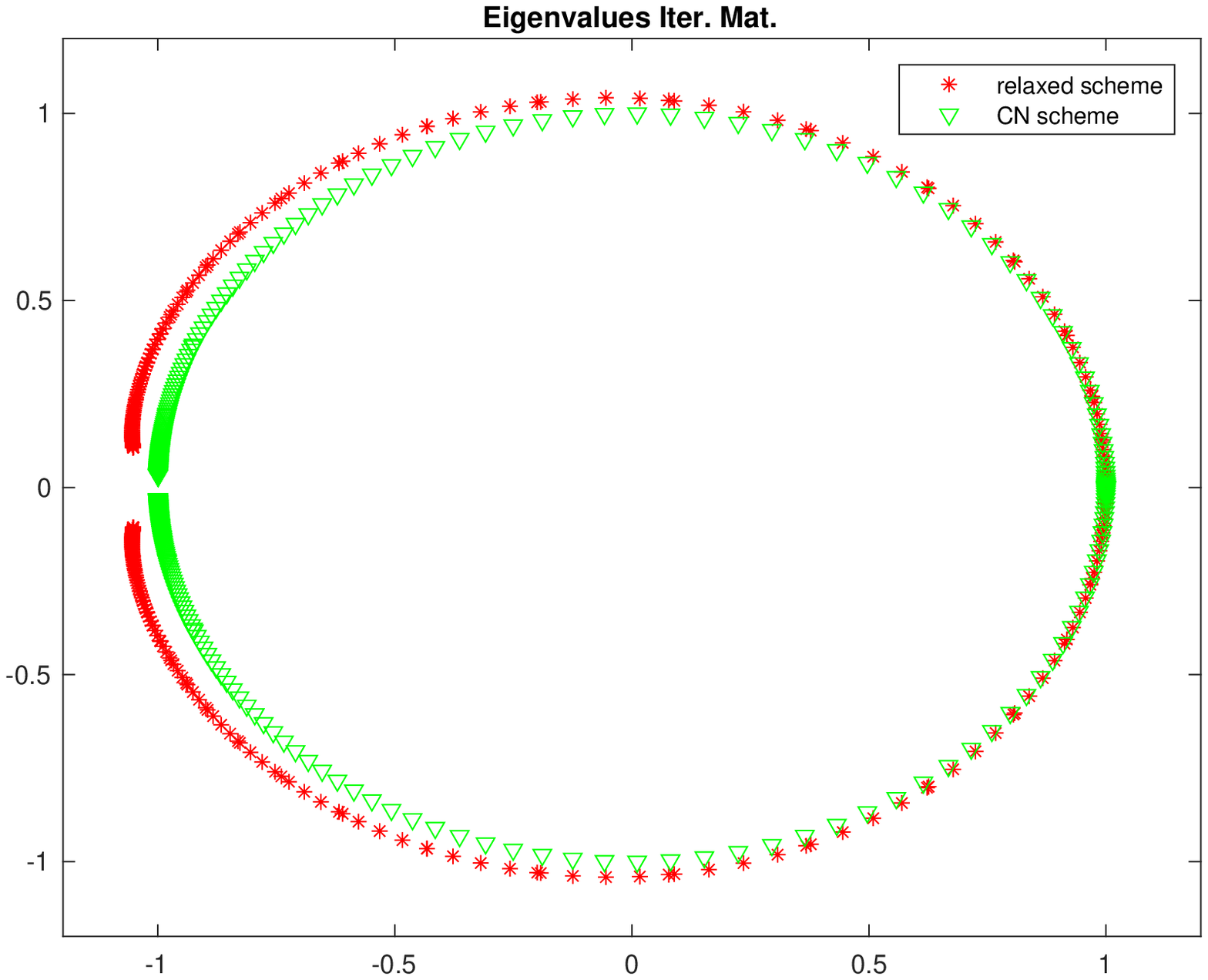}}
  \subfigure[$\Delta t\ =\ 5\times 10^{\,-1}$]{\includegraphics[width=0.48\textwidth]{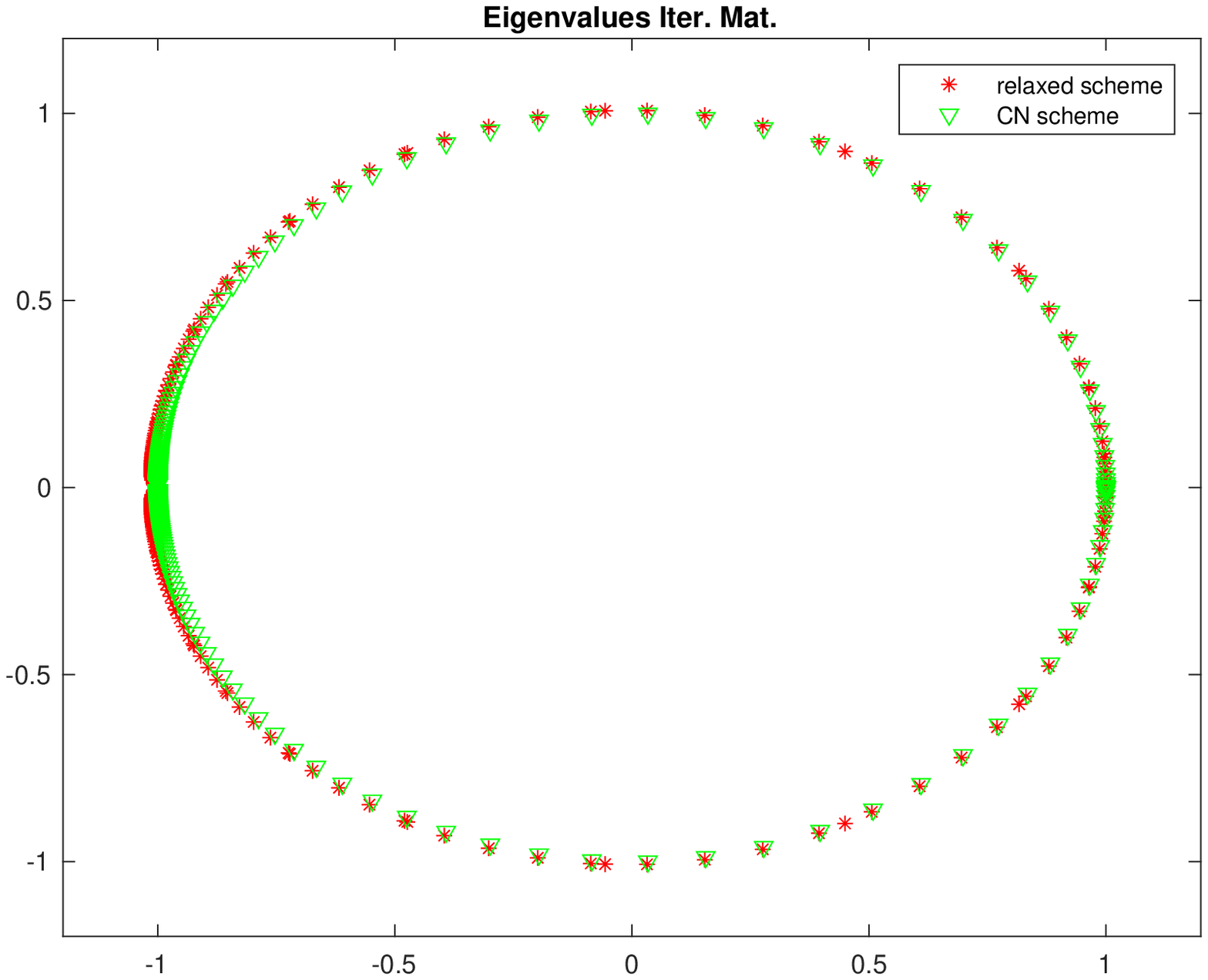}}
  \caption{\small\em Spectra of $\M_{\,r}$ and $\M_{\,\mathrm{CN}}$ matrices. For the spatial discretisation we use $\ell= \ 100$, $N\ =\ 400$ points and the relaxation parameter $\delta\ =\ 10^{-3}\,$. Various values of the time step are taken.}
  \label{Fig7:comp.spect}
\end{figure}

\begin{figure}
  \centering
  \subfigure[$\Delta t\ =\ 10^{\,-3}$]{\includegraphics[width=0.48\textwidth]{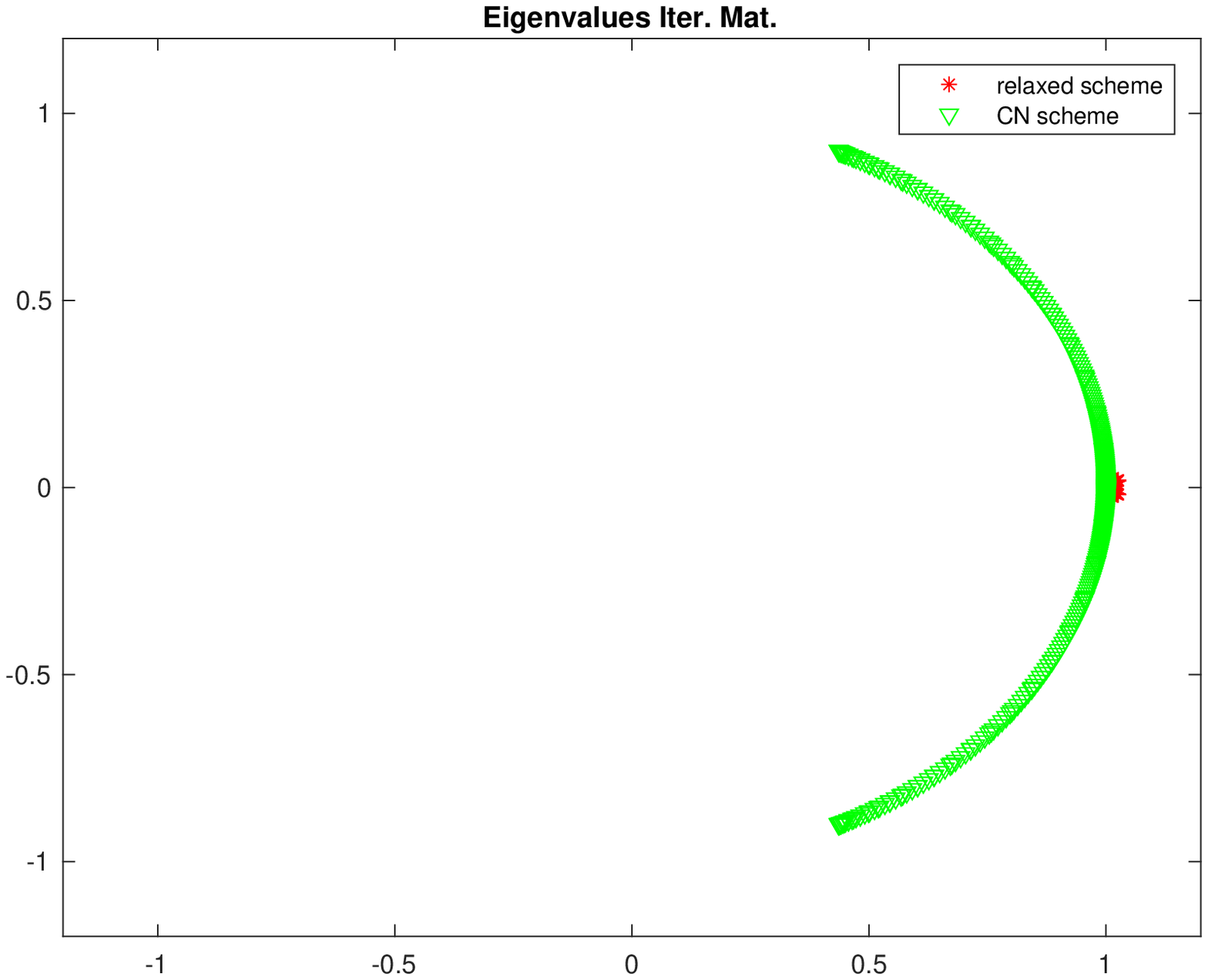}}
  \subfigure[$\Delta t\ =\ 10^{\,-2}$]{\includegraphics[width=0.48\textwidth]{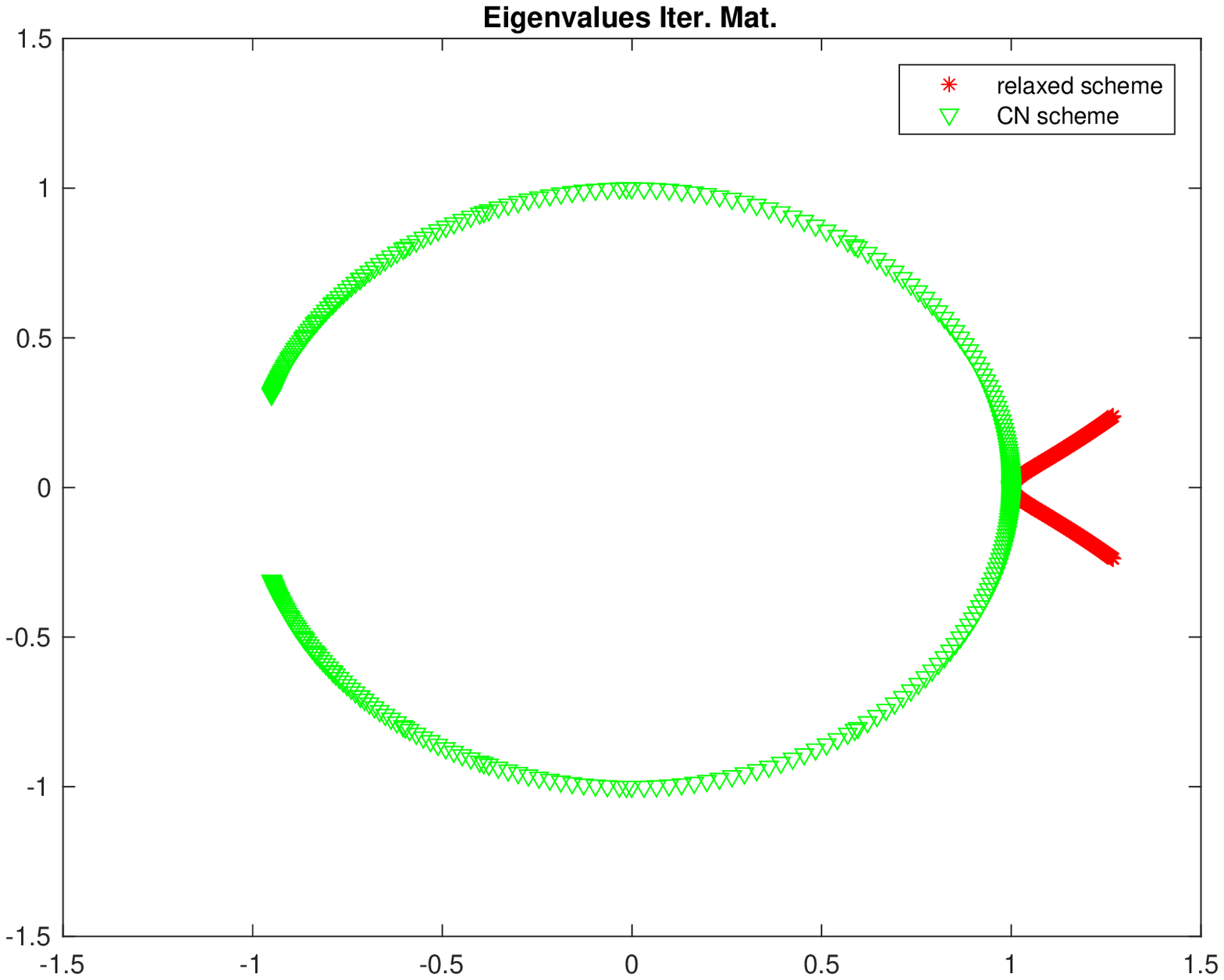}}
  \subfigure[$\Delta t\ =\ 10^{\,-1}$]{\includegraphics[width=0.48\textwidth]{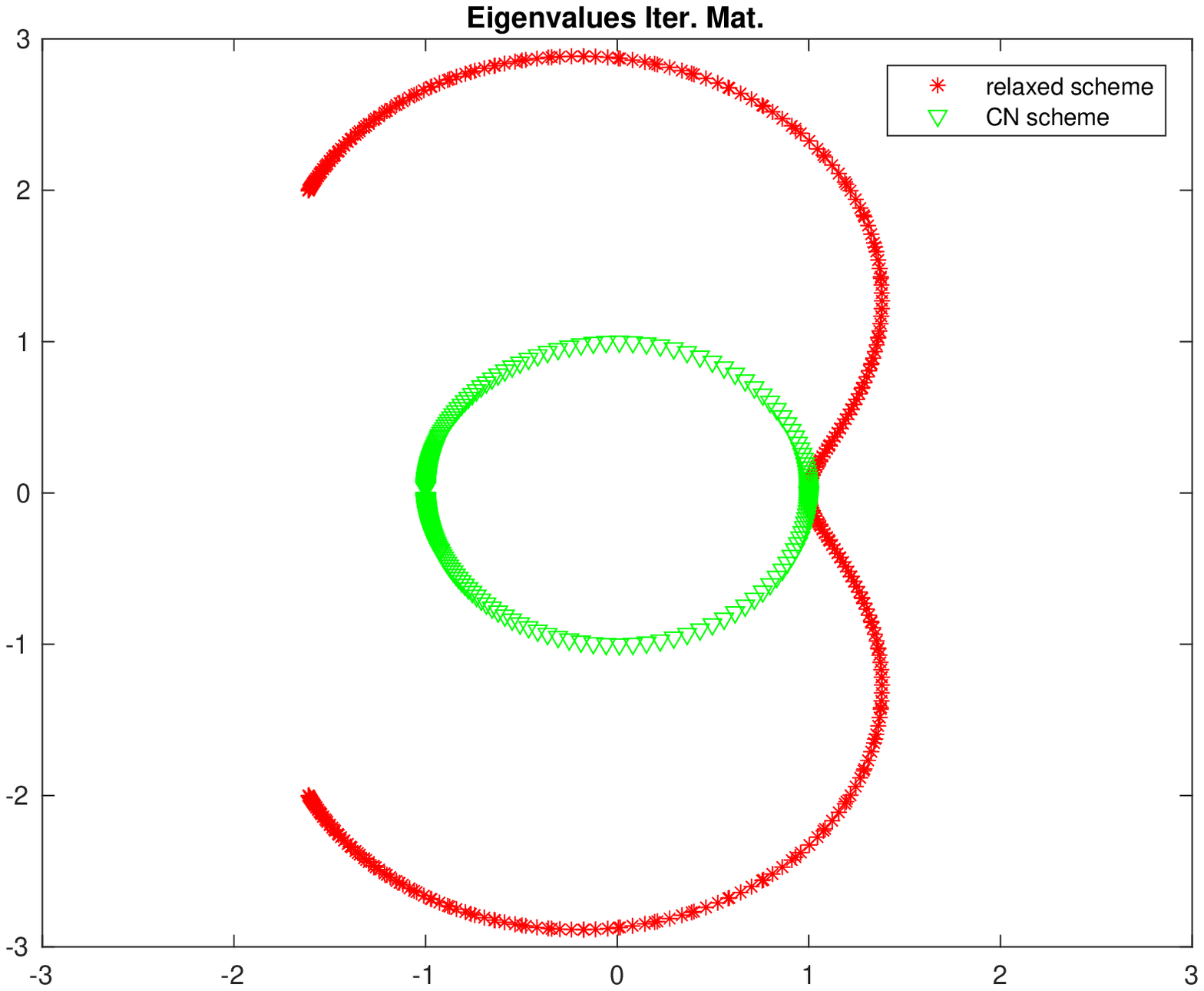}}
  \subfigure[$\Delta t\ =\ 5\times 10^{\,-1}$]{\includegraphics[width=0.48\textwidth]{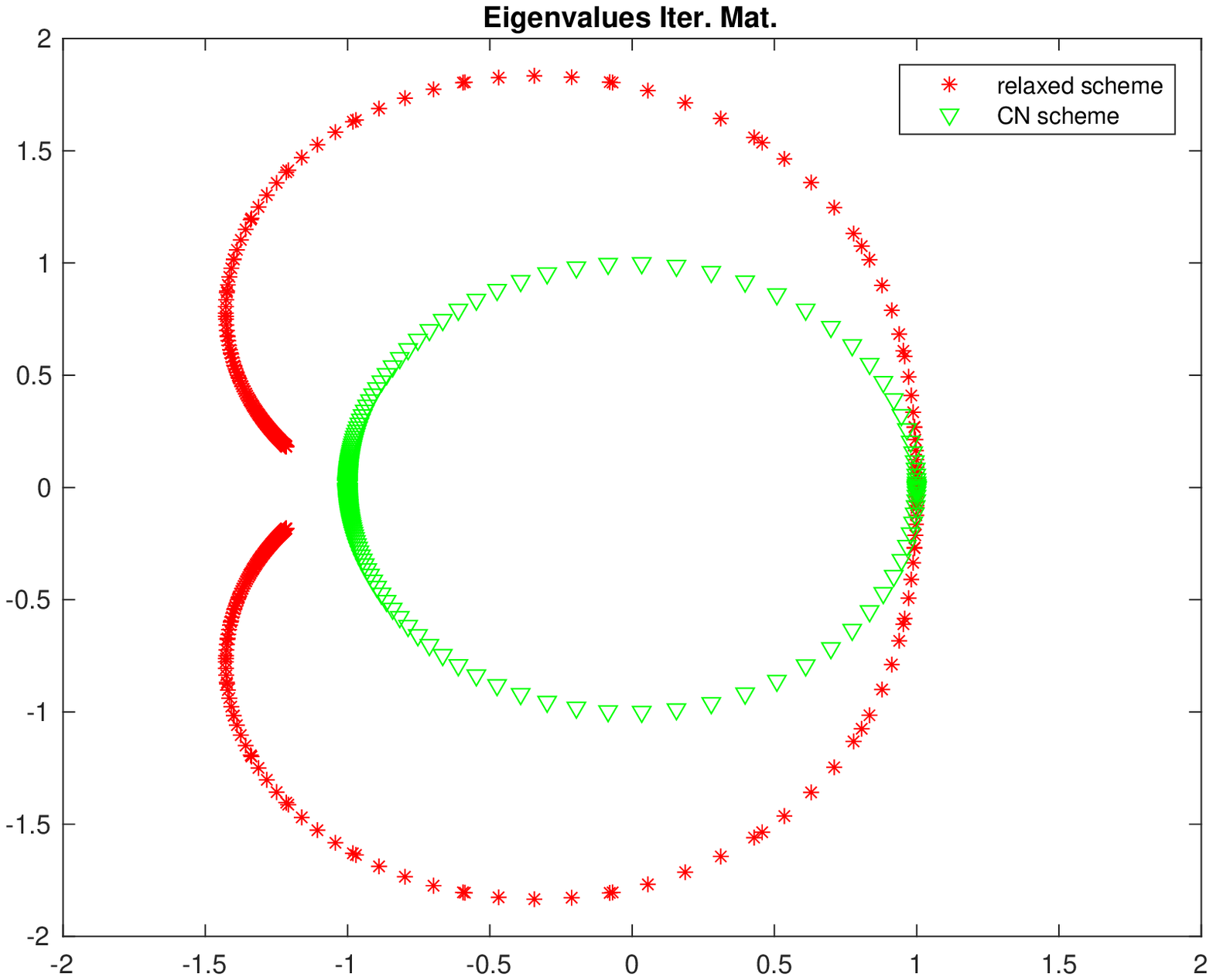}}
  \caption{\small\em Spectra of $\M_{\,r}$ and $\M_{\,\mathrm{CN}}$ matrices. For the spatial discretisation we use $\ell= \ 100$, $N\ =\ 400$ points and the relaxation parameter $\delta\ =\ 10^{-1}\,$. Various values of the time step are taken.}
  \label{Fig8:comp.spect}
\end{figure}

A first way to understand the effect of the relaxation in the stability of the time marching schemes is to consider the linear system \eqref{eq:kdv4}. We have used the 6\up{th} order compact schemes for the discretization of the spatial derivatives, as presented in the beginning of Section~\ref{sec:num}. We compare here in the complex plane the eigenvalues of the two matrices used in the numerical schemes: $\M_{\,r}$ and $\M_{\,\mathrm{CN}}\,$. We recall that these matrices are not of the same size so there is no reason that their relative spectra coincide. The eigenvalues of $\M_{\,\mathrm{CN}}$ are mapped on the unit circle (see Figure~\ref{Fig1:comp.spect}), we present hereafter comparison with those of $\M_{\,r}$ for different values of $N\,$, $\Delta t$ and $\delta$ (the relaxation parameter). At first, we take a very small value of $\delta\ =\ 10^{\,-17}\,$, however the results we obtain are identical all for larger but still small values of $\delta\,$, up to $10^{\,-5}\,$, as illustrated in Figure~\ref{Fig5:comp.spect}. For larger values of $\delta\,$, say $\delta\ =\ 10^{\,-4}\,$, $\delta\ =\ 10^{\,-3}\,$, $\delta\ =\ 10^{\,-1}\,$, eigenvalues of $\M_{\,r}$ are placed outside the unit disk making the relaxed scheme unstable, see Figures~\ref{Fig6:comp.spect}, \ref{Fig7:comp.spect} and \ref{Fig8:comp.spect}.

Below, in Table~\ref{Tab1:compminmax.spect}, we give the maximum and the minimum of the modulus of the eigenvalues of the $\M_{\,r}$ and $\M_{\,\mathrm{CN}}$ matrices for different values of the time step $\Delta t$ and relaxation parameter $\delta\,$. We can observe the influence of $\delta$ on the spectral radius of $\M_{\,r}\;$: the relaxed scheme becomes unstable, \ie $\rho\,(\M_{\,r})\ >\ 1\,$, for not sufficiently small values of $\delta\,$, say, \eg $\delta\ \geqslant\ 10^{\,-4}\,$; for small values of $\delta\,$, we have $\rho\,(\M_{\,r})\ =\ 1\,$, the eigenvalues of $\M_{\,r}$ are perfectly matched on the unit circle and the scheme is then unconditionally stable.

\begin{table}[!h]
  \centering
  \caption{\small\em Minimal and maximal values of eigenvalues of iteration matrices for $N\ =\ 400$ and $\ell\ =\ 100\,$. Here, 6\up{th} order Compact Schemes are used.}
  \bigskip
  \begin{tabular}{|c|c|c||c||c|}
  \hline
  $N$ & $\Delta t$ & $\delta$  & $\M_{\,\mathrm{CN}}$: $\sigma_{\,\mathrm{CN}}$, $\rho_{\,\mathrm{CN}}$ & $\M_{\,r}$: $\sigma_r$, $\rho_r$\\
  \hline
  400 &  $10^{-3}$ & $10^{-17}$ & $\sigma_{\,\mathrm{CN}}=1$, $\rho_{\,\mathrm{CN}}=1$ & $\sigma_{\,r}=1$, $\rho_{\,r} = 1$ \\
  \hline
  400 &  $10^{-2}$ & $10^{-17}$& $\sigma_{\,\mathrm{CN}}=1$, $\rho_{\,\mathrm{CN}}=1$& $\sigma_{\,r}=1$, $\rho_{\,r} = 1$ \\
  \hline
  400 &  $10^{-1}$ & $10^{-17}$& $\sigma_{\,\mathrm{CN}}=1$, $\rho_{\,\mathrm{CN}}=1$& $\sigma_{\,r}=1$, $\rho_{\,r} = 1$ \\
  \hline
  400 & $5\times 10^{-1}$ & $10^{-17}$& $\sigma_{\,\mathrm{CN}}=1$, $\rho_{\,\mathrm{CN}}=1$& $\sigma_{\,r}=1$, $\rho_{\,r} = 1$\\
  \hline  \hline
  400 &  $10^{-3}$ & $10^{-10}$ & $\sigma_{\,\mathrm{CN}}=1$, $\rho_{\,\mathrm{CN}}=1$ & $\sigma_{\,r}=1$, $\rho_{\,r} = 1.000000047$ \\
  \hline
  400 &  $10^{-2}$ & $10^{-10}$& $\sigma_{\,\mathrm{CN}}=1$, $\rho_{\,\mathrm{CN}}=1$& $\sigma_{\,r}=1$, $\rho_{\,r} = 1.00000007$ \\
  \hline
  400 &  $10^{-1}$ & $10^{-10}$& $\sigma_{\,\mathrm{CN}}=1$, $\rho_{\,\mathrm{CN}}=1$& $\sigma_{\,r}=1$, $\rho_{\,r} = 1.000000008$ \\
  \hline
  400 & $5\times 10^{-1}$ & $10^{-10}$& $\sigma_{\,\mathrm{CN}}=1$, $\rho_{\,\mathrm{CN}}=1$& $\sigma_{\,r}=1$, $\rho_{\,r} = 1.0000000015$\\
   \hline  \hline
  400 &  $10^{-3}$ & $10^{-8}$ & $\sigma_{\,\mathrm{CN}}=1$, $\rho_{\,\mathrm{CN}}=1$ & $\sigma_{\,r}=1$, $\rho_{\,r} = 1.000004775$ \\
  \hline
  400 &  $10^{-2}$ & $10^{-8}$& $\sigma_{\,\mathrm{CN}}=1$, $\rho_{\,\mathrm{CN}}=1$& $\sigma_{\,r}=1$, $\rho_{\,r} = 1.00000691$ \\
  \hline
  400 &  $10^{-1}$ & $10^{-8}$& $\sigma_{\,\mathrm{CN}}=1$, $\rho_{\,\mathrm{CN}}=1$& $\sigma_{\,r}=1$, $\rho_{\,r} = 1.000000798$ \\
  \hline
  400 & $5\times 10^{-1}$ & $10^{-8}$& $\sigma_{\,\mathrm{CN}}=1$, $\rho_{\,\mathrm{CN}}=1$& $\sigma_{\,r}=1$, $\rho_{\,r} = 1.0000001599$\\

  \hline\hline
  400 & $10^{-3}$ & $10^{-4}$&$\sigma_{\,\mathrm{CN}}=1$, $\rho_{\,\mathrm{CN}}=1$ &  $\sigma_{\,r}=1$, $\rho_{\,r} = 1.047217$\\
  \hline
  400 & $10^{-2}$ & $10^{-4}$& $\sigma_{\,\mathrm{CN}}=1$, $\rho_{\,\mathrm{CN}}=1$& $\sigma_{\,r}=1$, $\rho_{\,r} = 1.070529$\\
  \hline
  400 & $10^{-1}$ & $10^{-4}$& $\sigma_{\,\mathrm{CN}}=1$, $\rho_{\,\mathrm{CN}}=1$& $\sigma_{\,r}=1$, $\rho_{\,r} = 1.00796$\\
  \hline
  400 &  $5\times 10^{-1}$ & $10^{-4}$& $\sigma_{\,\mathrm{CN}}=1$, $\rho_{\,\mathrm{CN}}=1$& $\sigma_{\,r}=1$, $\rho_{\,r} = 1.001598$\\
  \hline\hline
  400 & $10^{-3}$ & $10^{-3}$&$\sigma_{\,\mathrm{CN}}=1$, $\rho_{\,\mathrm{CN}}=1$ &  $\sigma_{\,r}=1$, $\rho_{\,r} = 1.178956$\\
  \hline
  400 & $10^{-2}$ & $10^{-3}$& $\sigma_{\,\mathrm{CN}}=1$, $\rho_{\,\mathrm{CN}}=1$& $\sigma_{\,r}=1$, $\rho_{\,r} =1.56593$\\
  \hline
  400 & $10^{-1}$ & $10^{-3}$& $\sigma_{\,\mathrm{CN}}=1$, $\rho_{\,\mathrm{CN}}=1$& $\sigma_{\,r} = 1\,$, $\rho_{\,r} = 1.07633$\\
  \hline
  400 & $5\times 10^{-1}$ & $10^{-3}$& $\sigma_{\,\mathrm{CN}}=1$, $\rho_{\,\mathrm{CN}}=1$& $\sigma_{\,r}=1$, $\rho_{\,r} =1.015844$\\
  \hline\hline
  400 &  $10^{-3}$ & $10^{-1}$&$\sigma_{\,\mathrm{CN}}=1$, $\rho_{\,\mathrm{CN}}=1$ &  $\sigma_{\,\,r}=1$, $\rho_{\,r} = 1.02591$\\
  \hline
  400 &  $10^{-2}$ & $10^{-1}$& $\sigma_{\,\mathrm{CN}}=1$, $\rho_{\,\mathrm{CN}}=1$& $\sigma_{\,r}=1$, $\rho_{\,r} = 1.290614$\\
  \hline
  400 &  $10^{-1}$ & $10^{-1}$& $\sigma_{\,\mathrm{CN}}=1$, $\rho_{\,\mathrm{CN}}=1$& $\sigma_{\,r}=1$, $\rho_{\,r} = 2.90107786$\\
  \hline
  400 & $5\times 10^{-1}$ & $10^{-1}$& $\sigma_{\,\mathrm{CN}}=1$, $\rho_{\,\mathrm{CN}}=1$& $\sigma_{\,r}=1$, $\rho_{\,r} = 1.909388$\\
  \hline\hline
  \end{tabular}
  \bigskip  
  \label{Tab1:compminmax.spect}
\end{table}


\subsection{Simulation of a soliton}

To illustrate the effect of the time relaxation, we consider the simulation of a soliton on the domain $I\ =\ \bigl[\,0,\,\ell\,\bigr]\,$, for different values of $\delta$ and of $\Delta t\,$, namely
\begin{equation}\label{eq:kdv}
  u\,(x,\,t)\ =\ a\,\sech^{\,2}\Bigl(\frac{1}{2}\;\kappa\,(x\ -\ x_{\,0}\ -\ c\,t)\Bigr)\,,
\end{equation}
with $c\ =\ \frac{a}{3}$ and $\kappa\ =\ \sqrt{c}\,$. In our experiments presented below we use $a\ =\ 0.8\,$, $\ell\ =\ 100$ and $x_{\,0}\ =\ \dfrac{\ell}{2}\,$. The interval $I$ is discretized with $N\ =\ 400$ equidistant points and the space discretization is realized by using 6\up{th} compact schemes. Of course, due to the nonlinear \emph{orbital} stability properties of the solitons \cite{Weinstein1987}, the approximation to the solitary wave by the numerical solution of the relaxed system cannot be considered in a long time interval, making the validation delicate. Indeed, the classical error measures such as $L_{\,p}$ norms are irrelevant for orbits, meanwhile a numerical method approximates a classical solution and not an orbit.

\begin{figure}
  \centering
  \subfigure[$\Delta t\ =\ 5\times 10^{\,-3}\,$, $\delta\ =\ 10^{\,-17}\,$, $t\ =\ 139$]{\includegraphics[width=0.48\textwidth]{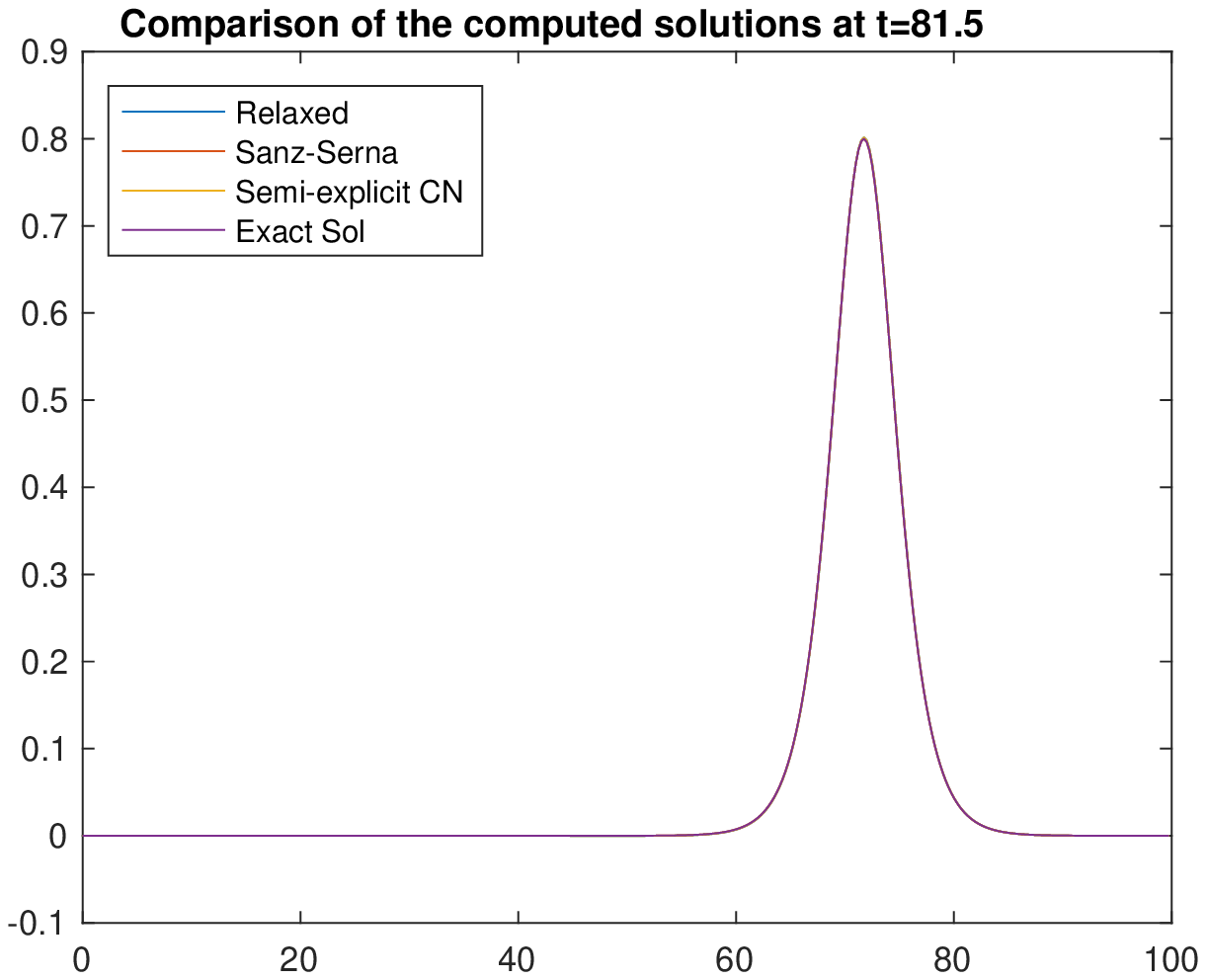}}
  \subfigure[$\Delta t\ =\ 5\times 10^{\,-3}\,$, $\delta\ =\ 10^{\,-17}\,$, $L_{\,2}$ error]{\includegraphics[width=0.48\textwidth]{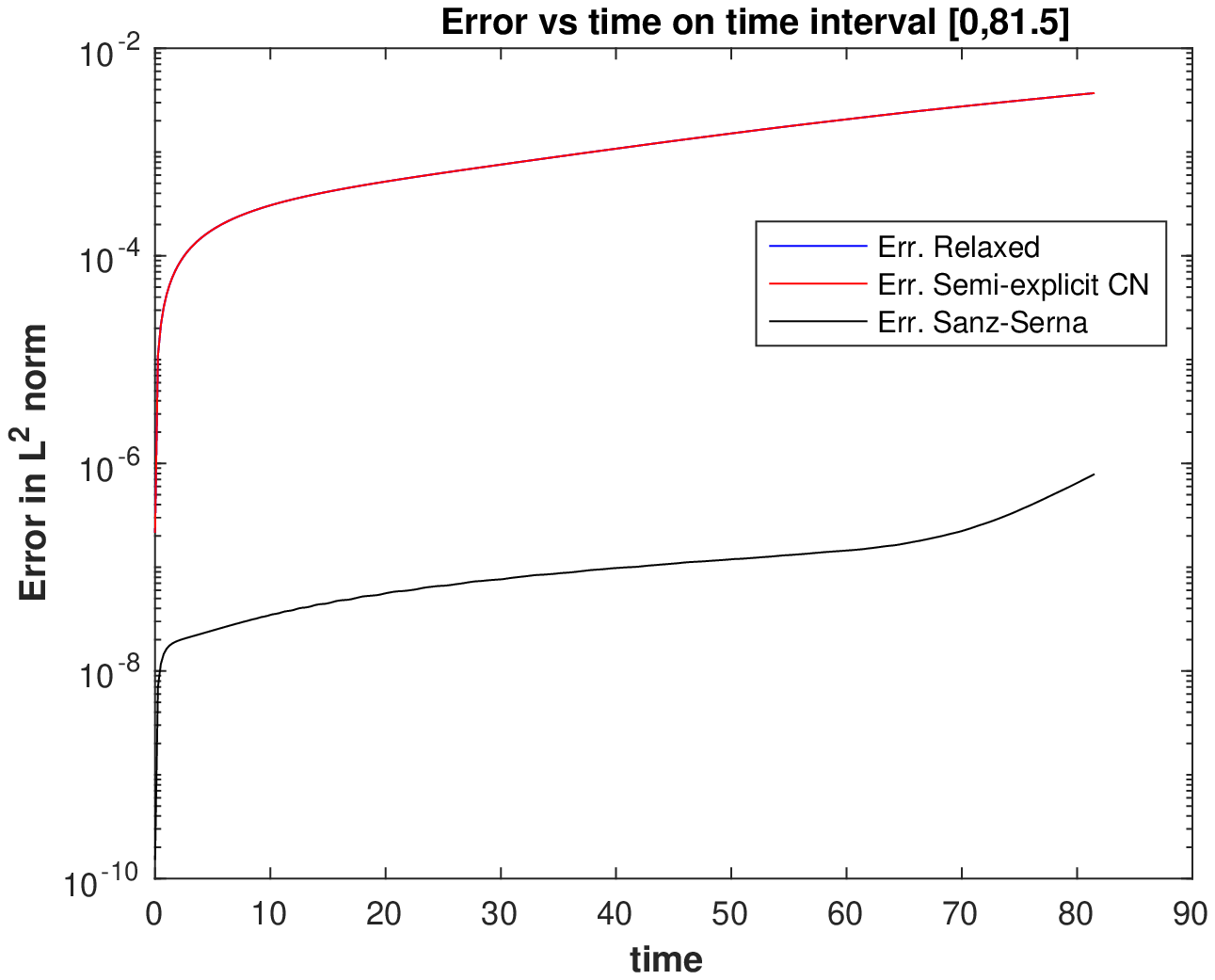}}
  \subfigure[$\Delta t\ =\ 10^{\,-1}\,$, $\delta\ =\ 10^{\,-4}\,$, $t\ =\ 120$]{\includegraphics[width=0.48\textwidth]{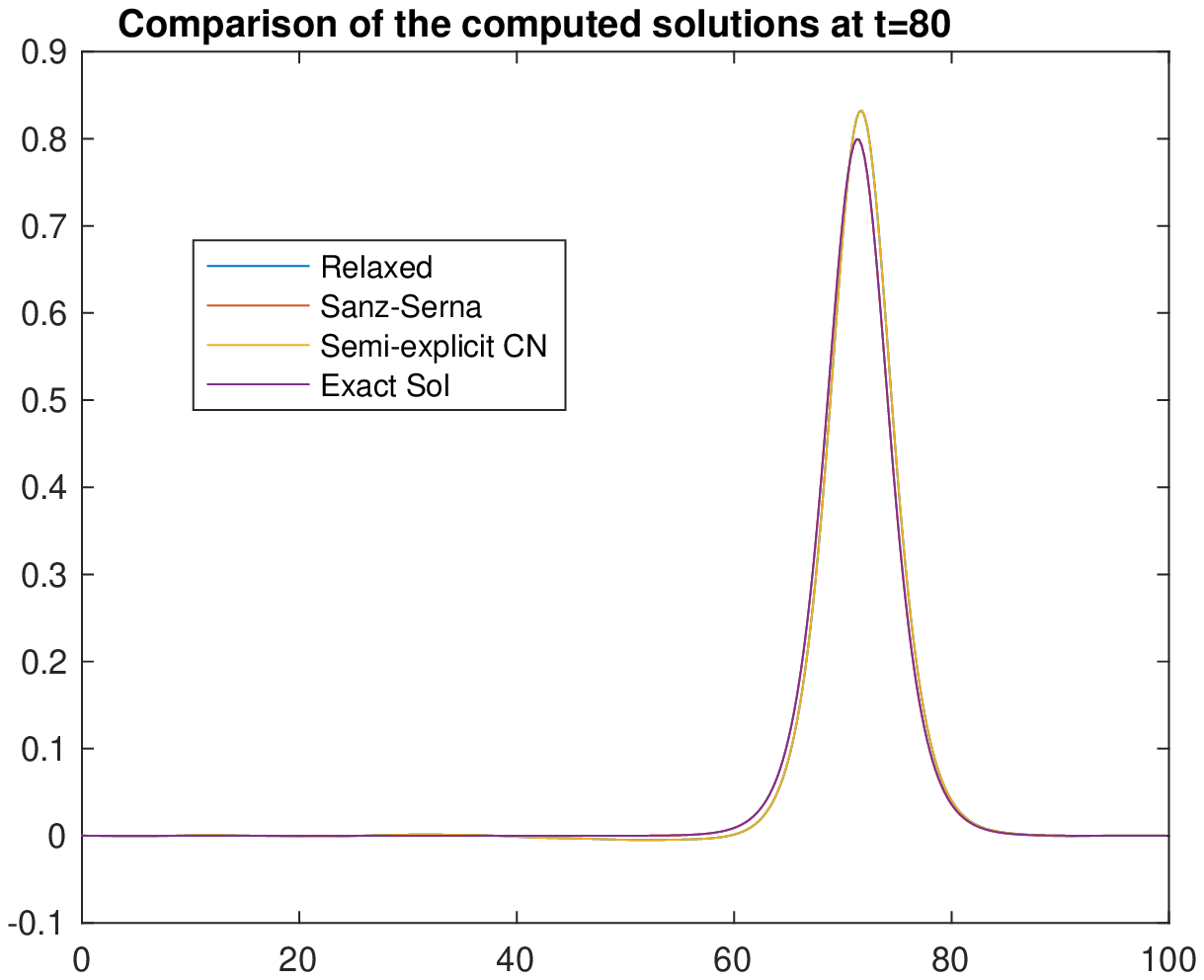}}
  \subfigure[$\Delta t\ =\ 10^{\,-1}\,$, $\delta\ =\ 10^{\,-4}\,$, $L_{\,2}$ error]{\includegraphics[width=0.48\textwidth]{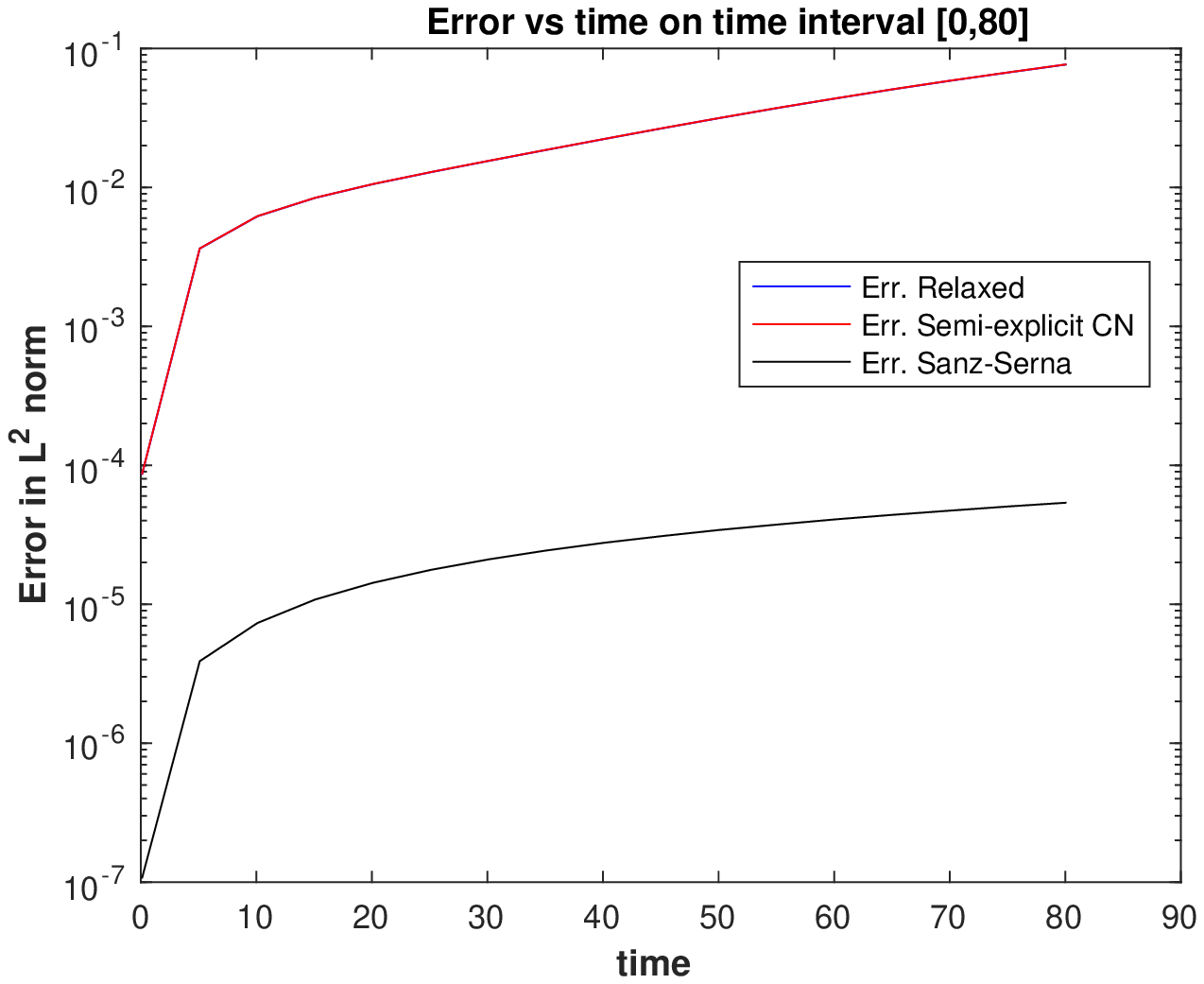}}
  \caption{\small\em Propagation of a solitary wave with relaxed schemes.}
  \label{fig:soliton1}
\end{figure}

We observe that the relaxation allows to approach the exact solution with an expected level of accuracy. For instance, when $\delta\ =\ 10^{\,-\,17}\,$, the solutions computed by the different schemes coincide on a fairly long time interval. With  $\delta\ =\ 10^{\,-\,4}$ and $\Delta t\ =\ 0.1\,$, the solution coincide on a smaller time interval as it is expected. More generally, we notice in our numerical experiments that the pairs $(\delta\,,\, \Delta t)$ that make the relaxed scheme stable in the linear case are operant for the nonlinear \acs{kdv} equation as well\footnote{However, we do not have a mathematical proof for this statement.}. The contrary is also observed, a numerical instability holds, for example, for $\Delta t\ =\ 0.005$ and $\delta\ =\ 0.001\,$, see Table~\ref{Tab1:compminmax.spect}. The numerical result for the soliton propagation are shown in Figure~\ref{fig:soliton1} for two values of the relaxation parameter $\delta\ =\ 10^{\,-\,17}$ and $\delta\ =\ 10^{\,-\,4}\,$. On the right panels we show the evolution of the $L_{\,2}$ error computed thanks to the exact Solution~\eqref{eq:kdv}.

\subsection{Second order relaxed times  schemes}

As illustrates above, relaxed schemes allow a correct numerical integration using only approximation to $\partial_{\,x}$ which is generally easily  available in a number of situations, while the approximation to $\partial_{\,x\,x\,x}$ can be very tricky to build. However, the relaxation scheme is only first order accurate in time and as the semi-implicit \textsc{Euler} scheme, it reveals not  to be suited for long time interval simulations. To overcome this drawback we propose here to reach a second order of accuracy by using a classical \textsc{Richardson} extrapolation. In two words, the numerical time integration of ODE
\begin{equation*}
  \od{u}{t}\ =\ F\,(u)\,,
\end{equation*}
by the forward \textsc{Euler} scheme defines the iterations
\begin{equation*}
  u^{\,k\,+\,1}\ =\ u^{\,k}\ +\ \Delta t\,F\,(u^{\,k})\ =\ G_{\,\Delta t}\,(u^{\,k})\,,
\end{equation*}
which are first order accurate approximations to $u\,(k\,\Delta t)\,$. The \textsc{Richardson} extrapolated sequence is defined by
\begin{eqnarray*}
  v_{\,1}\ &=\ G_{\,\Delta t}\,(u^{\,k})\,, \\
  v_{\,2\,,\,0}\ &=\ G_{\,\Delta t/2}\,(u^{\,k})\,, \\
  v_{\,2\,,\,1}\ &=\ G_{\,\Delta t /2}\,(v_{\,2\,,\,0})\,, \\
  u^{\,k\,+\,1}\ &=\ 2\,v_{\,2\,,\,1}\ -\ v_{\,1}\,,
\end{eqnarray*}
and is second order accurate in time. We will start here from a simple IMEX method, says backward \textsc{Euler} for the linear terms and forward for the nonlinear ones: in other words if $F\,(u)$ writes as $F\,(u)\ =\ -\,A\,u\ +\ H\,(u)\,$, then the propagator is formally defined as
\begin{equation*}
  G_{\,\Delta t}\,(u^{\,k})\ =\ (\Id\ +\ \Delta t\,A)^{\,-1}\cdot\left(u^{\,k}\ +\ \Delta t\,H\,(u^{\,k})\right)\,.
\end{equation*}
The IMEX relaxed scheme consists then in solving the following linear system at each step:
\begin{multline*}
\begin{pmatrix}
  \Id_{\,N} & 0 &-\Delta t\,\D_{\,x} \\
  -\,\Delta t\,\D_{\,x} & (\delta\ +\ \Delta t)\,\Id_{\,N} & 0 \\
  0 & -\,\Delta t\,\D_{\,x} & (\delta\ +\ \Delta t)\,\Id_{\,N} \\
\end{pmatrix}
\cdot
\begin{pmatrix}
  U^{\,(k\,+\,1})\\
  V^{\,(k\,+\,1})\\
  W^{\,(k\,+\,1})
\end{pmatrix}
\ =\  
  \begin{pmatrix}
    U^{\,(k)}\\
    V^{\,(k)}\\
    W^{\,(k)}
  \end{pmatrix}\ +\ 
  \begin{pmatrix}
    -\,\frac{\Delta t}{2}\; \D_{\,x}\, (U^{\,(k)})^{\,2} \\
    0 \\
    0
\end{pmatrix}\,.
\end{multline*}
The matrix of the system is noted by $M^{\,(\mathrm{IMEX})}_{\,\Delta t\,,\,\delta}\,$. The Extrapolated Relaxed schemes writes as

\begin{center}
\begin{minipage}[H]{12cm}
  \begin{algorithm}[H]
    \caption{: Extrapolated Relaxed Scheme}\label{ExtraRS}
    \begin{algorithmic}[1]
        \State $u^{\,(0)}$ given\\
            \For{$k\ =\ 0,\,1,\,\cdots$ until convergence}
             \State {\bf Solve} $ M^{\,(\mathrm{IMEX})}_{\,\Delta t/2,\,\delta}\,v_{\,1}\ =\ -\frac{\Delta t}{2}\,F\,(u^{\,(k})\,$,
               \State {\bf Solve} $ M^{\,(\mathrm{IMEX})}_{\,\Delta t/2,\,\delta}\,v_{\,2}\ =\ -\,\frac{\Delta t}{2}\,F\,(u_{\,1})\,$,
               \State {\bf Solve} $M^{\,(\mathrm{IMEX})}_{\,\Delta t,\,\delta}\,v_{\,3}\ =\ -\,\Delta t\,F\,(u^{\,(k)})\,$,
               \State {\bf Set} $u^{\,(k\,+\,1)}\ =\ 2\,u_{\,2}\ -\ u_{\,3}\,$.              
            \EndFor
    \end{algorithmic}
    \end{algorithm}
\end{minipage}
\end{center}
\bigskip

\begin{figure}
  \centering
  \subfigure[$\Delta t\ =\ 10^{\,-1}\,$, $\delta\ =\ 10^{\,-4}\,$, $t\ =\ 140$]{\includegraphics[width=0.48\textwidth]{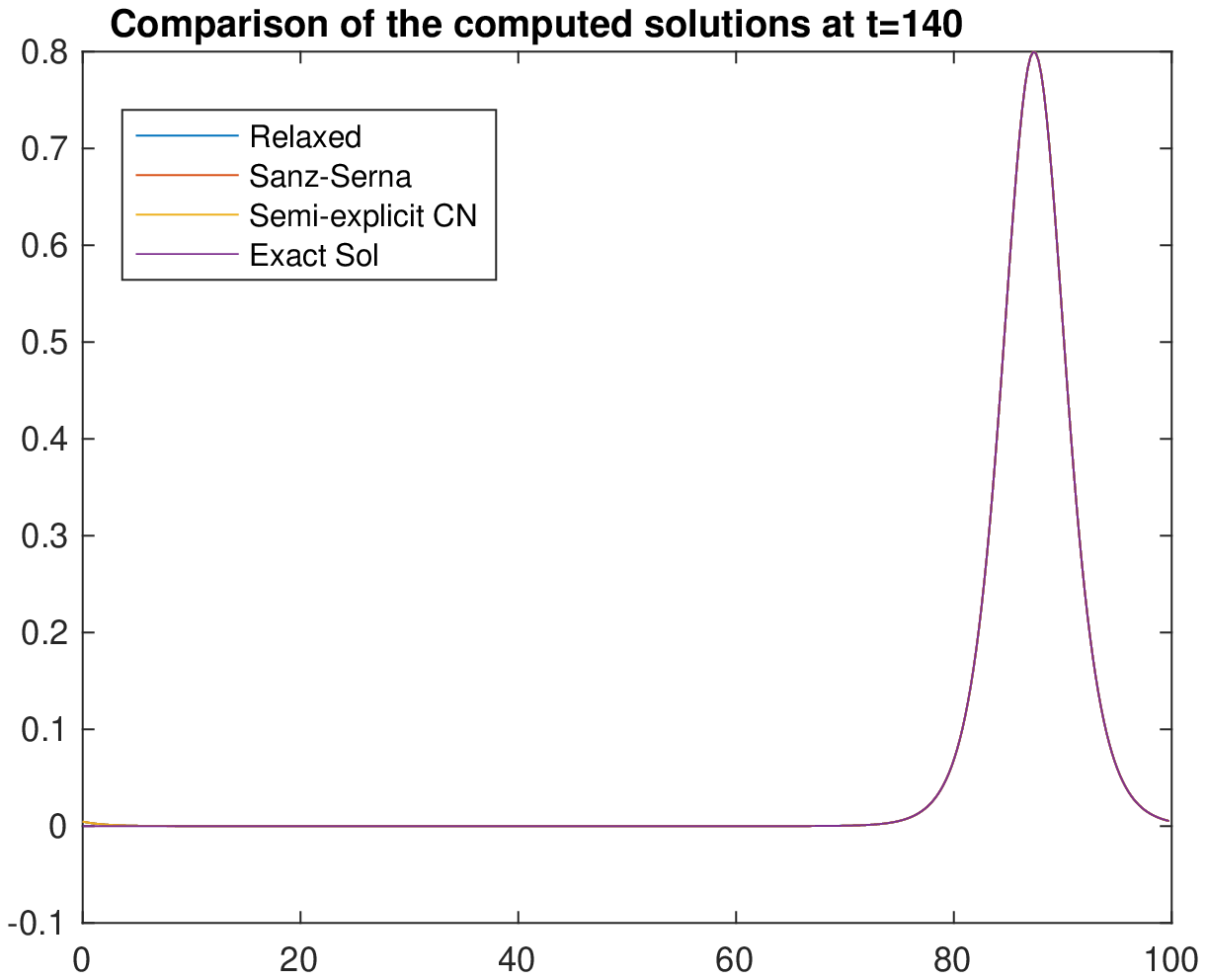}}
  \subfigure[$\Delta t\ =\ 10^{\,-1}\,$, $\delta\ =\ 10^{\,-4}\,$, $t\ =\ 140$]{\includegraphics[width=0.48\textwidth]{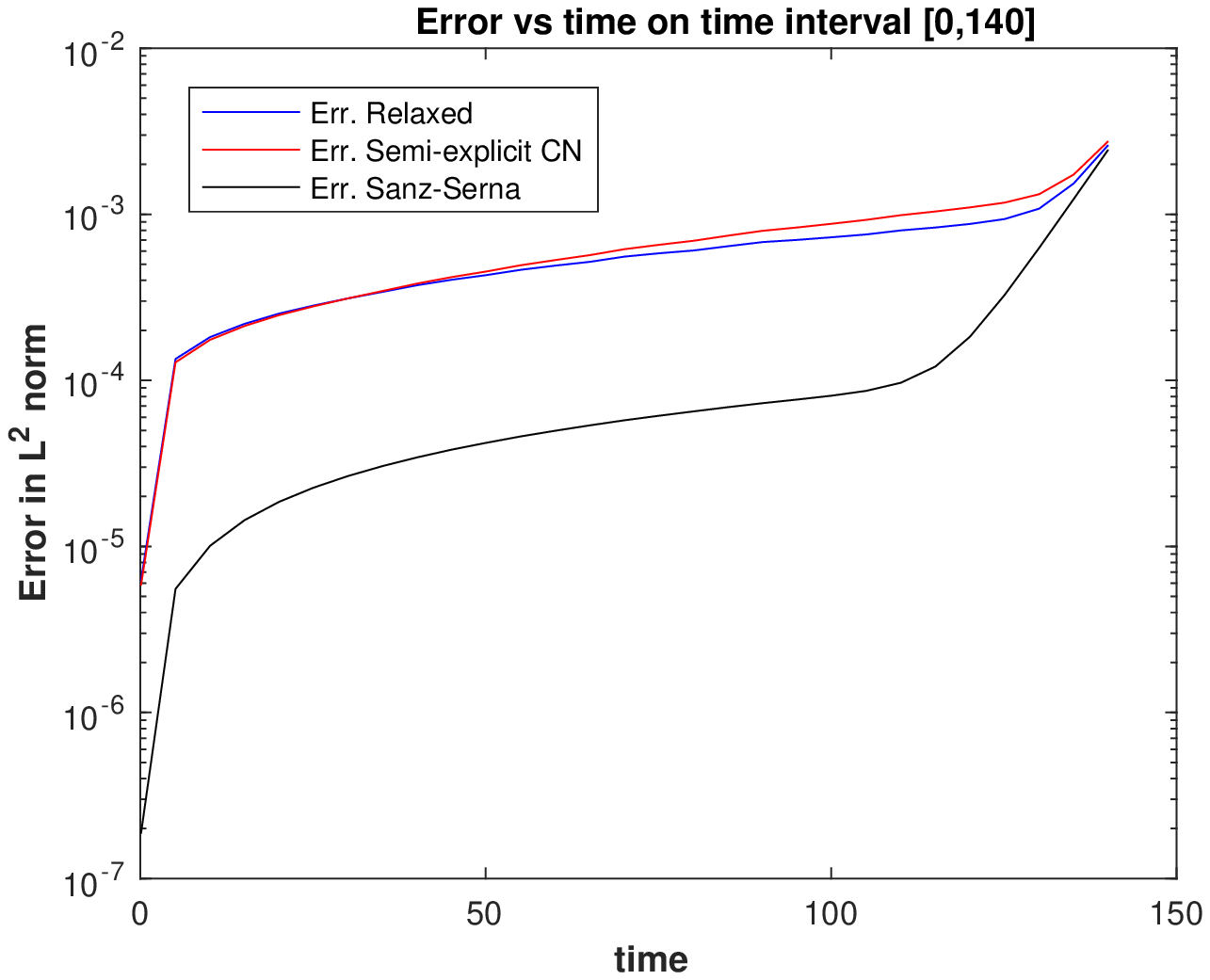}}
   \subfigure[$\Delta t\ =\ 10^{\,-2}\,$, $\delta\ =\ 10^{\,-4}\,$, $t\ =\ 140$]{ \includegraphics[width=0.48\textwidth]{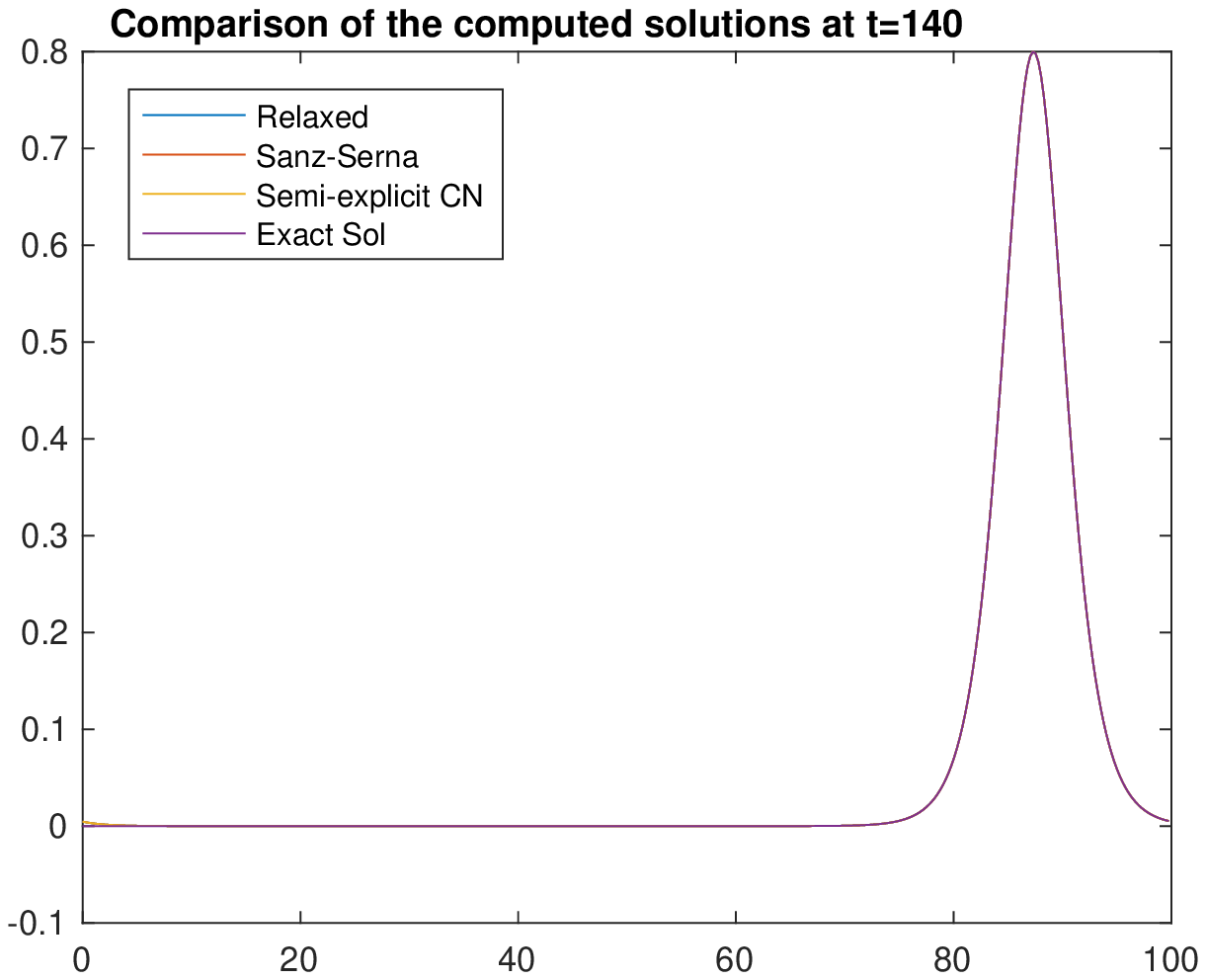}}
   \subfigure[$\Delta t\ =\ 10^{\,-2}$, $\delta\ =\ 10^{\,-4}\,$, $t\ =\ 140$]{ \includegraphics[width=0.48\textwidth]{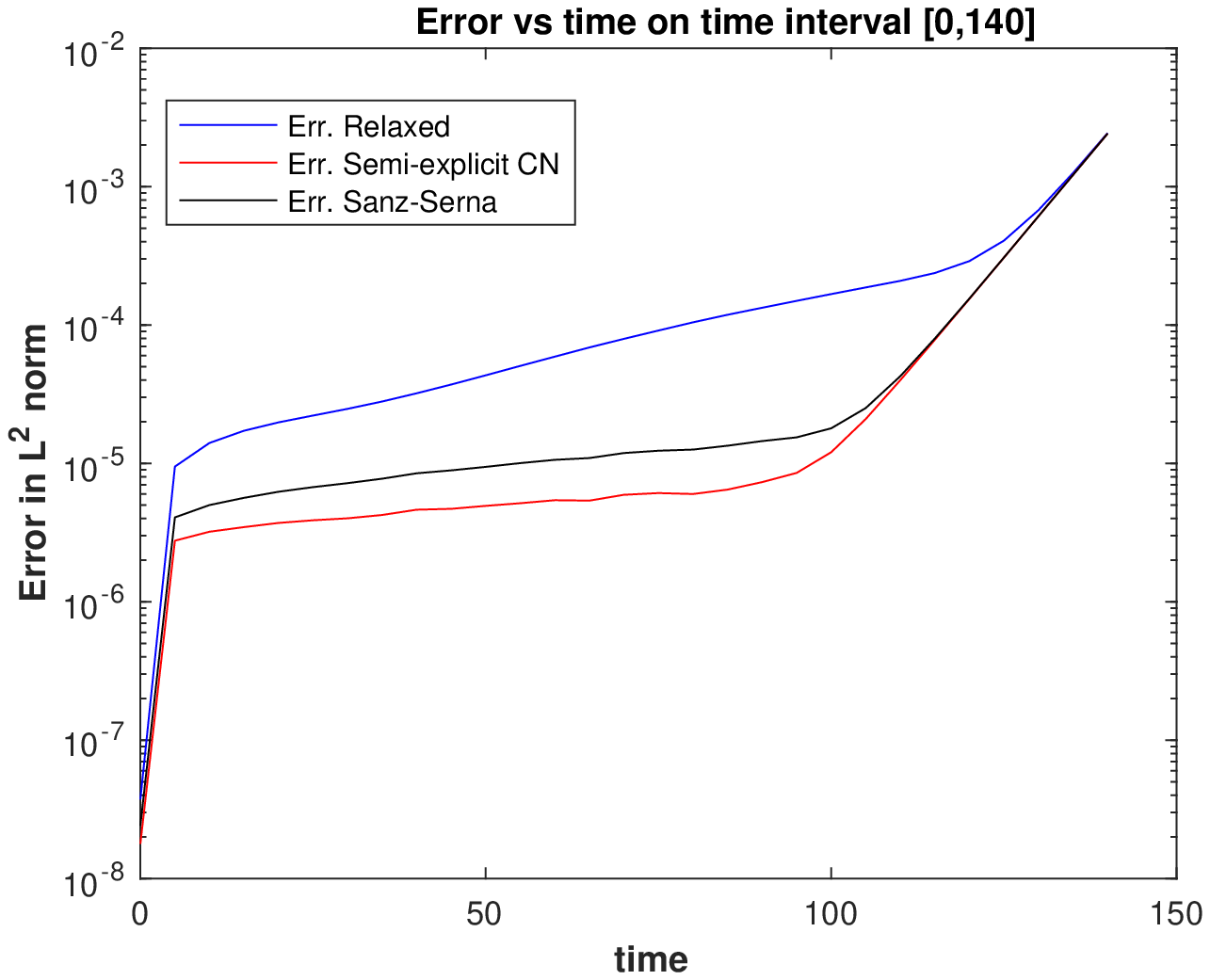}}
  \caption{\small\em Comparison of the exact and numerical solutions to the \acs{kdv} using \textsc{Sanz--Serna} and extrapolated Relaxed and IMEX schemes.}
  \label{fig:T21}
\end{figure}

\begin{figure}
  \centering
  \subfigure[$\Delta t\ =\ 10^{\,-1}\,$, $\delta\ =\ 10^{\,-4}\,$, $t\ =\ 140$]{\includegraphics[width=0.48\textwidth]{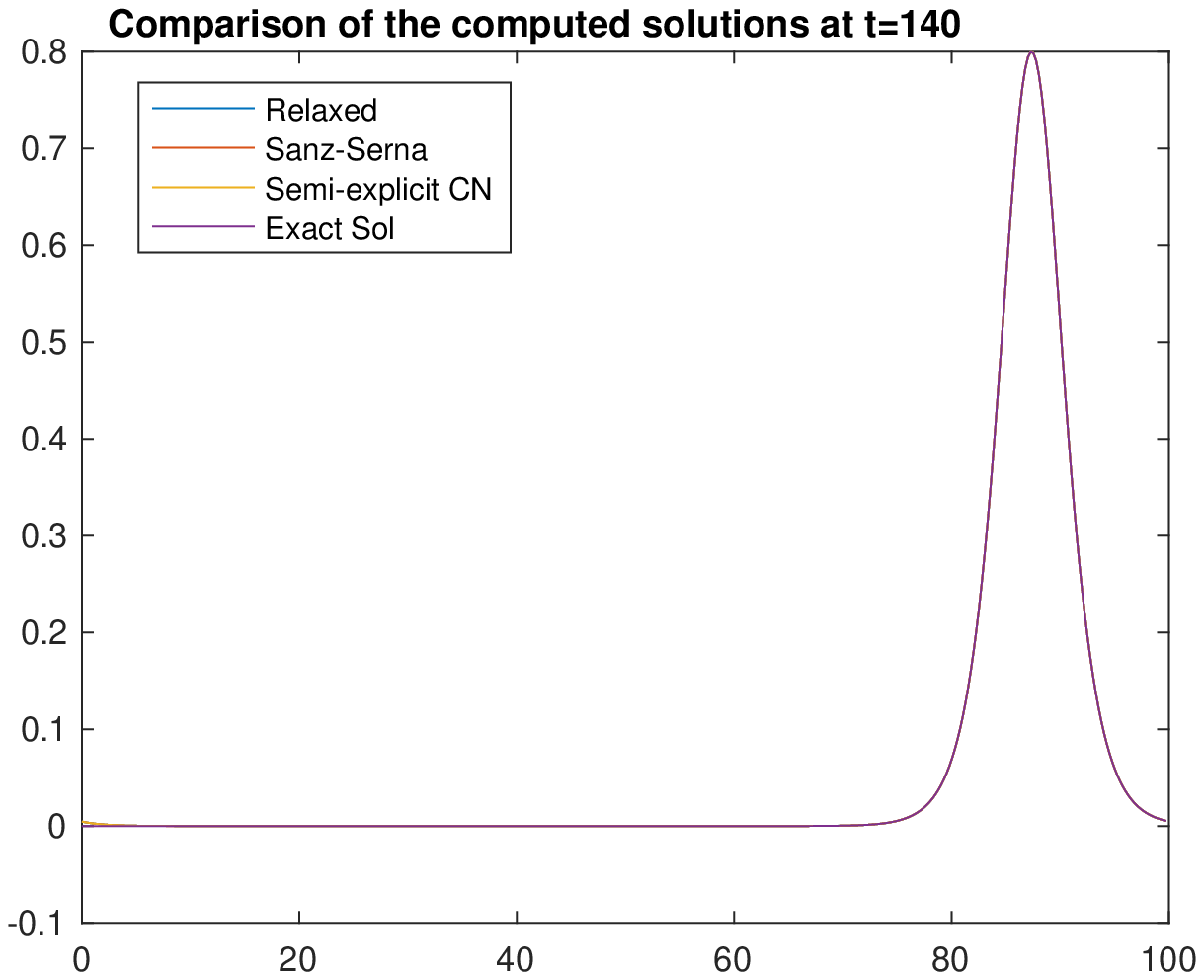}}
  \subfigure[$\Delta t\ =\ 10^{\,-1}\,$, $\delta\ =\ 10^{\,-4}\,$, $t\ =\ 140$]{\includegraphics[width=0.48\textwidth]{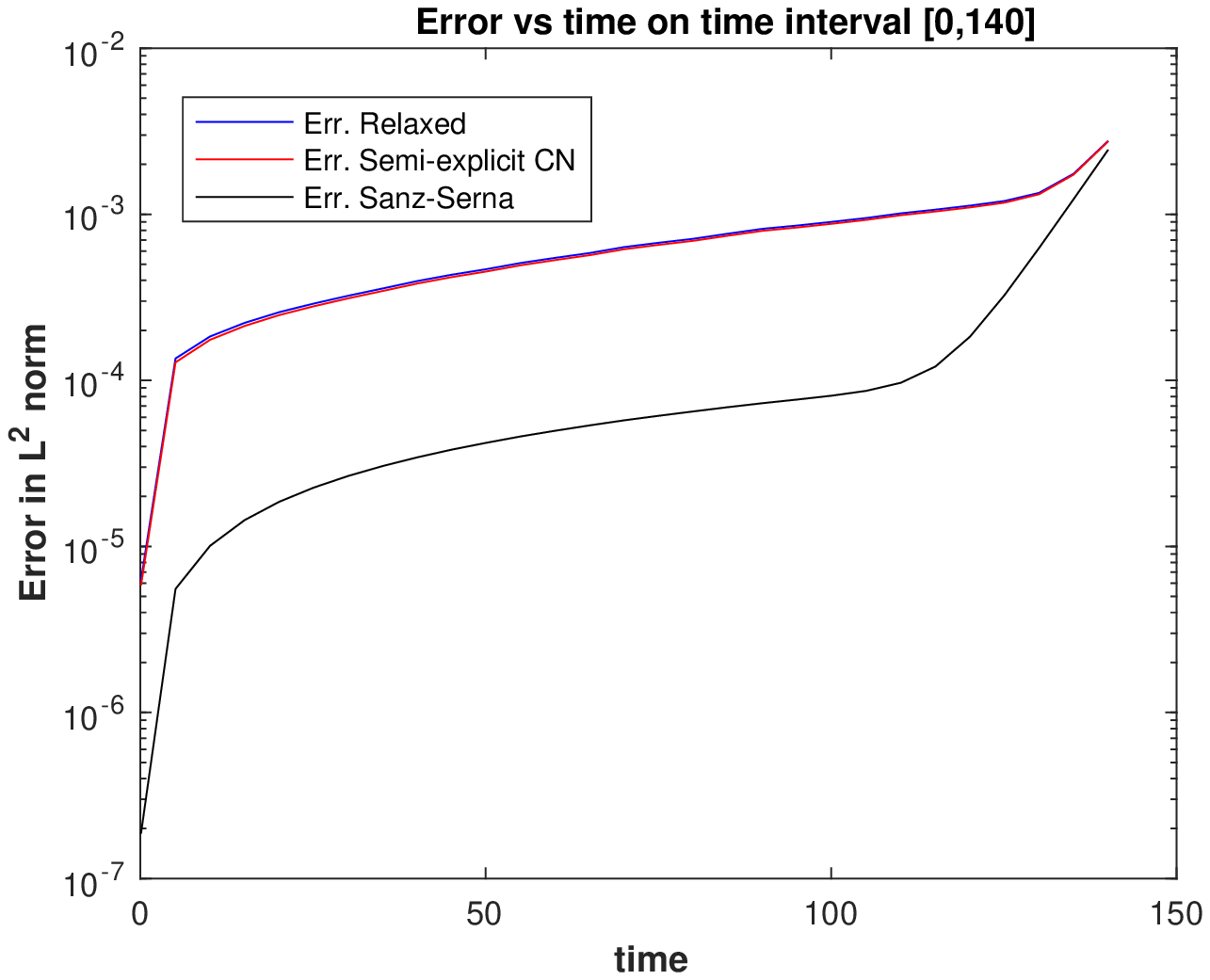}}
    \subfigure[$\Delta t\ =\ 10^{\,-2}\,$, $\delta\ =\ 10^{\,-4}\,$, $t\ =\ 140$]{ \includegraphics[width=0.48\textwidth]{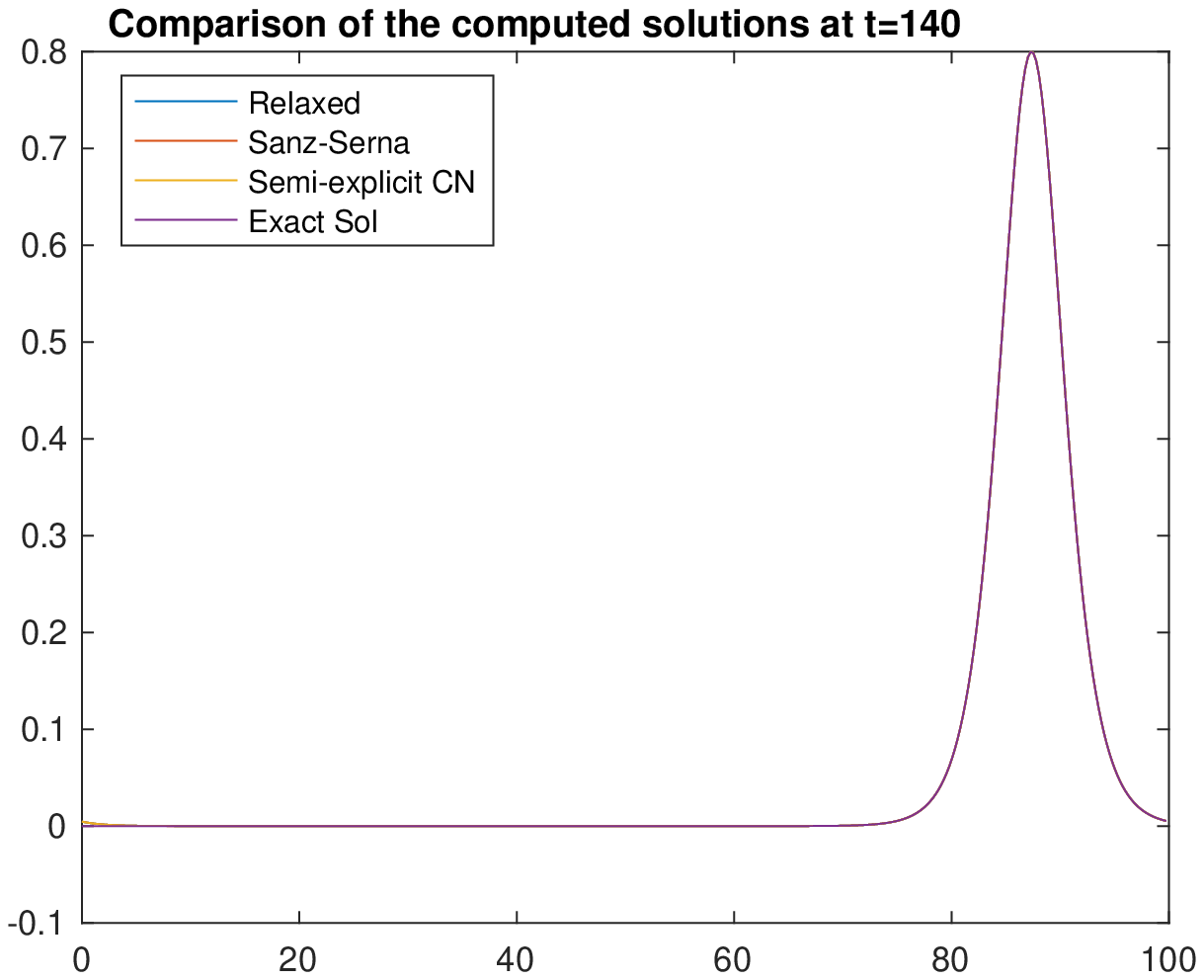}}
  \subfigure[$\Delta t\ =\ 10^{\,-2}$, $\delta\ =\ 10^{\,-4}\,$, $t\ =\ 140$]{ \includegraphics[width=0.48\textwidth]{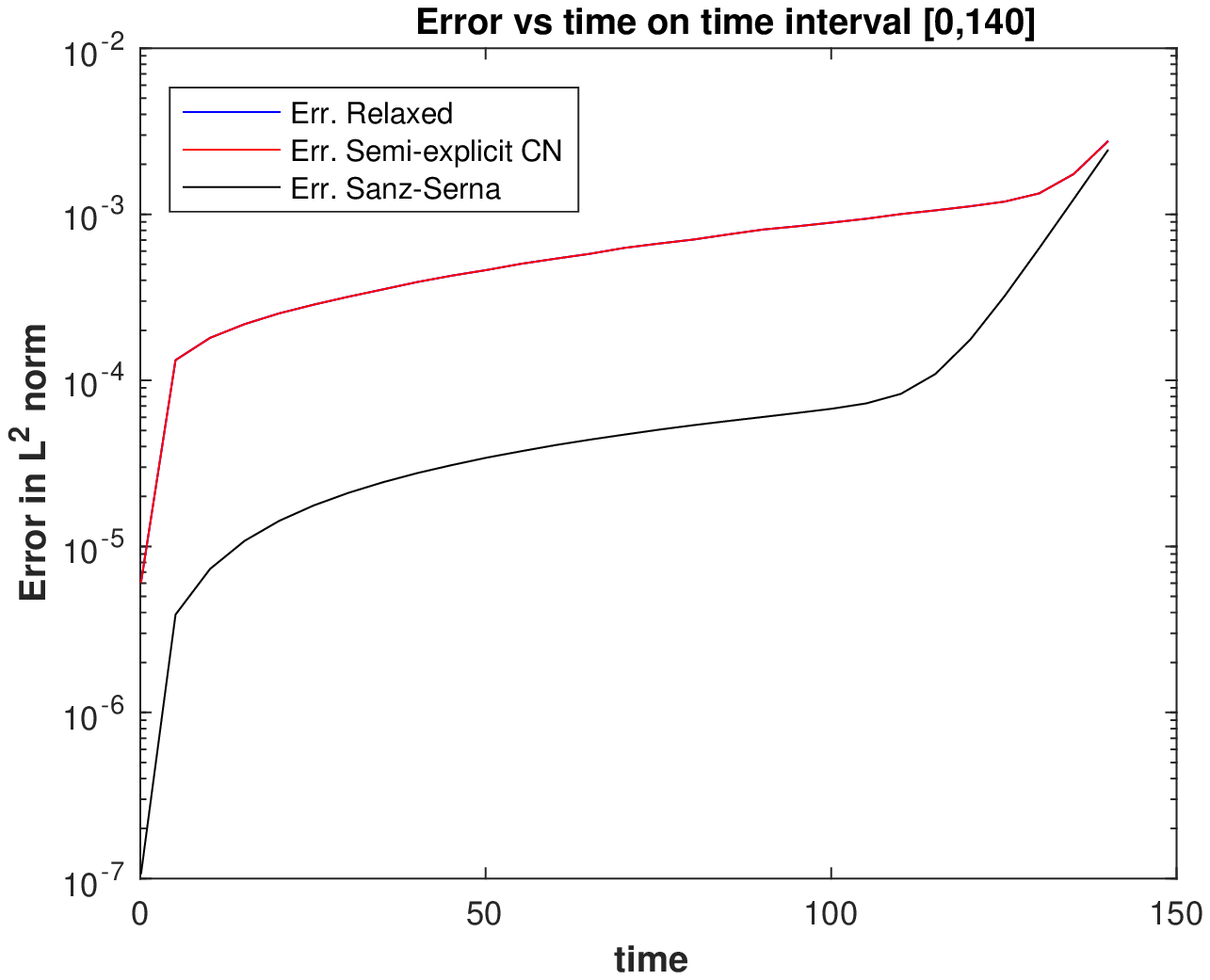}}
  \caption{\small\em Comparison of the exact and numerical solutions to the \acs{kdv} using \textsc{Sanz-Serna} and extrapolated Relaxed and IMEX schemes.}
  \label{fig:T23}
\end{figure}

Hereafter, in Figure~\ref{fig:T21} -- \ref{fig:T23}, we  present the comparison of the evolution of the soliton and its numerical approximations (\textsc{Sanz--Serna} and extrapolated Relaxed and IMEX schemes). We observe that the time extrapolation allows to approach the exact solution with a good accuracy on longer time intervals, when considering different values of $\delta$ and $\Delta t\,$; this is notable
particularly, \eg when taking $\delta\ =\ 10^{\,-4}$ and $\Delta t\ =\ 10^{\,-1}\,$.

\section{Discussion}
\label{sec:disc}

Above we presented some rationale behind time-relaxed formulations for several well-known dispersive wave equations. The main conclusions and perspectives of our study are outlined below.


\subsection{Conclusions}

In this study we presented an approximate reformulation for several dispersive wave equations. This formulation was inspired somehow by quasi-- (or pseudo--) compressibility methods to solve incompressible \textsc{Navier}--\textsc{Stokes} equations \cite{Kameyama2005}. So, by following this `philosophy' we proposed relaxed formulations for celebrated \acf{kdv}, \acf{bbm} and \textsc{Peregrine}'s system of equations. However, it is obvious that the same technique can be extended to many other scalar and vectorial dispersive wave equations. This formulation has a simple advantage to involve first order derivatives only while being in the form of a coupled evolution problem. This is the main difference with various local (or modified) reformulations used in continuous and discontinuous \textsc{Galerkin} methods \cite{Walkley2002, Walkley1999, Levy2004, YS}. Many standard numerical methods can be applied to solve the proposed relaxed formulation numerically. In the present study we applied compact finite differences (with spectral-like resolution) to discretize the problem in space along with a simple time stepping. The presented numerical tests and validations for the \acs{kdv} equation show that the relaxed \emph{discrete} formulation possesses a larger \acs{cfl}--type stability limit comparing to similar compact discretizations of the standard \acs{kdv} equation (without relaxation). This preliminary conclusion indicates that relaxed schemes might be good candidates for the numerical simulation of notoriously \emph{stiff} systems of equations such as the \acs{kdv}--\acs{kdv} \textsc{Boussinesq}-type system extensively studied numerically in \eg \cite{Bona2007}.


\subsection{Perspectives}

Above we presented some numerical illustrations for the classical \acs{kdv} equation only. As it was mentioned in the previous Section, this formulation was extended to other weakly-nonlinear models as well. The main goal consists in extending these time-relaxation numerical methods to \textsc{Boussinesq}-type systems of weakly nonlinear weakly dispersive equations such as the celebrated classical \textsc{Peregrine} \cite{Peregrine1967}, modified \textsc{Peregrine} \cite{Duran2011} or even fully nonlinear \acf{sgn} equations \cite{Serre1953} (some more conventional numerical strategies for these equations were outlined in \eg \cite{Dutykh2011a, Mitsotakis2014, Mitsotakis2014a}).

On a more theoretical side, we would like to obtain the estimations of the difference between the perturbed and unperturbed problems solutions. In other words, we would like to have theoretical arguments to state that the proposed relaxation process generates solutions which remain close to those of the original equation. This is the main point on our `theoretical' agenda.


\subsection*{Acknowledgments}
\addcontentsline{toc}{subsection}{Acknowledgments}

D.~\textsc{Dutykh} would like to acknowledge the hospitality of the Laboratory \textsc{LAMFA} and of the University of \textsc{Picardie} \textsc{Jules Verne} during his visit in November 2016. Reciprocally, J.-P.~\textsc{Chehab} acknowledges the hospitality of the Laboratory of Mathematics (LAMA UMR \#5127) and of the University \textsc{Savoie Mont Blanc} during his visit in December 2017.


\appendix
\section{Other types of dispersive equations}
\label{app:a}

In this Appendix we provide two other examples of relaxation for some widely used dispersive wave equations (scalar and system case).


\subsection{Benjamin--Bona--Mahony equation}
\label{sec:bbm}

The celebrated \acf{bbm} equation was derived first in \cite{Peregrine1966, bona} and can be recast in the following dimensionless form:
\begin{equation*}
  u_{\,t}\ +\ u\,u_{\,x}\ -\ u_{\,xxt}\ =\ 0\,.
\end{equation*}
Similarly to the \acs{kdv} case the variable $u\,(x,\,t)$ can be the free surface elevation or a horizontal fluid velocity either. It is possible to propose a similar relaxed formulation for the \acs{bbm} equation as well. Indeed, let us recast it first in a conservative form:
\begin{equation*}
  \bigl(u\ -\ u_{\,xx}\bigr)_{\,t}\ +\ \bigl(\half\, u^{\,2}\bigr)_x\ =\ 0\,.
\end{equation*}
The 2\up{nd} derivative can be lowered by introducing additional variables:
\begin{align*}
  \bigl(u\ -\ w\bigr)_{\,t}\ +\ \bigl(\half\, u^{\,2}\bigr)_{\,x}\ =&\ 0\,, \\
  u_{\,x}\ =&\ v\,, \\
  v_{\,x}\ =&\ w\,.
\end{align*}
It is convenient to introduce a new variable $\beta\,(x,\,t)\ \eqdef\ \bigl(u\ -\ w\bigr)\,(x,\,t)$ to simplify the first equation:
\begin{align*}
  \beta_{\,t}\ +\ \bigl(\half\, (\,\beta\ +\ w)^{\,2}\bigr)_{\,x}\ =&\ 0\,, \\
  (\beta\ +\ w)_{\,x}\ =&\ v\,, \\
  v_{\,x}\ =&\ w\,.
\end{align*}
The last step consists in adding relaxation terms to obtain an evolutionary system in \emph{all} variables:
\begin{align*}
  \beta_{\,t}\ +\ \bigl(\half\, (\,\beta\ +\ w)^{\,2}\bigr)_{\,x}\ =&\ 0\,, \\
  \delta\, v_{\,t}\ +\ v\ -\ (\,\beta\ +\ w)_x\ =&\ 0\,, \\
  \delta\, w_{\,t}\ +\ w\ -\ v_{\,x}\ =&\ 0\,,
\end{align*}
with $\delta\ \ll\ 1$ being again a small parameter.


\subsection{Peregrine system}

\begin{figure}
  \centering
  \includegraphics[width=0.99\textwidth]{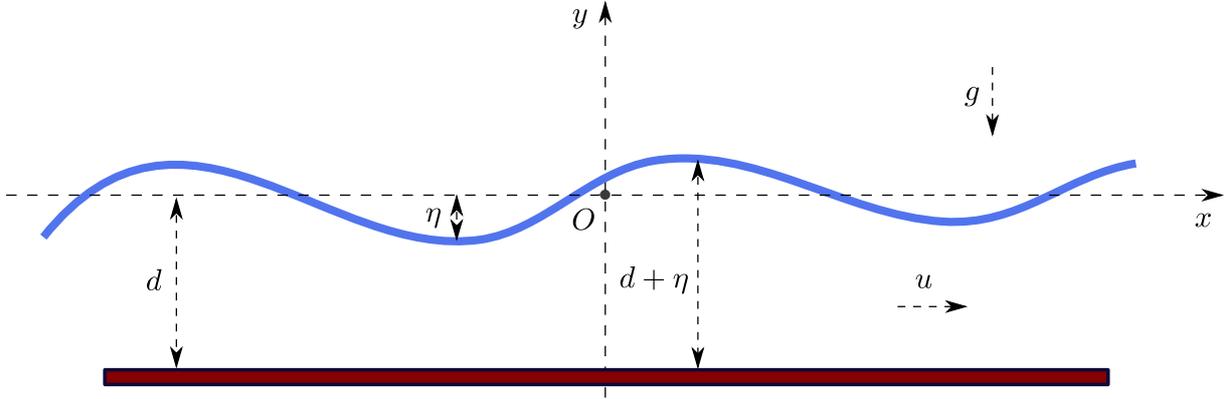}
  \caption{\small\em Sketch of the fluid domain with a free surface.}
  \label{fig:sketch}
\end{figure}

The \textsc{Peregrine} system was proposed by D.~H.~\textsc{Peregrine} (1967) in \cite{Peregrine1967}. In a particular case of a fluid layer of constant depth (\ie the even bottom case) this system reads
\begin{align}\label{eq:per1}
  \eta_{\,t}\ +\ \bigl[\,(d\ +\ \eta\,)\,u\,\bigr]_{\,x}\ =&\ 0\,, \\
  u_{\,t}\ +\ u\,u_{\,x}\ +\ g\,\eta_{\,x}\ -\ \third\, d^{\,2}\, u_{\,xxt} =&\ 0\,, \label{eq:per2}
\end{align}
where $d\ >\ 0$ is the constant fluid depth and $g\ >\ 0$ is the gravity acceleration. The variables $\eta\,(x,\,t)$ and $u\,(x,\,t)$ are the free surface elevation and the depth-averaged horizontal velocity correspondingly. The sketch of the fluid domain is shown in Figure~\ref{fig:sketch}.

As the first step, the \textsc{Peregrine} system \eqref{eq:per1}, \eqref{eq:per2} is rewritten in the following conservative form:
\begin{align*}
  \eta_{\,t}\ +\ \bigl[\,(d\ +\ \eta\,)\,u\,\bigr]_{\,x}\ =&\ 0\,, \\
  \bigl[\,u - \third\, d^{\,2}\,u_{\,xx}\bigr]_{\,t}\ +\ \bigl[\,\half\, u^{\,2}\ +\ g\,\eta\,\bigr]_{\,x}\ =&\ 0\,.
\end{align*}
The last conservative \textsc{Peregrine} system can be lowered and relaxed similar to the \acs{bbm} case (see the Section~\ref{sec:bbm}) by introducing new variables:
\begin{align*}
  \eta_{\,t}\ +\ \bigl[\,(d\ +\ \eta\,)\,(\rho\ +\ \third\, d^{\,2}\, w)\,\bigr]_{\,x}\ =&\ 0\,, \\
  \rho_{\,t}\ +\ \bigl[\,\half\, (\rho\ +\ \third\, d^{\,2}\, w)^{\,2}\ +\ g\,\eta\,\bigr]_{\,x}\ =&\ 0\,, \\
  \delta\, v_{\,t}\ +\ v\ -\ \bigl[\,\rho\ +\ \third\, d^{\,2}\, w\,\bigr]_{\,x}\ =&\ 0\,, \\
  \delta\, w_{\,t}\ +\ w\ -\ v_{\,x}\ =&\ 0\,,
\end{align*}
where $\delta\ \ll\ 1$ and $\rho\,(x,\,t)\ \eqdef\ u\,(x,\,t)\ -\ \third\, d^{\,2} w\,(x,\,t)$ is an auxiliary variable. If one takes the limit $\delta\ \to\ 0\,$, we shall recover a formulation similar to what is used in local (or modified) continuous \textsc{Galerkin} methods \cite{Walkley2002, Walkley1999}.

\begin{remark}
The celebrated \acs{kdv} and \acs{bbm} models can be written for the variable $u\,(x,\,t)$ being the free surface elevation $\eta\,(x,\,t)$ or the horizontal velocity (after a few changes of variables). In this way these equations appear in the theory of water waves. However, the \acs{kdv} and \acs{bbm}-type equations appear in many other physical settings as well (see \eg \cite{Schamel1973, Johnson1997, Grimshaw2002, Fedele2012c, Duran2013}).
\end{remark}


\bigskip\bigskip
\addcontentsline{toc}{section}{References}
\bibliographystyle{abbrv}
\bibliography{biblio}

\begin{thebibliography}{10}

\bibitem{Antonopoulos2010}
D.~C. Antonopoulos, V.~A. Dougalis, and D.~E. Mitsotakis.
\newblock {Galerkin approximations of the periodic solutions of Boussinesq
  systems}.
\newblock {\em Bulletin of Greek Math. Soc.}, 57:13--30, 2010.

\bibitem{Antuono2009}
M.~Antuono, V.~Y. Liapidevskii, and M.~Brocchini.
\newblock {Dispersive Nonlinear Shallow-Water Equations}.
\newblock {\em Stud. Appl. Math.}, 122(1):1--28, 2009.

\bibitem{bona}
T.~B. Benjamin, J.~L. Bona, and J.~J. Mahony.
\newblock {Model equations for long waves in nonlinear dispersive systems}.
\newblock {\em Philos. Trans. Royal Soc. London Ser. A}, 272:47--78, 1972.

\bibitem{Benkhaldoun2008}
F.~Benkhaldoun and M.~Sea{\"{i}}d.
\newblock {New finite-volume relaxation methods for the third-order
  differential equations}.
\newblock {\em Commun. Comput. Phys.}, 4:820--837, 2008.

\bibitem{Bona2007}
J.~L. Bona, V.~A. Dougalis, and D.~E. Mitsotakis.
\newblock {Numerical solution of KdV-KdV systems of Boussinesq equations: I.
  The numerical scheme and generalized solitary waves}.
\newblock {\em Mat. Comp. Simul.}, 74:214--228, 2007.

\bibitem{Bonneton2011}
P.~Bonneton, F.~Chazel, D.~Lannes, F.~Marche, and M.~Tissier.
\newblock {A splitting approach for the fully nonlinear and weakly dispersive
  Green-Naghdi model}.
\newblock {\em J. Comput. Phys.}, 230:1479--1498, 2011.

\bibitem{Boussinesq1877}
J.~V. Boussinesq.
\newblock {Essai sur la th{\'{e}}orie des eaux courantes}.
\newblock {\em M{\'{e}}moires pr{\'{e}}sent{\'{e}}s par divers savants {\`{a}}
  l'Acad. des Sci. Inst. Nat. France}, XXIII:1--680, 1877.

\bibitem{Chen2007}
H.~Chen, M.~Chen, and N.~Nguyen.
\newblock {Cnoidal Wave Solutions to Boussinesq Systems}.
\newblock {\em Nonlinearity}, 20:1443--1461, 2007.

\bibitem{CBB1}
R.~Cienfuegos, E.~Barth{\'{e}}l{\'{e}}my, and P.~Bonneton.
\newblock {A fourth-order compact finite volume scheme for fully nonlinear and
  weakly dispersive Boussinesq-type equations. Part I: Model development and
  analysis}.
\newblock {\em Int. J. Numer. Meth. Fluids}, 51:1217--1253, 2006.

\bibitem{Clamond2003}
D.~Clamond.
\newblock {Cnoidal-type surface waves in deep water}.
\newblock {\em J. Fluid Mech}, 489:101--120, jul 2003.

\bibitem{Duran2013}
A.~Duran, D.~Dutykh, and D.~Mitsotakis.
\newblock {On the Galilean Invariance of Some Nonlinear Dispersive Wave
  Equations}.
\newblock {\em Stud. Appl. Math.}, 131(4):359--388, nov 2013.

\bibitem{Duran2011}
A.~Dur{\'{a}}n, D.~Dutykh, and D.~Mitsotakis.
\newblock {Peregrine's System Revisited}.
\newblock In N.~Abcha, E.~N. Pelinovsky, and I.~Mutabazi, editors, {\em
  Nonlinear Waves and Pattern Dynamics}, pages 3--43. Springer International
  Publishing, Cham, 2018.

\bibitem{Dutykh2011a}
D.~Dutykh, D.~Clamond, P.~Milewski, and D.~Mitsotakis.
\newblock {Finite volume and pseudo-spectral schemes for the fully nonlinear 1D
  Serre equations}.
\newblock {\em Eur. J. Appl. Math.}, 24(05):761--787, 2013.

\bibitem{Dutykh2011e}
D.~Dutykh, T.~Katsaounis, and D.~Mitsotakis.
\newblock {Finite volume schemes for dispersive wave propagation and runup}.
\newblock {\em J. Comput. Phys.}, 230(8):3035--3061, apr 2011.

\bibitem{Dutykh2010e}
D.~Dutykh, T.~Katsaounis, and D.~Mitsotakis.
\newblock {Finite volume methods for unidirectional dispersive wave models}.
\newblock {\em Int. J. Num. Meth. Fluids}, 71:717--736, 2013.

\bibitem{Eskilsson2005}
C.~Eskilsson and S.~J. Sherwin.
\newblock {Discontinuous Galerkin Spectral/hp Element Modelling of Dispersive
  Shallow Water Systems}.
\newblock {\em J. Sci. Comput.}, 22:269--288, 2005.

\bibitem{Fedele2012c}
F.~Fedele and D.~Dutykh.
\newblock {Vortexons in axisymmetric Poiseuille pipe flows}.
\newblock {\em EPL}, 101(3):34003, feb 2013.

\bibitem{Grimshaw2002}
R.~Grimshaw.
\newblock {Internal Solitary Waves}.
\newblock In R.~Grimshaw, editor, {\em Environmental Stratified Flows}, pages
  1--27. Springer US, 2002.

\bibitem{Ismail2000}
M.~S. Ismail.
\newblock {A finite difference method for Korteweg-de Vries like equation with
  nonlinear dispersion}.
\newblock {\em International Journal of Computer Mathematics}, 74(2):185--193,
  2000.

\bibitem{Johnson1997}
R.~S. Johnson.
\newblock {\em {A modern introduction to the mathematical theory of water
  waves}}.
\newblock Cambridge University Press, Cambridge, 1997.

\bibitem{Kameyama2005}
M.~Kameyama, A.~Kageyama, and T.~Sato.
\newblock {Multigrid iterative algorithm using pseudo-compressibility for
  three-dimensional mantle convection with strongly variable viscosity}.
\newblock {\em J. Comput. Phys}, 206:162--181, 2005.

\bibitem{KdV}
D.~J. Korteweg and G.~de~Vries.
\newblock {On the change of form of long waves advancing in a rectangular
  canal, and on a new type of long stationary waves}.
\newblock {\em Phil. Mag.}, 39(5):422--443, 1895.

\bibitem{Lele1992}
S.~K. Lele.
\newblock {Compact finite difference schemes with spectral-like resolution}.
\newblock {\em J. Comp. Phys.}, 103(1):16--42, nov 1992.

\bibitem{Levy2004}
D.~Levy, C.-W. Shu, and J.~Yan.
\newblock {Local discontinuous Galerkin methods for nonlinear dispersive
  equations}.
\newblock {\em J. Comput. Phys.}, 196(2):751--772, 2004.

\bibitem{Mitsotakis2014a}
D.~Mitsotakis, D.~Dutykh, and J.~Carter.
\newblock {On the nonlinear dynamics of the traveling-wave solutions of the
  Serre system}.
\newblock {\em Wave Motion}, 70:166--182, apr 2017.

\bibitem{Mitsotakis2014}
D.~Mitsotakis, B.~Ilan, and D.~Dutykh.
\newblock {On the Galerkin/Finite-Element Method for the Serre Equations}.
\newblock {\em J. Sci. Comput.}, 61(1):166--195, feb 2014.

\bibitem{Peregrine1966}
D.~H. Peregrine.
\newblock {Calculations of the development of an undular bore}.
\newblock {\em J. Fluid Mech.}, 25(02):321--330, mar 1966.

\bibitem{Peregrine1967}
D.~H. Peregrine.
\newblock {Long waves on a beach}.
\newblock {\em J. Fluid Mech.}, 27:815--827, 1967.

\bibitem{Porubov2006}
A.~V. Porubov and G.~A. Maugin.
\newblock {Propagation of localized longitudinal strain waves in a plate in the
  presence of cubic nonlinearity}.
\newblock {\em Phys. Rev. E}, 74:46617, 2006.

\bibitem{Schamel1973}
H.~Schamel.
\newblock {A modified Korteweg-de Vries equation for ion acoustic wavess due to
  resonant electrons}.
\newblock {\em J. Plasma Phys.}, 9(03):377--387, mar 1973.

\bibitem{Serre1953}
F.~Serre.
\newblock {Contribution {\`{a}} l'{\'{e}}tude des {\'{e}}coulements permanents
  et variables dans les canaux}.
\newblock {\em La Houille blanche}, 8:374--388, 1953.

\bibitem{Stoker1957}
J.~J. Stoker.
\newblock {\em {Water Waves: The mathematical theory with applications}}.
\newblock Interscience, New York, 1957.

\bibitem{Walkley1999}
M.~Walkley and M.~Berzins.
\newblock {A finite element method for the one-dimensional extended Boussinesq
  equations}.
\newblock {\em Int. J. Num. Meth. Fluids}, 29(2):143--157, jan 1999.

\bibitem{Walkley2002}
M.~Walkley and M.~Berzins.
\newblock {A finite element method for the two-dimensional extended Boussinesq
  equations}.
\newblock {\em Int. J. Num. Meth. Fluids}, 39(10):865--885, aug 2002.

\bibitem{Weinstein1987}
M.~I. Weinstein.
\newblock {Existence and dynamic stability of solitary wave solutions of
  equations arising in long wave propagation}.
\newblock {\em Comm. Partial Diff. Eqns.}, 12(10):1133--1173, jan 1987.

\bibitem{YS}
J.~Yan and C.-W. Shu.
\newblock {A local discontinuous Galerkin method for KdV type equations}.
\newblock {\em SIAM J. Num. Anal.}, 40:769--791, 2002.

\end{thebibliography}
\bigskip\bigskip

\end{document}